\documentclass[aps,prd,reprint,superscriptaddress,showpacs]{revtex4-1}

\usepackage{mathtools} 
\usepackage{hyperref} 

\begin{document}


\title{Conservative 3+1 General Relativistic Boltzmann Equation}



\author{Christian Y. Cardall}
\affiliation{Physics Division, Oak Ridge National Laboratory, Oak Ridge, Tennessee 37831-6354, USA}
\affiliation{Department of Physics and Astronomy, University of Tennessee, Knoxville, Tennessee 37996-1200, USA}
\author{Eirik Endeve}
\affiliation{Computer Science and Mathematics Division, Oak Ridge National Laboratory, Oak Ridge, Tennessee 37831-6164, USA}
\author{Anthony Mezzacappa}
\affiliation{Department of Physics and Astronomy, University of Tennessee, Knoxville, Tennessee 37996-1200, USA}
\affiliation{Joint Institute for Computational Sciences, Oak Ridge National Laboratory, Oak Ridge, Tennessee 37831-6173, USA}


\date{\today}

\begin{abstract}
We present a new derivation of the conservative form of the general relativistic Boltzmann equation and specialize it to the 3+1 metric.
The resulting transport equation is intended for use in simulations involving numerical relativity, particularly in the absence of spherical symmetry.
The independent variables are lab frame coordinate basis spacetime position components and comoving frame curvilinear momentum space coordinates. 
With an eye towards astrophysical applications---such as core-collapse supernovae and compact object mergers---in which the fluid includes nuclei and/or nuclear matter at finite temperature, and in which the transported particles are neutrinos, we examine the relationship between lepton number and four-momentum exchange between neutrinos and the fluid.    
\end{abstract}


\maketitle


\section{Introduction \label{sec:Introduction}}

Neutrino transport is a necessary ingredient of core-collapse supernova simulations \cite{Mezzacappa2005ASCERTAINING-TH,Kotake2006Explosion-mecha,Kotake2012Multimessengers,Kotake2012Core-Collapse-S,Janka2012Explosion-Mecha,Burrows2012Perspectives-on,Janka2012Core-collapse-s}. 
Determining the fate of the stellar material---for instance, does an explosion happen, and if so, how?---requires calculation of the four-momentum and lepton number exchange between the fluid (which includes nuclei and/or nuclear matter at finite temperature) and the neutrinos that stream from and through it. 
For the purpose of studying the explosion mechanism, we take the traditional approach and consider only massless neutrinos described by classical distribution functions (phase space densities) $f(t,\mathbf{x},\mathbf{p})$ \footnote{Calculation of the emerging neutrino signals---of intrinsic interest as an observational probe of the core-collapse supernova environment, and of the properties of the neutrinos themselves---definitely requires treatment of the quantum effects induced by neutrino mass and flavor mixing \cite{Dasgupta2010Physics-and-Ast,Raffelt2011New-opportuniti,Dighe2011Signatures-of-s}.
Recent explorations suggest that flavor mixing does not impact the explosion mechansim \cite{Chakraborty2011No-Collective-N,Suwa2011Impacts-of-Coll,Dasgupta2012Role-of-collect,Sarikas2012Suppression-of-,Saviano2012Stability-analy,Sarikas2012Supernova-neutr,Pejcha2012Effect-of-colle}, 
but consensus on the impacts of flavor mixing in supernovae has a fickle history,
and future more definitive simulations that include neutrino transport with quantum kinetics could surprise us with flavor mixing effects on the explosion mechanism as well.}.
By way of introduction to the Boltzmann equation satisfied by $f(t,\mathbf{x},\mathbf{p})$, we discuss at some length the choices of reference frames for the spacetime and momentum space coordinates; and more briefly, conservative formulations and efforts to maintain consistency of lepton number and four-momentum exchange with the fluid.

The form of the Boltzmann equation governing distribution functions $f(t,\mathbf{x},\mathbf{p})$ depends on the choices of spacetime and momentum space coordinates.
For an overview of possible effects, consider the Boltzmann equation expressed as
\begin{equation}
\frac{\partial{f}}{\partial t} + \mathbf{v} \cdot \nabla_{\mathbf{x}} f + \mathbf{F} \cdot \nabla_{\mathbf{p}} f = \left(\frac{\partial{f}}{\partial t}\right)_{\mathrm{collisions}},
\label{eq:BoltzmannSimple}
\end{equation}
in which the number density of particles at position $\mathbf{x}$ with momentum $\mathbf{p}$ changes in time because (a) they stream away from $\mathbf{x}$ with velocity $\mathbf{v} = d\mathbf{x} / dt = \mathbf{p} / E(\mathbf{p})$ (where $E(\mathbf{p})$ is the particle energy); (b) their momenta are gradually altered by long-range (i.e. electromagnetic and/or gravitational) forces $\mathbf{F} = d{{\mathbf{p}}} / dt$; and (c) their momenta are abruptly altered by point-like collisions.
Naturally the precise expression of the gradients depends on the choice of coordinate systems. But less trivially, curvilinear spatial coordinates and rotating reference frames respectively would induce `ficticious' and Coriolis force contributions to $\mathbf{F}$. 
And as discussed further below, one may wish to reckon the components of $\mathbf{x}$ and $\mathbf{p}$ with respect to two different reference frames in relative motion; this further complicates the left-hand side with momentum-(component)-changing contributions to $\mathbf{F}$, associated for instance with angular aberrations and Doppler shifts. 
Sometimes called `observer corrections,' these terms involve the relative velocity of the two frames, and---in the case of accelerated relative motion---its gradients.
On the importance of such effects, see for instance Ref.~\cite{Lentz2012On-the-Requirem}.

Lindquist \cite{Lindquist1966Relativistic-Tr} introduced the use of distinct coordinate systems for spacetime and momentum space in the context of general relativity, through use of an orthonormal basis or `tetrad.'
Given a spacetime metric in a coordinate basis (also called a `natural' or `holonomic' basis), an orthonormal basis can be chosen at each spacetime point, in terms of which the metric locally takes the Minkowski form.  
Lindquist began by writing the general relativistic Boltzmann equation in terms of such an orthonormal basis.
He then re-expressed all time and space derivatives---$\partial f / \partial t$ and $\nabla_{\mathbf{x}} f$ in Eq.~(\ref{eq:BoltzmannSimple}), as well as the metric and tetrad derivatives entering $\mathbf{F}$---in terms of the original coordinate basis. However, all momentum components, and the momentum space gradient $\nabla_{\mathbf{p}} f$ in Eq.~(\ref{eq:BoltzmannSimple}), were left in terms of the orthonormal basis.
Thus the independent variables were the spacetime position components in the coordinate basis and the momentum components in the orthonormal basis. 

What purposes did this use of distinct coordinate systems for spacetime and momentum space serve?
An obvious advantage of an orthonormal basis in momentum space---not mentioned by Lindquist, and rarely (if ever) noted explicitly, perhaps because of its obviousness---is that, as a local Minkowski frame, it facilitates direct importation of flat-space results for particle interaction rates into the collision integral.
As for Lindquist's purposes, one was to make contact with familiar expressions from classical transport theory, and in some respects the use of an orthonormal momentum basis helped with this.
He also noted that the `tangent bundle' (spacetime plus the tangent space at every point of spacetime) is a natural way to conceptualize phase space in general relativity; in this context, in which momentum space is identified with the (flat) tangent space at each spacetime point, an orthonormal basis for momentum space also seems natural.   
But for Lindquist, a primary benefit of the `tetrad formalism' was not necessarily the orthonormality per se, but the mere fact that introduction of a second coordinate system allowed simultaneous exploitation of the differing symmetries of the spacetime and the associated momentum space he considered (i.e. spherical symmetry in position space, vs. axial symmetry about the radial direction in momentum space). 

While Lindquist used different bases for position and momentum space, he did not mention the possibility that these coordinate systems could be in relative motion.
Having in mind spherically symmetric gravitational collapse, it is perhaps not surprising that in specializing his Boltzmann equation he considered only comoving coordinates in both position space and momentum space.

The possibility of relative motion between the position space and momentum space coordinate systems apparently arose first in the context of special relativity.
Castor \cite{Castor1972Radiative-Trans} specialized Lindquist's comoving frame results in spherical symmetry \cite{Lindquist1966Relativistic-Tr} to flat spacetime, keeping only terms valid to $\mathcal{O}(V/c)$ in the fluid speed $V$.
Regarding this, Mihalas \cite{Mihalas1980Solution-of-the,Mihalas1981A-Comment-on-Ra} noted that the collision integrals were most easily evaluated in the comoving frame, for this is where the distributions of particle species in equilibrium (e.g. in the fluid, but also the transported species at high optical depth) were isotropic. 
But he expressed concern that it is only in an inertial frame that ``synchronism of clocks can be effected,'' and that use of inertial-frame time and space coordinates ``obviates the need to develop a metric for accelerated fluid frames (Castor 1972) which in general can be done only approximately.''
If anywhere, such concerns ultimately could be relevant only within the confines of special relativity, but they nevertheless motivated Mihalas to ``leave both the {\em space and time variables in the inertial frame}'' (emphasis in original; while there are an infinite number of inertial frames, ``the'' inertial frame he used was a `lab frame' at rest relative to the origin of some spherically symmetric physical system).
In Ref.~\cite{Mihalas1980Solution-of-the}, consistent with a focus on special relativity, he eschewed Lindquist's general relativistic machinery.
Instead, after writing down the lab frame transfer equation, he used the Lorentz transformation properties of the radiation variables to obtain a radiation transport equation in spherical symmetry, valid for $0 \le V/c \le 1$, whose independent variables were the lab frame time and radius and the comoving frame particle energy and direction angle.

The most important motivation to reckon spacetime coordinates in an Eulerian frame was one beyond the focus of Ref.~\cite{Mihalas1980Solution-of-the} on spherical symmetry: 
in multiple spatial dimensions, mesh-based treatments of the fluid with which the particles interact are often formulated in an Eulerian frame, because such phenomena as differential rotation, convection, and turbulence result in mesh stretching and entanglement in a Lagrangian approach.
Using to some extent general relativistic perspectives, flat spacetime transport equations in full dimensionality, expressed in terms of lab frame spacetime coordinates and comoving frame momentum space coordinates, were obtained by Buchler \cite{Buchler1983Radiation-trans,Buchler1986Radiation-trans} and by Kaneko et al. \cite{Kaneko1984The-comoving-fr} to $\mathcal{O}(V/c)$, and by Munier and Weaver for $0 \le V/c \le 1$ \cite{Munier1986Radiation-trans,Munier1986Radiation-trans2}.

Returning to Lindquist \cite{Lindquist1966Relativistic-Tr} as a point of departure, a fully general relativistic Boltzmann equation, expressed in terms of lab frame coordinate basis spacetime coordinates and orthonormal comoving frame momentum variables, was given independently by Riffert \cite{Riffert1986A-general-Euler} and by Mezzacappa and Matzner \cite{Mezzacappa1989Computer-simula}.
While they did not express it quite this way, the basic point in connection with Lindquist's approach is this:
the transformation from a coordinate basis to an orthonormal basis---that is, the local diagonalization of the metric to the special relativistic Minkowski form at each point in spacetime---is unique only up to rotations in space{\em time}, which includes {\em boosts} as well as spatial rotations.
In other words, even if one's coordinate basis is some sort of Eulerian lab frame, one is still free to choose Lindquist's tetrad of orthonormal basis vectors to be one carried by a Lagrangian observer, one comoving with the fluid and therefore (in general) undergoing accelerated motion relative to the lab frame.
This can be conceptualized as two sequential transformations, first from the lab frame coordinate basis to an orthonormal lab frame basis, followed by a Lorentz boost \cite{Mezzacappa1989Computer-simula}; 
or directly as a single transformation, in which the timelike orthonormal basis vector is the fluid four-velocity \cite{Riffert1986A-general-Euler} (see also Ref.~\cite{Misner1973Gravitation} and, in connection with comoving spacetime coordinates, Ref.~\cite{Schinder1988General-relativ}).
Mathematically, acceleration of the Lagrangian observer appears as velocity derivatives in the comoving frame connection coefficients entering $\mathbf{F}$ in Eq.~(\ref{eq:BoltzmannSimple}).

The presentation of Mezzacappa and Matzner \cite{Mezzacappa1989Computer-simula} in particular, couched as it is in terms of the geometric concepts of the 3+1 approach---and conceived with a view towards multidimensional simulations in which spacetime evolution, hydrodynamics, and Boltzmann neutrino radiation transport are all treated in full general relativity---can be regarded as the inspiration for the extension given in the present work.
Full elaboration of the general relativistic Boltzmann equation in lab frame spacetime coordinates and comoving frame momentum variables was only given in spherical symmetry in Refs.~\cite{Riffert1986A-general-Euler,Mezzacappa1989Computer-simula}, and the complexity was already nontrivial in that case.
Beyond the mere multiplication of dimensions, contributing to the complexity in the multidimensional case is a non-diagonal spatial metric that effectively must be diagonalized in the transformation to an orthonormal frame.
In the face of an inherently complicated equation, we strive to minimize pain and maximize utility to readers, by aiming for an appropriate balance between maintaining simplicity and clarity while keeping enough detail to make numerical implementation straightforward.

We now briefly discuss conservative formulations and efforts to maintain numerical consistency of lepton number and four-momentum exchange with the fluid.
In a conservative formulation, the left-hand side of Eq.~(\ref{eq:BoltzmannSimple}) is rewritten as a phase space divergence---that is, as the sum of spacetime and momentum space divergences.
Conservative formulations aid efforts to numerically maintain fidelity to global conservation laws (such as lepton number and energy) at modest resolution by transparently relating volume integrals to surface integrals.
A conservative formulation of the Boltzmann equation in spherical symmetry to $O(V/c)$ was given by Mezzacappa and Bruenn \cite{Mezzacappa1993Type-II-superno}.
Among other important discretization considerations, they generalized to the Boltzmann case a scheme for an explicit update of $f$ from the energy derivative term that was consistent with both lepton number and energy conservation \cite{Mezzacappa1993A-numerical-met} (originally developed by Bruenn \cite{Bruenn1985Stellar-core-co} for the case of flux-limited diffusion).
Still in spherical symmetry but in full general relativity, Liebend\"{o}rfer et al. \cite{Liebendorfer2004A-Finite-Differ} generalized this scheme to the angle derivative terms; and in order to avoid operator splitting of the observer corrections, converted these energy and angle updates into finite difference representations suitable for inclusion in the implicit solve used for the other terms in the Boltzmann equation.
Liebend\"{o}rfer et al. \cite{Liebendorfer2004A-Finite-Differ} also deduced discretizations of various terms in the number-conservative Boltzmann equation that provided consistency with the lab frame energy conservation law.
Cardall and Mezzacappa \cite{Cardall2003Conservative-fo} gave number- and four-momentum-conservative formulations of the Boltzmann equation in full general relativity and in the full dimensionality of phase space, but not in a 3+1 form useful in numerical relativity. 
First tests of a full 3D+3D Boltzmann solver have been reported \cite{Sumiyoshi2012Neutrino-Transf,Kotake2012Core-Collapse-S}, 
but without all the terms that appear as a result of fluid motion, which provide challenges to simultaneous energy-momentum and lepton number conservation.

Due to the reduced computational burdens incident to integrating out the momentum space angular variables, moments approaches have received recent attention as the field turns towards simulations that are 3D in position space. 
In a conformally flat ray-by-ray approximation, M\"uller et al. \cite{Muller2010A-New-Multi-dim} developed a scheme that renders non-conservative (in a sense discussed in Ref.~\cite{Endeve2012Conservative-Mu}) energy and momentum equations consistent with a conservative number equation, and have deployed it in 3D position space simulations retaining energy dependence \cite{Hanke2013SASI-Activity-i}.
(Another energy-dependent ray-by-ray simulation in 3D position space has been performed with a different approach, the `isotropic diffusion source approximation' \cite{Takiwaki2012Three-dimension}.) 
Shibata et al. \cite{Shibata2011Truncated-Momen} presented conservative energy and momentum angular moment transport equations in the 3+1 formulation of general relativity.
These have been implemented in simplified, energy-integrated (`grey') form in the 3D position space simulations of Kuroda et al. \cite{Kuroda2012Fully-General-R}.
In order to more fully address simultaneous energy and lepton number conservation in moments formalisms,
the energy derivative term in the Shibata et al. \cite{Shibata2011Truncated-Momen} equations was more fully elaborated into a useful 3+1 form by Cardall et al. \cite{Cardall2012Conservative-31}, who also considered in detail the term-by-term cancellations that must occur for consistency between conservative four-momentum transport equations and the conservative number transport equation; see also the partially relativistic limits discussed by Endeve et al. \cite{Endeve2012Conservative-Mu}.

In this paper we build on these developments by presenting conservative formulations of the Boltzmann equation in the 3+1 approach to general relativity.
We present a new derivation of the number-conservative formulation in Sec.~\ref{sec:ConservativeFormulation}.
The independent variables are lab frame coordinate basis spacetime position components and comoving frame curvilinear momentum space coordinates. 
In Sec.~\ref{sec:31_Specialization} we use `Eulerian decompositions' of the tetrad and its inverse in specializing to the 3+1 metric, and examine the relationship between the number-conservative and four-momentum conservative reformulations. 
We collect and discuss our results in Sec.~\ref{sec:Conclusion}, including overview tables of the many variables appearing in the formalism.

We conclude this introductory section by specifying some conventions we employ.
We take $\hbar = c = 1$ and use the metric signature $-+++$.
Greek indices run from $0$ to $3$ and latin indices run from $1$ to $3$.
We denote lab frame coordinate basis components with unadorned indices;
components measured in the orthonormal comoving frame with hatted indices;
and momentum space components in a curvilinear momentum space coordinate system with indices adorned with a tilde.

\section{Conservative formulation \label{sec:ConservativeFormulation}}

After specifying phase space coordinates, 
we present a new derivation of the conservative formulation of the general relativistic Boltzmann equation. 
See Ref.~\cite{Cardall2003Conservative-fo} for an alternative derivation using exterior calculus.

\subsection{Phase space coordinates \label{sec:PhaseSpaceCoordinates}}

The general relativistic Boltzmann equation \cite{Lindquist1966Relativistic-Tr,Ehlers1971General-Relativ,Israel1972The-Relativisti,Riffert1986A-general-Euler,Mezzacappa1989Computer-simula} equates the directional derivative of the distribution function along the phase flow to a collision integral:
\begin{equation}
\frac{df}{d\lambda} = C[f].
\label{eq:BoltzmannGeometric}
\end{equation}
That is, the change in $f$ along a phase space trajectory with affine parameter $\lambda$ is equal to the phase space density $C[f]$ of point-like collisions that add or remove particles from the trajectory.
The phase space measure is defined in such a way that $f$ and $C[f]$ are both invariant scalars.

For practical computations it is necessary to introduce phase space coordinates, which we initially take to be spacetime position coordinates $x^{\hat{\mu}}$ and momentum space coordinates $p^{\hat{\imath}}$ measured in an orthonormal frame comoving with the fluid.
We denote components in the comoving frame with hatted indices. 
We take only the spatial momentum components $p^{\hat{\imath}}$ as independent variables, due to the mass shell relation 
\begin{equation}
m^2 = -p_{\hat{\mu}} p^{\hat{\mu}} = -\eta_{\hat{\mu}\hat{\nu}} p^{\hat{\mu}} p^{\hat{\nu}},
\label{eq:MassShell}
\end{equation}
where $m$ is the particle mass and 
$\left(\eta_{\hat\mu \hat\nu} \right) = \mathrm{diag}[-1,1,1,1]$ is the Minkowski metric.
In terms of these phase space coordinates, Eq.~(\ref{eq:BoltzmannSimple}) becomes
\begin{equation}
\frac{dx^{\hat\mu}}{d\lambda} \frac{\partial f}{\partial x^{\hat\mu}} 
+ \frac{dp^{\hat\imath}}{d\lambda} \frac{\partial f}{\partial p^{\hat\imath}}
= C[f].
\label{eq:BoltzmannCoordinates} 
\end{equation}
For electrically neutral particles, the geodesic equations specifying particle trajectories are
\begin{eqnarray}
\frac{dx^{\hat\mu}}{d\lambda} &=& p^{\hat\mu}, \label{eq:Geodesic_x} \\
\frac{dp^{\hat\mu}}{d\lambda} &=& - {\Gamma^{\hat\mu}}_{\hat\nu \hat\rho} \, p^{\hat\nu} p^{\hat\rho},
  \label{eq:Geodesic_p}
\end{eqnarray}
so that
\begin{equation}
p^{\hat\mu} \frac{\partial f}{\partial x^{\hat\mu}} 
- {\Gamma^{\hat\imath}}_{\hat\nu \hat\mu} \, p^{\hat\nu} p^{\hat\mu} \frac{\partial f}{\partial p^{\hat\imath}} 
= C[f].
\label{eq:BoltzmannComoving}
\end{equation}
(For electrically charged particles the coupling to the electromagnetic field can come along for the ride in all that follows; see e.g. Ref.~\cite{Cardall2003Conservative-fo}.)
As we shall see (cf. Eqs.~(\ref{eq:TimelikeBasisVector})-(\ref{eq:ConnectionComoving}) below), when spacetime is curved, and/or curvilinear lab frame spacetime coordinates are used, and/or the comoving frame undergoes accelerated motion, the comoving frame connection coefficients ${\Gamma^{\hat\mu}}_{\hat\nu \hat\rho}$ do not vanish, despite the fact that the metric is $\eta_{\hat\mu \hat\nu}$ in an orthonormal frame.
This is because the metric can be brought into Minkowski form only locally, with a transformation that varies from point to point in spacetime.

Ultimately we want independent variables other than the orthonormal comoving frame components $x^{\hat\mu}$ and $p^{\hat\imath}$.
A global coordinate basis for spacetime position typically is needed for numerical simulations, often an Eulerian one when the particle transport is coupled to multidimensional hydrodynamics.
As for momentum space, Cartesian momentum components normally are replaced by spherical-type coordinates, i.e. momentum magnitude (or energy, for massless particles) and two angle variables.

Given a lab frame coordinate basis (denoted by unadorned indices) in which the metric components are $g_{\mu\nu}$, the coordinate transformation ${L^\mu}_{\hat\mu}$ to the comoving frame is such that
\begin{equation}
{L^\mu}_{\hat\mu} {L^\nu}_{\hat\nu}\, g_{\mu\nu} = \eta_{\hat\mu \hat\nu}.
\label{eq:MetricDiagonalization}
\end{equation}
We note for later use that this implies
\begin{equation}
\det\left({L^\mu}_{\hat\mu} \right) = (-g)^{-1/2},
\label{eq:TransformationDeterminant}
\end{equation}
where $g = \det\left( g_{\mu\nu}\right)$.
Because the comoving frame components of the fluid four-velocity $u^\mu$ are $\left( u^{\hat \mu}\right) = \left( u^{\hat 0}, u^{\hat \imath} \right)^T = \left(1, 0, 0, 0 \right)^T$,
the transformation ${L^\mu}_{\hat\mu}$ will indeed be to the comoving frame provided
\begin{equation}
{L^\mu}_{\hat 0} = {L^\mu}_{\hat 0} u^{\hat 0} = {L^\mu}_{\hat \mu} u^{\hat \mu} = u^\mu.
\label{eq:TimelikeBasisVector}
\end{equation}
In contrast, ${L^\mu}_{\hat\imath}$ are not unique, but are fixed by Eq.~(\ref{eq:MetricDiagonalization}) only up to spatial rotations.

A lab frame coordinate basis also gives us a handle on the comoving frame connection coefficients.
In the lab frame coordinate basis, the connection coefficients are torsion-free (symmetric in the lower indices), and given in terms of the metric by 
\begin{equation}
{{\Gamma^\mu}}_{\nu\rho} 
= \frac{1}{2} g^{\mu \sigma} \left(
\frac{\partial g_{\sigma \nu}}{\partial x^\rho}
+ \frac{\partial g_{\sigma \rho}}{\partial x^\nu}
- \frac{\partial g_{\nu \rho}}{\partial x^\sigma}
\right).
\label{eq:ConnectionLab}
\end{equation}
The connection coefficients do not transform as tensors, but as (see for instance Ref.~\cite{Misner1973Gravitation} or the Appendix of Ref.~\cite{Riffert1986A-general-Euler})
\begin{equation}
{\Gamma^{\hat{\mu}}}_{\hat{\nu}\hat{\rho}}
=  {L^{\hat{\mu}}}_{{\mu}} {L^\nu}_{\hat{\nu}} {L^\rho}_{\hat{\rho}} \,{\Gamma^\mu}_{\nu\rho} 
+ {L^{\hat{\mu}}}_{{\mu}} {L^\rho}_{\hat{\rho}} 
\frac{\partial {L^\mu}_{\hat{\nu}}}{\partial x^\rho},
\label{eq:ConnectionComoving}
\end{equation}
where ${L^{\hat\mu}}_{\mu}$ is the inverse of ${L^\mu}_{\hat\mu}$.
Note that the comoving frame connection coefficients are not manifestly symmetric in their lower indices; in comoving frame covariant derivatives such as $\nabla_{\hat\rho} u^{\hat\mu} = \partial_{\hat\rho} u^{\hat\mu} + {\Gamma^{\hat{\mu}}}_{\hat{\nu}\hat{\rho}} u^{\hat\nu}$, the final index of ${\Gamma^{\hat{\mu}}}_{\hat{\nu}\hat{\rho}}$ should match that of the derivative operators.

Consider next the use of curvilinear coordinates in momentum space (denoted by indices adorned with a tilde), for which we use the technology of differential geometry in a manner similar to the way it is used in spacetime. 
We introduce a three-metric $\lambda_{\tilde\imath \tilde\jmath}$ giving the momentum space line element 
\begin{equation}
d\Phi^2 = \lambda_{\tilde\imath \tilde\jmath}\, dp^{\tilde\imath}\, dp^{\tilde \jmath},
\label{eq:LineElementMomentum}
\end{equation}
where $d\Phi$ is the `proper distance' between two points in momentum space.
The transformation
\begin{equation}
{P^{\tilde \imath}}_{\hat \imath} = \frac{\partial p^{\tilde \imath}}{\partial p^{\hat\imath}}
\label{eq:MomentumTransformation}
\end{equation}
between curvilinear and Cartesian momentum space coordinates is such that, similar to Eq.~(\ref{eq:MetricDiagonalization}),  
\begin{equation}
{P^{\tilde\imath}}_{\hat\imath} {P^{\tilde\jmath}}_{\hat\jmath}\, \lambda_{\tilde\imath \tilde\jmath} = \delta_{\hat\imath \hat\jmath},
\label{eq:MomentumMetricDiagonalization}
\end{equation}
where $\delta_{\hat\imath \hat\jmath}$ is the Kronecker $\delta$.
Similar to Eqs.~(\ref{eq:MetricDiagonalization}) and (\ref{eq:TransformationDeterminant}), this implies
\begin{equation}
\det\left({P^{\tilde\imath}}_{\hat\imath} \right) = \lambda^{-1/2},
\label{eq:MomentumTransformationDeterminant}
\end{equation}
where $\lambda = \det\left( \lambda_{\tilde\imath \tilde\jmath} \right)$.
The inverse of ${P^{\tilde \imath}}_{\hat \imath}$ is
\begin{equation}
{P^{\hat \imath}}_{\tilde \imath} = \frac{\partial p^{\hat \imath}}{\partial p^{\tilde\imath}}.
\label{eq:InverseMomentumTransformation}
\end{equation}
We also define momentum space connection coefficients ${\Pi^{\tilde\imath}}_{\tilde\jmath \tilde k}$. 
These can be given in terms of $\lambda_{\tilde\imath \tilde\jmath}$ by an analogue of Eq.~(\ref{eq:ConnectionLab}).
Alternatively, an analogue of Eq.~(\ref{eq:ConnectionComoving}) gives
\begin{eqnarray}
{\Pi^{\tilde\imath}}_{\tilde\jmath \tilde k}
&=&  {P^{\tilde \imath}}_{\hat \imath} {P^{\hat \jmath}}_{\tilde \jmath} {P^{\hat k}}_{\tilde k} \,{\Pi^{\hat\imath}}_{\hat\jmath \hat k}
+ {P^{\tilde \imath}}_{\hat \imath} {P^{\hat k}}_{\tilde k}
\frac{\partial {P^{\hat\imath}}_{\tilde{\jmath}}}{\partial p^{\hat k}} \\
&=& {P^{\tilde \imath}}_{\hat \imath} {P^{\hat k}}_{\tilde k}
\frac{\partial {P^{\hat\imath}}_{\tilde{\jmath}}}{\partial p^{\hat k}},
\label{eq:ConnectionMomentum}
\end{eqnarray}
thanks to the vanishing of the Cartesian connection coefficients ${\Pi^{\hat\imath}}_{\hat\jmath \hat k}$ in flat momentum space \footnote{This expression does not at first appear symmetric in $\tilde\jmath$ and $\tilde k$ as expected of ${\Pi^{\tilde\imath}}_{\tilde\jmath \tilde k}$; but note that when the curvilinear $p^{\tilde\imath}$ are regarded as the independent variables, ${P^{\hat k}}_{\tilde k}
\left( \partial {P^{\hat\imath}}_{\tilde{\jmath}} / \partial p^{\hat k}\right) = \partial {P^{\hat\imath}}_{\tilde{\jmath}} / \partial p^{\tilde k} = \partial^2 p^{\hat\imath} / \partial p^{\tilde k}\partial p^{\tilde \jmath}$ }. 
With these connection coefficients we define a `momentum space covariant derivative' operator $\mathcal{D}_{\tilde \imath}$, giving for instance 
\begin{equation}
\mathcal{D}_{\tilde k} {R^{\tilde\imath}} = \frac{\partial {R^{\tilde\imath}} }{\partial p^{\tilde k}}  + {\Pi^{\tilde\imath}}_{\tilde\jmath \tilde k} {R^{\tilde\jmath}}
\end{equation} 
for some momentum space three-vector $R^{\tilde\imath}$.

We emphasize that the metric $\lambda_{\tilde\imath \tilde\jmath}$, coordinate transformations ${P^{\tilde \imath}}_{\hat \imath}$ and ${P^{\hat \imath}}_{\tilde \imath}$, and connection coefficients ${\Pi^{\tilde\imath}}_{\tilde\jmath \tilde k}$ in momentum space are completely separate from those of spacetime.
They operate only in (flat) momentum space, and we introduce them only to facilitate formal handling of generic momentum space curvilinear coordinates.

As a concrete example of curvilinear momentum space coordinates---just an example, meant in no way to specialize our formalism to particular momentum space coordinates---consider for instance
\begin{eqnarray}
p^{\hat 1} &=& p \cos\vartheta, \\
p^{\hat 2} &=& p \sin\vartheta \cos\varphi, \\
p^{\hat 3} &=& p \sin\vartheta \sin\varphi,
\end{eqnarray} 
a version of momentum space spherical coordinates.
Ordering the curvilinear variables $\left( p^{\tilde\imath} \right) = (p, \vartheta, \varphi)^T$, Eq.~(\ref{eq:InverseMomentumTransformation}) is computed by inspection as
\begin{equation}
\left( {P^{\hat \imath}}_{\tilde \imath} \right) = \begin{pmatrix}
\cos\vartheta && - p \sin\vartheta&&  0\\
 \sin\vartheta \cos\varphi && p \cos\vartheta \cos\varphi && -p \sin\vartheta\sin\varphi  \\
 \sin\vartheta \sin\varphi && p \cos\vartheta \sin\varphi  && p \sin\vartheta \cos\varphi 
\end{pmatrix}.
\end{equation}
The transformation ${P^{\tilde \imath}}_{\hat \imath}$ of Eq.~(\ref{eq:MomentumTransformation}) is then the matrix inverse
\begin{equation}
\left( {P^{\tilde \imath}}_{\hat \imath} \right) = \frac{1}{p}
\begin{pmatrix}
p \cos\vartheta && p \sin\vartheta \cos\varphi && p \sin\vartheta \sin\varphi \\
-\sin\vartheta && \cos\vartheta \cos\varphi && \cos\vartheta \sin\varphi  \\
0 && -\sin\varphi / \sin\vartheta && \cos\varphi / \sin\vartheta 
\end{pmatrix}.
\end{equation}
Finally,
\begin{equation}
d\Phi^2 = dp^2 + p^2 \left( d\vartheta^2 + \sin^2 \vartheta \, d\varphi^2 \right)
\end{equation}
is the line element in this example, from which the momentum space metric components $\lambda_{\tilde\imath \tilde\jmath}$ can be read.

With transformations ${L^{\mu}}_{\hat\mu}$ and ${P^{\tilde\imath}}_{\hat\imath}$ to our desired phase space coordinates in hand, we use them to rewrite the derivatives in Eq.~(\ref{eq:BoltzmannComoving}), such that
\begin{equation}
p^{\hat\mu} {L^{\mu}}_{\hat\mu} \frac{\partial f}{\partial x^{\mu}} 
- {\Gamma^{\hat\imath}}_{\hat\nu \hat\mu} \, p^{\hat\nu} p^{\hat\mu} {P^{\tilde\imath}}_{\hat\imath} \frac{\partial f}{\partial p^{\tilde\imath}} 
= C[f].
\label{eq:BoltzmannChosen}
\end{equation}
Thus the dependence of the distribution function is $f=f\left(x^\mu, p^{\tilde\imath}\right)$, with lab frame coordinate basis spacetime position coordinates $x^\mu$ and comoving frame momentum space curvilinear coordinates $p^{\tilde\imath}$ as  independent variables.
The partial derivatives with respect to $x^\mu$ are with $p^{\tilde\imath}$ held fixed, and vice-versa.

\subsection{Divergences \label{sec:Divergences}}

By a `conservative formulation' we mean one expressed in terms of divergences with respect to our chosen phase space coordinates.
In obtaining a conservative Boltzmann equation we make use of covariant derivatives in both spacetime and momentum space.
Accordingly, we begin by writing Eq.~(\ref{eq:BoltzmannChosen}) as
\begin{equation}
p^{\hat\mu} {L^{\mu}}_{\hat\mu} \nabla_\mu f 
- {\Gamma^{\hat\imath}}_{\hat\nu \hat\mu} \, p^{\hat\nu} p^{\hat\mu} {P^{\tilde\imath}}_{\hat\imath} \mathcal{D}_{\tilde\imath} f 
= C[f],
\label{eq:BoltzmannCovariantDerivatives}
\end{equation}
using the fact that $f$ is a scalar.
Then the product rule satisfied by covariant derivatives gives
\begin{eqnarray}
\nabla_\mu \left( {L^{\mu}}_{\hat\mu}\, p^{\hat\mu}  f \right)
+ &\mathcal{D}_{\tilde\imath}& \left(
- {P^{\tilde\imath}}_{\hat\imath} {\Gamma^{\hat\imath}}_{\hat\nu \hat\mu} \, p^{\hat\nu} p^{\hat\mu}   f \right)  \nonumber \\
- f \,\nabla_\mu \left( {L^{\mu}}_{\hat\mu}\, p^{\hat\mu} \right) &+& f \,  \mathcal{D}_{\tilde\imath} \left(
{P^{\tilde\imath}}_{\hat\imath} {\Gamma^{\hat\imath}}_{\hat\nu \hat\mu}\, p^{\hat\nu} p^{\hat\mu} \right) = C[f].
\label{eq:ConservativeExtra}
\end{eqnarray}
The first two terms of this equation are very close to the conservative form we seek, and our task is to show how the last two `extra' terms on the left-hand side (almost) cancel.

Consider first the `extra' spatial derivative term (the third on the left-hand side of Eq.~(\ref{eq:ConservativeExtra})). 
It is 
\begin{equation}
- f \,\nabla_\mu \left( {L^{\mu}}_{\hat\mu} p^{\hat\mu} \right) = - f p^{\hat\mu} \left( \frac{\partial {L^{\mu}}_{\hat\mu}}{\partial x^\mu} + {\Gamma^\mu}_{\nu\mu} {L^{\nu}}_{\hat\mu} \right),
\label{eq:SpacetimeExtra}
\end{equation}
as the partial derivative is taken with $p^{\tilde\imath}$ held fixed.
For later use we keep in mind that (e.g. Ref.~\cite{Misner1973Gravitation})
\begin{equation}
{\Gamma^\mu}_{\nu\mu} = \frac{1}{\sqrt{-g}} \frac{\partial \sqrt{-g}}{\partial x^\nu},
\label{eq:GammaContraction}
\end{equation}
but we will not need to use this in Eq.~(\ref{eq:SpacetimeExtra}). 

Processing of the `extra' momentum derivative term $ f \,  \mathcal{D}_{\tilde\imath} \left(
{P^{\tilde\imath}}_{\hat\imath} {\Gamma^{\hat\imath}}_{\hat\nu \hat\mu}\, p^{\hat\nu} p^{\hat\mu} \right)$---the fourth on the left-hand side of Eq.~(\ref{eq:ConservativeExtra})---is interesting but a bit detailed, and we leave the specifics to Appendix~\ref{app:ExtraMomentumTerm}. 
There are three factors affected by the derivative.
Only in one of them, $\mathcal{D}_{\tilde\imath} {P^{\tilde\imath}}_{\hat\imath}$, do the momentum space connection coefficients come into play, as this is the only free index (and also the only tilde index).
It turns out that 
\begin{equation}
\mathcal{D}_{\tilde\imath} {P^{\tilde\imath}}_{\hat\imath} = 0, 
\label{eq:CovariantDPVanish}
\end{equation}
which we show in Appendix~\ref{app:ExtraMomentumTerm} using Eq.~(\ref{eq:ConnectionMomentum}).
The other two factors affected by the derivative are the momenta  $p^{\hat\nu}$ and $p^{\hat\mu}$ in
\begin{eqnarray}
f {P^{\tilde\imath}}_{\hat\imath}\, \mathcal{D}_{\tilde\imath}  \left(
 {\Gamma^{\hat\imath}}_{\hat\nu \hat\mu}\, p^{\hat\nu} p^{\hat\mu} \right)
&=& f {P^{\tilde\imath}}_{\hat\imath} \frac{\partial}{\partial p^{\tilde\imath}} \left(
  {\Gamma^{\hat\imath}}_{\hat\nu \hat\mu}\, p^{\hat\nu} p^{\hat\mu} \right) \\
&=& f  \frac{\partial}{\partial p^{\hat\imath}} \left( {\Gamma^{\hat\imath}}_{\hat\nu \hat\mu} \, p^{\hat\nu} p^{\hat\mu} \right) \\
 &=&  f \, {\Gamma^{\hat\imath}}_{\hat\nu \hat\mu} \left( \frac{\partial p^{\hat\nu}}{\partial p^{\hat\imath}} p^{\hat\mu} + p^{\hat\nu} \frac{\partial p^{\hat\mu}}{\partial p^{\hat\imath}} \right), \nonumber \\
 & &
\label{eq:LastTwoMomentumDerivatives}  
\end{eqnarray}
in which we have used Eq.~(\ref{eq:MomentumTransformation}).
In the momentum derivatives, $p^{\hat 0}$ must be considered a function of $p^{\hat\imath}$ through Eq.~(\ref{eq:MassShell}).
The result for the first term of Eq.~(\ref{eq:LastTwoMomentumDerivatives}) is worked out in Appendix~\ref{app:ExtraMomentumTerm} (see Eq.~(\ref{eq:FirstMomentumDerivativeTermResultAppendix})):
\begin{eqnarray}
f  &{\Gamma^{\hat\imath}}&_{\hat\nu \hat\mu} \frac{\partial p^{\hat\nu}}{\partial p^{\hat\imath}} p^{\hat\mu} \nonumber \\
 &=& -f \,{\Gamma^{\hat\imath}}_{\hat\nu \hat\mu} \,p^{\hat\nu} p^{\hat\mu} {P^{\tilde\imath}}_{\hat\imath} \left(-p_{\hat 0}\right) \mathcal{D}_{\tilde\imath}\left[\frac{1}{\left(-p_{\hat 0}\right)}\right].
\label{eq:FirstMomentumDerivativeTermResult} 
\end{eqnarray}
The result for the second term of Eq.~(\ref{eq:LastTwoMomentumDerivatives}) is also worked out in Appendix~\ref{app:ExtraMomentumTerm} (see Eq.~(\ref{eq:SecondMomentumDerivativeTermResultAppendix})):
\begin{equation}
f \, {\Gamma^{\hat\imath}}_{\hat\nu \hat\mu} p^{\hat\nu}  \frac{\partial p^{\hat\mu}}{\partial p^{\hat\imath}} 
 = f p^{\hat\nu} \left( {L^\nu}_{\hat{\nu}} \,{\Gamma^\mu}_{\nu\mu} +  \frac{\partial {L^\mu}_{\hat{\nu}}}{\partial x^\mu} \right). 
\label{eq:SecondMomentumDerivativeTermResult} 
\end{equation}
These results are obtained with the help of the geodesic Eq.~(\ref{eq:Geodesic_p}), along with a lowered-index version; the fact that the mass shell relation of Eq.~(\ref{eq:MassShell}) implies $0 = \mathrm{d} \left( p_{\hat\mu} p^{\hat\mu} \right) = 2\, p_{\hat\mu}\, \mathrm{d}p^{\hat\mu}$, where here ``$\mathrm{d}$'' represents any derivative operator; matrix identities involving derivatives, traces, logarithms, and determinants; and Eqs.~(\ref{eq:TransformationDeterminant}), (\ref{eq:ConnectionComoving}), and (\ref{eq:GammaContraction}).

We are ready to exhibit our conservative formulation of the Boltzmann equation, using Eqs.~(\ref{eq:SpacetimeExtra})-(\ref{eq:SecondMomentumDerivativeTermResult}) in Eq.~(\ref{eq:ConservativeExtra}).
Note that Eq.~(\ref{eq:SecondMomentumDerivativeTermResult}) cancels with Eq.~(\ref{eq:SpacetimeExtra}), and that Eq.~(\ref{eq:FirstMomentumDerivativeTermResult}) can be combined with the first momentum derivative term in Eq.~(\ref{eq:ConservativeExtra}).
The result is the conservative form we seek:
\begin{equation}
S_N + M_N = C[f],
\label{eq:ConservativeCovariant}
\end{equation}
where
\begin{equation}
S_N = \nabla_\mu \left( {L^{\mu}}_{\hat\mu}\, p^{\hat\mu}   f \right)
\label{eq:SpacetimeDivergenceCovariant}
\end{equation}
is the spacetime divergence, and
\begin{equation}
M_N = \left(-p_{\hat 0}\right) \mathcal{D}_{\tilde\imath} \left[
- \frac{ 1 }{\left(-p_{\hat 0}\right)} {P^{\tilde\imath}}_{\hat\imath} {\Gamma^{\hat\imath}}_{\hat\nu \hat\mu} \, p^{\hat\nu} p^{\hat\mu}   f \right]
\label{eq:MomentumDivergenceCovariant}
\end{equation}
is the momentum space divergence.
In terms of partial derivatives, the divergences can be expressed
\begin{eqnarray}
S_N &=& \frac{1}{\sqrt{-g}} \frac{\partial}{\partial x^\mu} \left(\sqrt{-g}\, {L^{\mu}}_{\hat\mu}\, p^{\hat\mu}   f \right), 
\label{eq:SpacetimeDivergencePartial} \\
M_N &=& \frac{\left(-p_{\hat 0}\right)}{\sqrt{\lambda}} \frac{\partial}{\partial p^{\tilde\imath}} \left[
- \frac{ \sqrt{\lambda} }{\left(-p_{\hat 0}\right)} {P^{\tilde\imath}}_{\hat\imath} {\Gamma^{\hat\imath}}_{\hat\nu \hat\mu} \, p^{\hat\nu} p^{\hat\mu}   f \right],
\label{eq:MomentumDivergencePartial}
\end{eqnarray}
using Eq.~(\ref{eq:GammaContraction}) and its momentum space analogue.
In comparing with Eq. (132) of Ref.~\cite{Cardall2003Conservative-fo}, note that $\left(-p_{\hat 0}\right) = E (\mathbf{p})$ and that $\sqrt{\lambda} = \left[\det\left({P^{\tilde\imath}}_{\hat\imath} \right)\right]^{-1} = \det\left({P^{\hat\imath}}_{\tilde\imath} \right) = \det\left( \partial\mathbf{p}/ \partial\mathbf{u} \right)$; see Eq.~(\ref{eq:MomentumTransformationDeterminant}).

The conservative nature of Eqs.~(\ref{eq:ConservativeCovariant}), (\ref{eq:SpacetimeDivergencePartial})-(\ref{eq:MomentumDivergencePartial})  
is seen by considering the integral over the phase space volume.
The invariant momentum space volume element in our momentum space curvilinear coordinates is
\begin{equation}
d{\tilde{\mathbf{p}}} = \frac{\sqrt{\lambda}}{(-p_{\hat 0})}  \, \frac{dp^{\tilde 1}  \, dp^{\tilde 2}  \, dp^{\tilde 3}}{(2\pi)^3}. 
\end{equation}
The factor of $\sqrt{\lambda} = \det\left({P^{\hat\imath}}_{\tilde\imath} \right)$ is the Jacobian of the transformation from Cartesian momentum coordinates, and $\left( -p_{\hat 0} \right)^{-1}$ arises from restricting an invariant 4-momentum volume element to the $p^{\hat 0} > 0$ branch of the mass shell (e.g. Ref.~\cite{Ehlers1971General-Relativ}). 
We use units in which $\hbar = c = 1$; relative to works in which instead $h = c =1$,  this leads to the factor of $(2\pi)^3$ in the denominator.
When Eq.~(\ref{eq:ConservativeCovariant}) is multiplied by $d{\tilde{\mathbf{p}}}$ and integrated over momentum space, 
the momentum space divergence term of Eq.~(\ref{eq:MomentumDivergencePartial}) manifestly yields a vanishing surface integral, leaving the number balance equation
\begin{equation}
\frac{1}{\sqrt{-g}} \frac{\partial}{\partial x^\mu}
\left( \sqrt{-g} \,N^\mu \right) = \int C[f] \, d{\tilde{\mathbf{p}}},
\label{eq:NumberConservation}
\end{equation}
where 
\begin{equation}
N^\mu = \int  f \,p^\mu \, d{\tilde{\mathbf{p}}}
\label{eq:NumberVector}
\end{equation}
is the number flux vector (e.g. Ref.~\cite{Ehlers1971General-Relativ}), expressed here in terms of the lab frame coordinate basis (note $p^\mu = {L^\mu}_{\hat{\mu}} \,p^{\hat{\mu}}$).
Turning to position space, recall that three-volume elements in spacetime are one-forms whose index specifies an orientation. 
The spatial three-volume reckoned by an Eulerian observer (one associated with the lab frame coordinate basis) is 
\begin{equation}
\left( d\mathbf{x} \right)_\mu =  \sqrt{-g} \,\varepsilon_{\mu 1 2 3}\, dx^1 \, dx^2 \, dx^3,
\end{equation}
where $\varepsilon_{\mu\nu\rho\sigma}$ is the alternating symbol with $\varepsilon_{0 1 2 3} = +1$.
Considering the integrand of Eq.~(\ref{eq:NumberVector}), we see that
\begin{eqnarray}
d\mathsf{N} &=& f \, p^\mu \,d{\tilde{\mathbf{p}}} \left(d\mathbf{x}\right)_\mu \\
&=& f \!\left(x^\mu, p^{\tilde\imath} \right)\left( {L^0}_{\hat\mu} \, p^{\hat\mu}\right) \left[ \frac{ \sqrt{\lambda} \, dp^{\tilde 1}  \, dp^{\tilde 2}  \, dp^{\tilde 3}}{(-p_{\hat 0})(2\pi)^3}\right]\nonumber \\
& & \times \left( \sqrt{-g} \, dx^1 \, dx^2 \, dx^3 \right) 
\end{eqnarray}
is the number of particles measured by an Eulerian observer in a differential phase space volume surrounding $(x^i, p^{\tilde\imath})$ at time $t = x^0$.
The spatial number density $N^0$ of particles measured by an Eulerian observer is 
\begin{equation}
N^0\!\left(x^\mu \right) = \int f \!\left(x^\mu, p^{\tilde\imath} \right)\left( {L^0}_{\hat\mu} \, p^{\hat\mu}\right) \left[ \frac{ \sqrt{\lambda} \, dp^{\tilde 1}  \, dp^{\tilde 2}  \, dp^{\tilde 3}}{(-p_{\hat 0})(2\pi)^3}\right],
\end{equation}
and multiplying this spatial density by $\sqrt{-g} \, dx^1 \, dx^2 \, dx^3$ and integrating over all space gives that total number $\mathsf{N}$ of particles in the universe represented by the distribution function $f\!\left(x^\mu, p^{\tilde\imath} \right)$:
\begin{equation}
\mathsf{N}(t) = \int d\mathsf{N} = \int N^0\!\left(x^\mu \right) \sqrt{-g} \, dx^1 \, dx^2 \, dx^3.
\end{equation}
Assuming a vanishing particle density at spatial infinity, multiplying Eq.~(\ref{eq:NumberConservation}) by $\sqrt{-g} \, dx^1 \, dx^2 \, dx^3$ and integrating over all space manifestly yields a vanishing surface integral, leaving
\begin{equation}
\frac{d\mathsf{N}(t)}{dt} = \int C[f] \left[ \frac{ \sqrt{\lambda} \, dp^{\tilde 1}  \, dp^{\tilde 2}  \, dp^{\tilde 3}}{(-p_{\hat 0})(2\pi)^3}\right] \left( \sqrt{-g} \, dx^1 \, dx^2 \, dx^3 \right) 
\end{equation}
for the time rate of change of the total number of particles in the universe represented by the distribution function $f\!\left(x^\mu, p^{\tilde\imath} \right)$.
It is the ready conversion of volume integrals to surface integrals via the divergence theorem---allowing the time rate of change of a `conserved' variable in a volume to be related to a flux through the surface of the volume---that is the hallmark of a conservative formulation.

\section{3+1 Specialization \label{sec:31_Specialization}}
 
After briefly reviewing the 3+1 formulation of general relativity, we introduce `Eulerian decompositions' of the tetrad of comoving frame orthonormal basis vectors;
a decomposition of the momentum space coordinate transformation;
express the conservative Boltzmann equation in terms of the resulting variables;
and examine the relationship between lepton number and four-momentum exchange.
 
\subsection{Geometry description \label{sec:GeometryDescription}}
 
Numerical relativity often is built upon the 3+1 formulation of general relativity.
Pedagogical introductions include Refs.~\cite{Misner1973Gravitation,York1983The-initial-val,Gourgoulhon200731-Formalism-an}; 
see also, very briefly, Sec.~IIIA of Ref.~\cite{Cardall2012Conservative-31}.

 
In this approach---which provides our lab frame coordinate basis---one considers a foliation of spacetime into spacelike slices, i.e. three-dimensional hypersurfaces $\Sigma_t$ labeled by coordinate time.
The metric components $g_{\mu\nu}$ can be read off the line element
\begin{eqnarray}
ds^2 &=& g_{\mu\nu}\, dx^\mu dx^\nu \\
&=& -\alpha^2 dt^2 + \gamma_{i j} \left(dx^i + \beta^i\,dt \right) \left(dx^j + \beta^j\,dt \right).
\label{eq:LineElement}
\end{eqnarray}
The lapse function $\alpha$ determines the orthogonal proper time $\alpha \, dt$ separating slices $\Sigma_t$ and $\Sigma_{t+dt}$.
The shift vector $\beta^i$ is tangent to the spacelike slice and characterizes the velocity of the spatial coordinates as seen by an observer at rest in the slice.
The three-metric $\gamma_{ij}$, also tangent to the spacelike slice, encodes its internal geometry.
The equation
\begin{equation}
\sqrt{-g} = \alpha\sqrt{\gamma}
\label{eq:MetricDeterminant}
\end{equation}
expresses the determinant $g$ of the four-metric in terms of the lapse function and the determinant $\gamma$ of the three-metric.
Another symmetric tensor tangent to the spacelike slice, the extrinsic curvature $K_{ij}$, describes the warp of the spacelike slices as embedded in spacetime.
 
In the 3+1 approach, solution of the Einstein equations is transformed into a Cauchy problem: 
specify initial data (satisfying certain constraints from the Einstein equations) on an initial spacelike slice;
and with coordinate freedom fixed and spatial boundary conditions specified, evolve the geometry of the spacelike slices forward in time.
The dynamical variables are the independent components of $\gamma_{ij}$ and $K_{ij}$, which satisfy first-order-in-time evolution equations.
The coordinate freedom of general relativity corresponds to freedom to specify the lapse $\alpha$ and shift $\beta^i$, which determine the slicing of spacetime and the motion of spatial coordinates respectively. 
In this paper we regard $\gamma_{ij}$ and $K_{ij}$ as given, for instance as having been obtained by numerical solution (often of even further transformed systems, as for instance in Baumgarte-Shapiro-Shibata-Nakamura and related approaches); see e.g. Refs.~\cite{Gourgoulhon200731-Formalism-an,Alcubierre2008Introduction-to,Baumgarte2010Numerical-Relat}.
 
`Eulerian decomposition' of tensors into `Eulerian projections' orthogonal and tangent to the spacelike slice will prove useful. 
(This served us well in Ref.~\cite{Cardall2012Conservative-31} on a moments approach to radiation transport.)
The unit normal $n^\mu$ to a spacelike slice---which is also the four-velocity of an `Eulerian observer'---has lab frame coordinate basis components
\begin{eqnarray}
\left( n^\mu \right) &=& (1/\alpha, -\beta^i / \alpha)^T, \label{eq:UnitNormalU} \\
\left( n_\mu \right) &=& (-\alpha,0,0,0). \label{eq:UnitNormalD} 
\end{eqnarray}
The orthogonal projector is
\begin{equation}
\gamma_{\mu\nu} = g_{\mu\nu} + n_\mu n_\nu.
\label{eq:OrthogonalProjector}
\end{equation}
From Eqs.~(\ref{eq:LineElement}) and (\ref{eq:UnitNormalD}) it follows that the spatial part of $\gamma_{\mu\nu}$ equals the three-metric $\gamma_{ij}$, motivating use of the same base symbol.
Contraction of an arbitrary vector with $n^\mu$ yields the portion orthogonal to a spacelike slice, and contraction with $\gamma_{\mu\nu} = g_{\mu\nu} + n_\mu n_\nu$ yields the portion tangent to the spacelike slice.
Indeed a trivial calculation confirms that $\gamma_{\mu\nu} n^\nu = 0$.

\subsection{Tetrad decomposition \label{sec:TetradDecomposition}}

In Sec.~\ref{sec:ConservativeFormulation} we mostly thought of ${L^\mu}_{\hat\mu}$ as a transformation matrix between the lab frame coordinate basis and the orthonormal comoving basis, 
but we can focus instead on the fact that it represents the tetrad of comoving orthonormal basis vectors. 
In this perspective, the upper index $\mu$ in ${L^\mu}_{\hat\mu}$ labels the (lab frame coordinate basis) {\em component} of the comoving orthonormal basis vector, while
the lower index $\hat\mu$ is not a component but simply a {\em label} identifying the particular vector in the tetrad.

We already saw something of this in Eq.~(\ref{eq:TimelikeBasisVector}), and we can extend the thinking to the comoving spatial unit vectors. 
We saw in Eq.~(\ref{eq:TimelikeBasisVector}) that the components of the Lagrangian observer's four-velocity in the comoving frame are
\begin{equation}
\left( u^{\hat\mu} \right) = \left( 1, 0, 0, 0 \right)^T,
\label{eq:FluidVelocityU}
\end{equation}
so that 
\begin{equation}
{L^\mu}_{\hat 0} = {L^\mu}_{\hat 0} u^{\hat 0} = {L^\mu}_{\hat \mu} u^{\hat \mu} = u^\mu.
\label{eq:TimelikeVector}
\end{equation}
That is, ${L^\mu}_{\hat 0}$ is the Lagrangian observer's timelike basis vector, expressed in components measured by an Eulerian observer.
Similarly, the Lagrangian observer can choose spatial basis vectors
\begin{eqnarray}
\left( w^{\hat\mu} \right) &=& \left( 0, 1, 0, 0 \right)^T, \\
\left( y^{\hat\mu} \right) &=& \left( 0, 0, 1, 0 \right)^T, \\
\left( z^{\hat\mu} \right) &=& \left( 0, 0, 0, 1 \right)^T,
\end{eqnarray}
whose lab frame coordinate basis components are
\begin{eqnarray}
{L^\mu}_{\hat 1} &=& {L^\mu}_{\hat 1} w^{\hat 1} = {L^\mu}_{\hat \mu} w^{\hat \mu} = w^\mu, \label{eq:SpacelikeVector_1} \\
{L^\mu}_{\hat 2} &=& {L^\mu}_{\hat 2} y^{\hat 2} = {L^\mu}_{\hat \mu} y^{\hat \mu} = y^\mu, \label{eq:SpacelikeVector_2} \\
{L^\mu}_{\hat 3} &=& {L^\mu}_{\hat 3} z^{\hat 3} = {L^\mu}_{\hat \mu} z^{\hat \mu} = z^\mu. \label{eq:SpacelikeVector_3}
\end{eqnarray}
For the time being we will use the individual expressions $u^\mu, w^\mu, y^\mu, z^\mu$ to emphasize that we are working with a tetrad of vectors.
Later, for the sake of more compact expressions using index notation, we will return to notation like ${L^\mu}_{\hat\mu}$, and ask that the reader keep expressions like Eqs.~(\ref{eq:TimelikeVector}), (\ref{eq:SpacelikeVector_1})-(\ref{eq:SpacelikeVector_3}) in mind, and regard hatted indices like $\hat\mu$ only as labels identifying a particular vector in the tetrad in such instances.

In a moments approach to radiation transport \cite{Cardall2012Conservative-31}, the only comoving basis vector we had to deal with was ${L^\mu}_{\hat 0} = u^\mu$, and we found it useful to consider its Eulerian decomposition into parts orthogonal and tangent to the spacelike slice:
\begin{equation}
u^\mu = \Lambda \left(n^\mu + v^\mu\right).
\label{eq:ThreeVelocity}
\end{equation}
The orthogonality requirement on $v^\mu$,
\begin{equation}
n_\mu v^\mu = 0,
\label{eq:ThreeVelocityOrthogonality}
\end{equation}
implies (see Eq.~(\ref{eq:UnitNormalD})) that $v^\mu$ is spacelike and has components
\begin{equation}
\left(v^\mu \right)= \left(0, v^i \right)^T
\label{eq:v_Spacelike}
\end{equation}
in the lab frame coordinate basis.
The interpretation of $v^\mu$ as the three-velocity of a Lagrangian observer as measured by an Eulerian observer is confirmed by squaring Eq.~(\ref{eq:ThreeVelocity}) to find the expected Lorentz factor
\begin{equation}
\Lambda = \left( 1 - v^\mu v_\mu \right)^{-1/2} = \left( 1 - v^i v_i \right)^{-1/2}
\label{eq:BoostFactor}
\end{equation}
for a boost between frames with relative three-velocity $v^\mu$. 
(Recall that $u_\mu u^\mu = n_\mu n^\mu = -1$.) 

Here we also find it useful to make Eulerian decompositions of the spatial comoving basis vectors $w^\mu, y^\mu, z^\mu$.
We write
\begin{eqnarray}
w^\mu &=& A\, n^\mu + a^\mu, 
\label{eq:wyz_1} \\
y^\mu &=& B\, n^\mu + b^\mu, 
\label{eq:wyz_2} \\
z^\mu &=& C\, n^\mu + c^\mu,
\label{eq:wyz_3}
\end{eqnarray}
specifying that
\begin{equation}
0 = n_\mu a^\mu =  n_\mu b^\mu = n_\mu c^\mu,
\label{eq:SpatialBasisOrthogonality}
\end{equation}
or, in the lab frame coordinate basis,
\begin{equation}
\left(a^\mu \right)= \left(0, a^i \right)^T, \ \ \left(b^\mu \right)= \left(0, b^i \right)^T, \ \ \left(c^\mu \right)= \left(0, c^i \right)^T
\label{eq:abc_Spacelike}
\end{equation}
in order that $a^\mu, b^\mu, c^\mu$ be tangent to the spacelike slice.

The metric transformation in Eq.~(\ref{eq:MetricDiagonalization}) translates into orthonormality conditions on $u^\mu, w^\mu, y^\mu, z^\mu$.
As a physical four-velocity, the timelike basis vector $u^\mu$ already satisfies
\begin{equation} 
u_\mu u^\mu = {L^\mu}_{\hat 0} {L^\nu}_{\hat 0}\, g_{\mu\nu} = \eta_{\hat 0 \hat 0} = -1.
\end{equation}
The off-diagonal time-space elements of Eq.~(\ref{eq:MetricDiagonalization}),
\begin{equation}
{L^\mu}_{\hat 0} {L^\nu}_{\hat \imath}\, g_{\mu\nu} = \eta_{\hat 0 \hat\imath} = 0,
\end{equation}
give orthogonality relations
\begin{equation}
0 = u_\mu w^\mu =  u_\mu y^\mu = u_\mu z^\mu,
\label{eq:TimeSpaceOrthogonality}
\end{equation}
which turn out to imply
\begin{eqnarray}
A &=& v_\mu a^\mu = v_i a^i, 
\label{eq:ABC_1}\\
B &=& v_\mu b^\mu = v_i b^i, 
\label{eq:ABC_2} \\
C &=& v_\mu c^\mu = v_i c^i,
\label{eq:ABC_3}
\end{eqnarray}
where we have used Eq.~(\ref{eq:ThreeVelocity}).
The diagonal spatial elements of Eq.~(\ref{eq:MetricDiagonalization}),
\begin{equation}
{L^\mu}_{\hat \imath} {L^\nu}_{\hat \imath}\, g_{\mu\nu} = \eta_{\hat \imath \hat\imath} = +1 \ \ \ \mathrm{(no\ summation)},
\end{equation}
impose the normalization conditions
\begin{equation}
+1 = w_\mu w^\mu =  y_\mu y^\mu = z_\mu z^\mu,
\end{equation}
which imply
\begin{eqnarray}
\gamma_{ij}\, a^i a^j &=& a_i a^i = a_\mu a^\mu = 1 + A^2, 
\label{eq:abc_Normalization_1} \\
\gamma_{ij}\, b^i b^j &=& b_i b^i = b_\mu b^\mu = 1 + B^2, 
\label{eq:abc_Normalization_2} \\
\gamma_{ij}\, c^i c^j &=& c_i c^i = c_\mu c^\mu = 1 + C^2
\label{eq:abc_Normalization_3}
\end{eqnarray}
for the normalization of $a^\mu, b^\mu, c^\mu$ (the portions of $w^\mu, y^\mu, z^\mu$ tangent to the spacelike slice).
Finally, the off-diagonal spatial elements of Eq.~(\ref{eq:MetricDiagonalization}),
\begin{equation}
{L^\mu}_{\hat \imath} {L^\nu}_{\hat \jmath}\, g_{\mu\nu} = \eta_{\hat \imath \hat\jmath} = 0 \ \ \ (\hat\imath \ne \hat\jmath),
\end{equation}
give orthogonality relations
\begin{equation}
0 = w_\mu y^\mu =  w_\mu z^\mu = y_\mu z^\mu,
\label{eq:SpatialOrthogonality}
\end{equation}
which imply
\begin{eqnarray}
\gamma_{ij}\, a^i b^j &=& a_i b^i = a_\mu b^\mu = A B, 
\label{eq:abc_NonOrthogonality_1} \\
\gamma_{ij}\, a^i c^j &=& a_i c^i = a_\mu c^\mu = A C, 
\label{eq:abc_NonOrthogonality_2} \\
\gamma_{ij}\, b^i c^j &=& b_i c^i = b_\mu c^\mu = B C.
\label{eq:abc_NonOrthogonality_3}
\end{eqnarray}
Thus orthogonality of $w^\mu, y^\mu, z^\mu$ requires that their parts $a^\mu, b^\mu, c^\mu$ tangent to the spacelike slice are {\em not} orthogonal---for $v^i \ne 0$, that is. For $v^i \rightarrow 0$, the orthonormal comoving tetrad $u^\mu, w^\mu, y^\mu, z^\mu$ becomes an orthonormal lab frame tetrad: the timelike basis vector $u^\mu \rightarrow n^\mu$ (Eqs.~(\ref{eq:ThreeVelocity}), (\ref{eq:BoostFactor})), becoming orthogonal to the spacelike slice; and $A, B, C \rightarrow 0$ (Eqs.~(\ref{eq:ABC_1})-(\ref{eq:ABC_3})), so that the orthonormal triad $w^\mu, y^\mu, z^\mu \rightarrow a^\mu, b^\mu, c^\mu$ (Eqs.~(\ref{eq:wyz_1})-(\ref{eq:wyz_3})), with the tangents $a^\mu, b^\mu, c^\mu$ to the spacelike slice becoming orthonormal (Eqs.~(\ref{eq:abc_Normalization_1})-(\ref{eq:abc_Normalization_3}), (\ref{eq:abc_NonOrthogonality_1})-(\ref{eq:abc_NonOrthogonality_3})).
(Of course, this $v^i \rightarrow 0$ lab frame {\em orthonormal} basis generally does not coincide with the non-orthonormal lab frame {\em coordinate} basis, as is evident from the fact that Eq.~(\ref{eq:LineElement}) is not the Minkowski metric for generic $g_{\mu\nu}$.)  

Specification of the comoving frame tetrad $u^\mu, w^\mu, y^\mu, z^\mu$ is not yet complete.
The timelike comoving basis vector---the Lagrangian observer four-velocity $u^\mu$---is determined by the evolution of the fluid and geometry.
As for the spatial comoving basis vectors $w^\mu, y^\mu, z^\mu$, we have given their Eulerian decompositions in Eqs.~(\ref{eq:wyz_1})-(\ref{eq:wyz_3}).
We can regard $A, B, C$ as determined by Eqs.~(\ref{eq:ABC_1})-(\ref{eq:ABC_3}).
In light of Eq.~(\ref{eq:abc_Spacelike}), 
this leaves nine unknowns---$a^i, b^i, c^i$---but only six equations so far for them (Eqs.~(\ref{eq:abc_Normalization_1})-(\ref{eq:abc_Normalization_3}), (\ref{eq:abc_NonOrthogonality_1})-(\ref{eq:abc_NonOrthogonality_3})).
The remaining three degrees of freedom correspond to the choice of spatial orientation of the comoving orthonormal frame.

One possibility for pinning down the spatial orientation would be to insist that the Lagrangian observer Fermi-Walker transport her orthonormal triad of spatial basis vectors, disallowing spatial rotations in the comoving frame \cite{Misner1973Gravitation}.
This simplifies life for the Lagrangian observer by removing Coriolis force contributions to the comoving frame connection coefficients ${\Gamma^{\hat\mu}}_{\hat\nu \hat\rho}$. 
It would also provide a means of eliminating the appearance of the time derivatives of the tetrads we shall encounter in the momentum space divergence in Eq.~(\ref{eq:MomentumDivergencePartial}).
However, it would add to the list of evolution equations to be solved, and complicate the connection between the particle number and four-momentum conservative reformulations of the Boltzmann equation (see Sec.~\ref{sec:Reconciliation}).

We choose instead to specify three algebraic conditions that simplify the diagonalization of $\gamma_{ij}$ implicit in Eq.~(\ref{eq:MetricDiagonalization}).
It could be required, for instance, that certain lab frame coordinate basis components of $b^i$ and $c^i$ vanish:
\begin{eqnarray} 
b^1 &=& 0, \label{eq:b1_Vanish} \\
c^1 &=& c^2 = 0. \label{eq:c1c2_Vanish}
\end{eqnarray}
The remaining components of $a^i, b^i, c^i$ are then determined by (a) a single equation to be solved for the single non-vanishing component $c^3$ of $c^i$; (b) two equations to be solved for the components $b^2$ and $b^3$ of $b^i$; and (c) three equations to be solved for all three components of $a^i$. 
The resulting equations are not particularly illuminating, and we relegate them to Appendix~\ref{app:TetradSolution}.
We do note that in the special case of diagonal $\gamma_{ij}$ and $v^i = 0$, these equations reduce as expected to $a^1 = \left(\gamma_{11} \right)^{-1/2}, b^2 = \left(\gamma_{22} \right)^{-1/2}, c^3 = \left(\gamma_{33} \right)^{-1/2}$, with all other components vanishing.

For use in more compact index notation, it is convenient to collect the Eulerian decompositions in Eqs.~(\ref{eq:ThreeVelocity}), (\ref{eq:wyz_1})-(\ref{eq:wyz_3}) as
\begin{equation}
{L^\mu}_{\hat\mu} = n^\mu \mathcal{L}_{\hat\mu} + {\ell^\mu}_{\hat\mu},
\label{eq:TetradDecomposition}
\end{equation}
where
\begin{eqnarray}
\left( \mathcal{L}_{\hat\mu} \right) &=& \left( \Lambda, A, B, C \right), 
\label{eq:TetradTimelike} \\
\left( {\ell^\mu}_{\hat\mu} \right) &=& \left( \Lambda v^\mu, a^\mu, b^\mu, c^\mu \right),
\label{eq:TetradSpatial}
\end{eqnarray}
keeping in mind that ${\ell^0}_{\hat\mu} = 0$ according to Eqs.~(\ref{eq:v_Spacelike}) and (\ref{eq:abc_Spacelike}).
The choices in Eqs.~(\ref{eq:b1_Vanish}) and (\ref{eq:c1c2_Vanish}) make ${\ell^i}_{\hat\imath}$ (viewed as a matrix) lower triangular.
(We caution that the use of the base symbol $\ell$ in Ref.~\cite{Cardall2012Conservative-31} differs from its use here.)

Having discussed the interpretation of 
\begin{equation}
\left({L^\mu}_{\hat\mu}\right) = \left( u^\mu, w^\mu, y^\mu, z^\mu \right)
\label{eq:IndexedBasis}
\end{equation}
as a tetrad of comoving frame orthonormal basis vectors, it is easy to see that
\begin{equation}
\left({L^{\hat\mu}}_{\mu}\right) = \left( -u_\mu, w_\mu, y_\mu, z_\mu \right)^T
\label{eq:IndexedBasisInverse}
\end{equation}
is the inverse, for the orthonormality conditions render Eqs.~(\ref{eq:IndexedBasis}) and (\ref{eq:IndexedBasisInverse}) 
consistent with
\begin{equation}
{L^{\hat\mu}}_{\mu} {L^{\mu}}_{\hat\nu} = {\delta^{\hat\mu}}_{\hat\nu}.
\label{eq:TetradDeltaHat}
\end{equation}
The Eulerian decomposition of the inverse is
\begin{equation}
{L^{\hat\mu}}_{\mu} = \mathcal{L}^{\hat\mu} \,n_\mu + {\ell^{\hat\mu}}_{\mu},
\label{eq:TetradDecompositionInverse}
\end{equation}
with
\begin{eqnarray}
\left( \mathcal{L}^{\hat\mu} \right) &=& \left( -\Lambda, A, B, C \right)^T, 
\label{eq:TetradTimelikeInverse} \\
\left( {\ell^{\hat\mu}}_{\mu} \right) &=& \left( -\Lambda v_\mu, a_\mu, b_\mu, c_\mu \right)^T.
\label{eq:TetradSpatialInverse}
\end{eqnarray}
In addition to Eq.~(\ref{eq:TetradDeltaHat}), we have also 
\begin{equation}
{L^{\mu}}_{\hat\mu} {L^{\hat\mu}}_{\nu} = {\delta^{\mu}}_{\nu}.
\label{eq:TetradDelta}
\end{equation}
Inserting Eqs.~(\ref{eq:TetradDecomposition}) and (\ref{eq:TetradDecompositionInverse}) in Eq.~(\ref{eq:TetradDelta}) and taking appropriate contractions with $n^\mu$ and $\gamma_{\mu\nu}$, we find
\begin{eqnarray}
\mathcal{L}_{\hat\mu} \mathcal{L}^{\hat\mu} &=& -1, 
\label{eq:DeltaDecomposition_nn} \\
\mathcal{L}_{\hat\mu} {\ell^{\hat\mu}}_\mu &=& 0, 
\label{eq:DeltaDecomposition_ng}\\ 
{\ell^\mu}_{\hat\mu} \mathcal{L}^{\hat\mu} &=& 0, 
\label{eq:DeltaDecomposition_gn}\\
{\ell^\mu}_{\hat\mu}{\ell^{\hat\mu}}_\nu &=&{\gamma^\mu}_\nu
\label{eq:DeltaDecomposition_gg}
\end{eqnarray}
for various contractions of the Eulerian projections of ${L^\mu}_{\hat\mu}$ and ${L^{\hat\mu}}_{\mu}$.
These expressions of tetrad orthonormality in terms of the Eulerian projections will prove useful in Sec.~\ref{sec:Reconciliation}.

\subsection{Momentum decomposition \label{sec:MomentumDecomposition}}

In a spirit similar to the decomposition of ${L^{\mu}}_{\hat\mu}$---and for reasons that will be become more apparent in Secs.~\ref{sec:Boltzmann} and \ref{sec:Reconciliation}---we will find it useful to decompose the momentum space transformation ${P^{\tilde \imath}}_{\hat \imath}$ and its inverse ${P^{\hat \imath}}_{\tilde \imath}$ (see Eqs.~(\ref{eq:MomentumTransformation}) and (\ref{eq:InverseMomentumTransformation})) into pieces parallel and orthogonal to the particle three-momentum $p^{\hat\imath}$.
In particular we write
\begin{equation}
{P^{\tilde \imath}}_{\hat \imath} =  \frac{Q^{\tilde\imath} \, p_{\hat\imath}}{p} + {U^{\tilde \imath}}_{\hat \imath}.
\label{eq:MomentumDecomposition} 
\end{equation}
Here
\begin{equation}
Q^{\tilde\imath} = \frac{{P^{\tilde \imath}}_{\hat \imath} \, p^{\hat\imath}}{p}
\label{eq:Projection_Q}
\end{equation}
is the projection of ${P^{\tilde \imath}}_{\hat \imath}$ parallel to $p^{\hat\imath}$, and
\begin{equation}
p = \sqrt{p^{\hat\imath} p_{\hat\imath}}
\label{eq:MomentumMagnitude}
\end{equation}
is the three-momentum magnitude.
The projection orthogonal to $p^{\hat\imath}$ is 
\begin{equation}
{U^{\tilde \imath}}_{\hat \imath} = {P^{\tilde \imath}}_{\hat \jmath} {k^{\hat \jmath}}_{\hat \imath}, 
\label{eq:Projection_U}
\end{equation}
in which
\begin{equation}
{k^{\hat \jmath}}_{\hat \imath} = {\delta^{\hat \jmath}}_{\hat \imath} - \frac{p^{\hat\jmath} p_{\hat\imath}}{p^2}
\label{eq:MomentumSpaceProjector}
\end{equation}
is a momentum space orthogonal projector.
Similar to Eq.~(\ref{eq:MomentumDecomposition}), the decomposition of the inverse is
\begin{equation}
{P^{\hat \imath}}_{\tilde \imath} =  \frac{ p^{\hat\imath} \, Q_{\tilde\imath}}{p} + {U^{\hat \imath}}_{\tilde \imath}.
\label{eq:InverseMomentumDecomposition} 
\end{equation}
As with ${L^{\mu}}_{\hat\mu}$ and its projections in Eqs.~(\ref{eq:TetradDelta})-(\ref{eq:DeltaDecomposition_gg}), the fact that
\begin{equation}
{P^{\hat \imath}}_{\tilde \imath} {P^{\tilde \imath}}_{\hat \jmath}= {\delta^{\hat\imath}}_{\hat\jmath}
\label{eq:MomentumTransformDelta}
\end{equation} 
implies the following relations among the projections:
\begin{eqnarray}
Q_{\tilde\imath} Q^{\tilde\imath} &=& 1,
\label{eq:DeltaDecomposition_pp} \\
{U^{\hat \imath}}_{\tilde \imath} Q^{\tilde\imath} &=& 0,
\label{eq:DeltaDecomposition_pk} \\
Q_{\tilde\imath} {U^{\tilde \imath}}_{\hat \imath} &=& 0,
\label{eq:DeltaDecomposition_kp} \\
{U^{\hat \imath}}_{\tilde \imath}  {U^{\tilde \imath}}_{\hat \jmath} &=& {k^{\hat\imath}}_{\hat\jmath},
\label{eq:DeltaDecomposition_kk}
\end{eqnarray}
which follow from inserting Eqs.~(\ref{eq:MomentumDecomposition}) and (\ref{eq:InverseMomentumDecomposition}) into Eq.~(\ref{eq:MomentumTransformDelta}) and taking appropriate contractions with $p^{\hat\imath}$ and ${k^{\hat\imath}}_{\hat\jmath}$.

\subsection{Boltzmann equation \label{sec:Boltzmann}}

We now give, in 3+1 form, the divergences in the number-conservative formulation of the Boltzmann equation in Eq.~(\ref{eq:ConservativeCovariant}).
Key to our derivation is use of the Eulerian decompositions of the tetrad ${L^\mu}_{\hat\mu}$ and its inverse ${L^{\hat\mu}}_\mu$ given in Eqs.~(\ref{eq:TetradDecomposition}) and (\ref{eq:TetradDecompositionInverse}).
These appear in the spacetime divergence in Eq.~(\ref{eq:SpacetimeDivergencePartial}) and in the comoving frame connection coefficients in the momentum space divergence in Eq.~(\ref{eq:MomentumDivergencePartial}). 
The expression for the comoving frame connection coefficients in Eq.~(\ref{eq:ConnectionComoving}) can be recast as 
\begin{equation}
{\Gamma^{\hat{\mu}}}_{\hat{\nu}\hat{\rho}} = {L^{\hat\mu}}_\nu {L^\rho}_{\hat\rho} \left(\nabla_\rho {L^\nu}_{\hat\nu} \right).
\label{eq:RicciRotation}
\end{equation}
This is the expression for the `Ricci rotation coefficients' given by Lindquist \cite{Lindquist1966Relativistic-Tr}. 
Riffert \cite{Riffert1986A-general-Euler} emphasized Eq.~(\ref{eq:ConnectionComoving}), claiming greater simplicity, but their equivalence is apparent.
For our purposes, Eq.~(\ref{eq:RicciRotation}) is more convenient, as it allows us to obtain a 3+1 formulation of the Boltzmann equation while almost completely avoiding explicit encounters with connection coefficients (see Appendix~\ref{app:ComovingConnection}). 

Of the two terms on the left-hand side of Eq.~(\ref{eq:ConservativeCovariant}), the spacetime divergence given by Eq.~(\ref{eq:SpacetimeDivergencePartial}) is by far the simpler.
Using Eqs.~(\ref{eq:MetricDeterminant}), (\ref{eq:TetradDecomposition}), and (\ref{eq:UnitNormalU}), this spacetime divergence is
\begin{equation}
S_N = \frac{\left( -p_{\hat 0} \right)}{\alpha\sqrt{\gamma}}\left[ \frac{\partial \left(D_N \right)}{\partial t} + \frac{\partial \left( F_N \right)^i }{\partial x^i} \right],
\label{eq:Spacetime_N_31}
\end{equation}
where 
\begin{eqnarray}
D_N &=& \frac{\sqrt{\gamma}}{\left( -p_{\hat 0} \right)} \, \mathcal{L}_{\hat\mu} \, p^{\hat\mu} f, 
\label{eq:Density_N} \\
\left( F_N \right)^i &=& \frac{\sqrt{\gamma}}{\left( -p_{\hat 0} \right)} \left( \alpha\, {\ell^i}_{\hat\mu} - \beta^i \mathcal{L}_{\hat\mu} \right) p^{\hat\mu} f
\label{eq:Flux_N}
\end{eqnarray}
are respectively the conserved number density and number flux.

Using the results for the comoving frame connection coefficients given in Appendix~\ref{app:ComovingConnection}, the momentum space divergence in Eq.~(\ref{eq:MomentumDivergencePartial}) can be expressed
\begin{eqnarray}
M_N &=& \frac{1}{\alpha\sqrt{\gamma}} \frac{\left( -p_{\hat 0} \right)}{\sqrt{\lambda}} \frac{\partial}{\partial p^{\tilde\imath}} \left\{
\sqrt{\lambda} \, \frac{Q^{\tilde\imath} \left(-p_{\hat 0}\right)}{p}\! \left[ \left( R_N \right)^{\hat 0} + \left( O_N \right)^{\hat 0} \right] \right. \nonumber \\
&& \left. + \sqrt{\lambda} \, {U^{\tilde \imath}}_{\hat \imath} \left[ \left( R_N \right)^{\hat \imath} + \left( O_N \right)^{\hat \imath} \right] \right\},
\label{eq:Momentum_N_31}
\end{eqnarray}
where
\begin{eqnarray}
\left( R_N \right)^{\hat\rho} &=& \frac{\alpha\sqrt{\gamma}}{\left( -p_{\hat 0} \right)}\, p^{\hat\nu} p^{\hat\mu} f \nonumber \\
& & \times \left[  \mathcal{L}^{\hat\rho}\, {\ell^j}_{\hat\nu} \left( \frac{ \mathcal{L}_{\hat\mu}  }{\alpha} \frac{\partial \alpha}{\partial x^j} 
-  {\ell^k}_{\hat\mu} \, K_{jk} \right) 
 \right. \nonumber \\
& & \left. - {\ell^{\hat\rho j}} \! \left( \! \frac{\mathcal{L}_{\hat\nu} \mathcal{L}_{\hat\mu}  }{\alpha} \frac{\partial \alpha}{\partial x^j} 
\!-\! \frac{\ell_{k\hat\nu} \, \mathcal{L}_{\hat\mu}}{\alpha} \frac{\partial \beta^k}{\partial x^j} 
\!-\!  \frac{{\ell^k}_{\hat\nu} \,{\ell^i}_{\hat\mu}}{2} \frac{\partial \gamma_{ki} }{\partial x^j} \!\right) \!\right]\!  \nonumber \\
& & \label{eq:Redshift_N}
\end{eqnarray}
describes momentum shifts (that is, redshift and angular aberration in momentum space spherical coordinates) due to gravity as embodied in the spacetime geometry, and 
\begin{eqnarray}
\left( O_N \right)^{\hat\rho} &=& \frac{\sqrt{\gamma}}{\left( -p_{\hat 0} \right)} \, p^{\hat\nu} p^{\hat\mu} f \nonumber \\
& & \times \left\{ \mathcal{L}^{\hat\rho} \left[  \mathcal{L}_{\hat\mu} \frac{\partial \mathcal{L}_{\hat\nu}}{\partial t} + \left( \alpha\, {\ell^j}_{\hat\mu} - \beta^j \mathcal{L}_{\hat\mu}\right) \frac{\partial \mathcal{L}_{\hat\nu}}{\partial x^j} \right] \right. \nonumber \\
& & \left. 
- \ell^{\hat\rho k} \left[ \mathcal{L}_{\hat\mu} \frac{\partial \ell_{k \hat\nu}}{\partial t}  +  \left(\alpha\, {\ell^j}_{\hat\mu} - \beta^j \mathcal{L}_{\hat\mu}\right) \frac{\partial \ell_{k \hat\nu}}{\partial x^j}
\right] \right\} \nonumber \\
& &
\label{eq:Observer_N}
\end{eqnarray}
are `observer corrections' due to the acceleration of a Lagrangian observer (and partially entangled with the geometry as well), arising from the tetrad ${L^\mu}_{\hat\mu}$ connecting the orthonormal comoving frame to the lab frame coordinate basis.
It is through the decomposition of ${P^{\tilde \imath}}_{\hat \imath}$ in Eq.~(\ref{eq:MomentumDecomposition})---along with Eq.~(\ref{eq:GeodesicIdentity})---that the first term of Eq.~(\ref{eq:Momentum_N_31}) contains ${\Gamma^{\hat{0}}}_{\hat{\nu}\hat{\mu}}$, expressed here in terms of $\left( R_N \right)^{\hat 0}$ and $\left( O_N \right)^{\hat 0}$.
This will prove beneficial in relating the number-conservative formulation of the Boltzmann equation to the four-momentum-conservative formulation.

\subsection{Lepton number and four-momentum reconciliation \label{sec:Reconciliation}}

In addition to the number-conservative reformulation of the Boltzmann equation given by Eqs.~(\ref{eq:ConservativeCovariant})-(\ref{eq:MomentumDivergencePartial}), there is also a four-momentum-conservative reformulation \cite{Cardall2003Conservative-fo}. 
It has the same basic structure, with spacetime and momentum space divergences, but with an additional index:
\begin{equation}
\left(S_T\right)^\rho + \left(M_T\right)^\rho = p^\rho C[f].
\label{eq:MomentumConservative}
\end{equation}
Each of the components of this equation is redundant with Eq.~(\ref{eq:ConservativeCovariant}) for number exchange with the fluid; but the right-hand side yields source terms for the fluid equations, and the structure of the left-hand side illuminates issues faced in accounting for energy and momentum conservation in the discretized case.
We examine the forms taken by $\left(S_T\right)^\rho$ and $\left(M_T\right)^\rho$ in our chosen phase space coordinates, and consider how the four-momentum-conservative reformuation in Eq.~(\ref{eq:MomentumConservative}) is related to the number-conservative reformulation in Eq.~(\ref{eq:ConservativeCovariant}).

We begin at a high level, working with covariant derivatives in spacetime.
Comparing the right-hand sides of Eqs.~(\ref{eq:MomentumConservative}) and Eq.~(\ref{eq:ConservativeCovariant}), we see that the left-hand side of Eq.~(\ref{eq:MomentumConservative}) must follow from multiplying the left-hand side of Eq.~(\ref{eq:ConservativeCovariant}) by $p^\rho = {L^\rho}_{\hat{\rho}}\,p^{\hat{\rho}}$.
Using $S_N$ as given in Eq.~(\ref{eq:SpacetimeDivergenceCovariant}), we have
\begin{equation}
p^\rho  S_N = \left(S_T\right)^\rho + \left(E_S\right)^\rho,
\end{equation} 
where
\begin{equation}
\left(S_T\right)^\rho = \nabla_\mu \left( {L^{\rho}}_{\hat\rho} {L^{\mu}}_{\hat\mu}\, p^{\hat\rho} p^{\hat\mu}   f \right),
\label{eq:SpacetimeDivergence_T}
\end{equation}
and 
\begin{equation}
\left(E_S\right)^\rho = - f {L^{\mu}}_{\hat\mu}\, p^{\hat\mu} p^{\hat\rho} \left( \nabla_\mu {L^{\rho}}_{\hat\rho} \right)
\label{eq:SpacetimeExtra_T}
\end{equation}
is the `extra' term that results from pulling $p^\rho$ inside the divergence.
Turning to the momentum space divergence and using $M_N$ as given in Eq.~(\ref{eq:MomentumDivergencePartial}), we have
\begin{equation}
p^\rho  M_N = \left(M_T\right)^\rho + \left(E_M\right)^\rho,
\end{equation} 
where
\begin{equation}
\left(M_T\right)^\rho = \frac{\left(-p_{\hat 0}\right)}{\sqrt{\lambda}} \frac{\partial}{\partial p^{\tilde\imath}} \left[
- \frac{ \sqrt{\lambda} }{\left(-p_{\hat 0}\right)} {P^{\tilde\imath}}_{\hat\imath} {\Gamma^{\hat\imath}}_{\hat\nu \hat\mu} \, p^{\hat\nu} p^{\hat\mu} {L^{\rho}}_{\hat\rho} p^{\hat\rho}  f \right],
\label{eq:MomentumDivergence_T}
\end{equation}
and
\begin{eqnarray}
\left(E_M\right)^\rho &=& f \,{\Gamma^{\hat\imath}}_{\hat\nu \hat\mu} \, p^{\hat\nu} p^{\hat\mu} {L^{\rho}}_{\hat\rho} \,{P^{\tilde\imath}}_{\hat\imath} \frac{\partial p^{\hat\rho}}{\partial p^{\tilde\imath}} 
\label{eq:ExtraMomentum_T_n_1}\\
&=& f \,{\Gamma^{\hat\imath}}_{\hat\nu \hat\mu} \, p^{\hat\nu} p^{\hat\mu} {L^{\rho}}_{\hat\rho} \, \frac{\partial p^{\hat\rho}}{\partial p^{\hat\imath}},
\end{eqnarray}
in which we have used Eq.~(\ref{eq:MomentumTransformation}) in the second line.
Using Eqs.~(\ref{eq:MomentumDerivativeComponents}), (\ref{eq:MassShellDerivative}), and (\ref{eq:Geodesic_p}), this `extra' momentum space derivative term becomes
\begin{equation}
\left(E_M\right)^\rho = f \,{L^{\rho}}_{\hat\rho} \,{\Gamma^{\hat\rho}}_{\hat\nu \hat\mu} \, p^{\hat\nu} p^{\hat\mu}.  
\label{eq:ExtraMomentum_T_n_2}
\end{equation} 
This can be expressed as
\begin{equation}
\left(E_M\right)^\rho = f {L^\mu}_{\hat\mu} \left(\nabla_\mu {L^\rho}_{\hat\nu} \right)p^{\hat\nu} p^{\hat\mu}
\end{equation}
by virtue of Eqs.~(\ref{eq:RicciRotation}) and (\ref{eq:TetradDelta}).
Comparing with Eq.~(\ref{eq:SpacetimeExtra_T}), we see that 
\begin{equation}
\left(E_S\right)^\rho + \left(E_M\right)^\rho = 0.
\end{equation}
Thus Eq.~(\ref{eq:MomentumConservative}) is verified with the divergences $\left(S_T\right)^\rho$ and $\left(M_T\right)^\rho$ given by Eqs.~(\ref{eq:SpacetimeDivergence_T}) and (\ref{eq:MomentumDivergence_T}) respectively.

The relationship between Eq.~(\ref{eq:ConservativeCovariant}) for particle number exchange with the fluid and Eq.~(\ref{eq:MomentumConservative}) for four-momentum exchange also can be examined in terms of the more detailed expressions for $S_N$ and $M_N$ we have given in Eqs.~(\ref{eq:Spacetime_N_31})-(\ref{eq:Observer_N}).
In doing so it is convenient to split the four-momentum equations into an energy equation and a three-momentum equation.
The equation for energy exchange with the fluid is obtained by multiplying Eq.~(\ref{eq:ConservativeCovariant}) by the lab frame neutrino energy, i.e. the projection of $p^\rho$ orthogonal to the spacelike slice:
\begin{equation}
- n_\rho\, p^\rho =  -n_\rho\, {L^\rho}_{\hat\rho}\, p^{\hat\rho} = \mathcal{L}_{\hat\rho} \, p^{\hat\rho},
\label{eq:LabEnergy}
\end{equation}
where we have used the Eulerian decomposition of ${L^\rho}_{\hat\rho}$ given by Eq.~(\ref{eq:TetradDecomposition}).
An equation for momentum exchange with the fluid is obtained by multiplying Eq.~(\ref{eq:ConservativeCovariant}) by the lab frame neutrino three-momentum, i.e. the projection of $p^\rho$ tangent to the spacelike slice:
\begin{equation}
\gamma_{j \rho}\, p^\rho =  \gamma_{j \rho}\, {L^\rho}_{\hat\rho} \, p^{\hat\rho} = \ell_{j \hat\rho}\, p^{\hat\rho},
\label{eq:LabMomentum}
\end{equation}
where once again we have used the Eulerian decomposition of ${L^\rho}_{\hat\rho}$ given by Eq.~(\ref{eq:TetradDecomposition}).
Let us represent the right-hand sides of both Eq.~(\ref{eq:LabEnergy}) and (\ref{eq:LabMomentum}) by
\begin{equation}
\mathcal{L}_{\hat\rho} \, p^{\hat\rho}, \ \ell_{j \hat\rho}\, p^{\hat\rho} \rightarrow Z_{\hat\rho}\, p^{\hat\rho}.
\end{equation}
Thus using $Z_{\hat\rho}\, p^{\hat\rho}$ to represent both the lab frame energy $\mathcal{L}_{\hat\rho} \, p^{\hat\rho}$ and the lab frame momentum $\ell_{j \hat\rho}\, p^{\hat\rho}$, we derive in parallel the energy- and momentum-conservative reformulations of the Boltzmann equation.

We begin with the spacetime divergence.
First, using Eq.~(\ref{eq:Spacetime_N_31})-(\ref{eq:Flux_N}), we have
\begin{equation}
 Z_{\hat\rho} \, p^{\hat\rho} S_N = \frac{\left( -p_{\hat 0} \right)}{\alpha\sqrt{\gamma}}\left[ \frac{\partial \left(D_{T,Z} \right)}{\partial t} + \frac{\partial \left( F_{T,Z} \right)^i }{\partial x^i} + E_{S,Z} \right].
\label{eq:Spacetime_T_31_Z}
\end{equation}
Here
\begin{equation}
D_{T,Z} 
= Z_{\hat\rho} \, p^{\hat\rho} D_N
\label{eq:Density_T_Z}
\end{equation}
is the `conserved' density;
for $Z_{\hat\rho}\, p^{\hat\rho} \rightarrow \mathcal{L}_{\hat\rho} \, p^{\hat\rho}$ we have the `conserved' energy density
\begin{equation}
D_{T,n} 
= \mathcal{L}_{\hat\rho} \, p^{\hat\rho} D_N,
\label{eq:Density_T_n}
\end{equation}
while for $Z_{\hat\rho}\, p^{\hat\rho} \rightarrow \ell_{j \hat\rho}\, p^{\hat\rho}$ we have the `conserved' momentum density
\begin{equation}
\left( D_{T,\gamma} \right)_j 
= \ell_{j \hat\rho} \, p^{\hat\rho} D_N.
\label{eq:Density_T_g}
\end{equation}
Next in Eq.~(\ref{eq:Spacetime_T_31_Z}) is the flux 
\begin{equation}
\left( F_{T,Z} \right)^i 
= Z_{\hat\rho} \, p^{\hat\rho} \left( F_N \right)^i 
\label{eq:Flux_T_Z},
\end{equation}
which for $Z_{\hat\rho}\, p^{\hat\rho} \rightarrow \mathcal{L}_{\hat\rho} \, p^{\hat\rho}$ is the energy flux
\begin{equation}
\left( F_{T,n} \right)^i 
=\mathcal{L}_{\hat\rho} \, p^{\hat\rho} \left( F_N \right)^i, 
\label{eq:Flux_T_n}
\end{equation}
and for $Z_{\hat\rho}\, p^{\hat\rho} \rightarrow \ell_{j \hat\rho}\, p^{\hat\rho}$ is the momentum flux
\begin{equation}
{\left( F_{T,\gamma} \right)^i}_j 
=\ell_{j \hat\rho} \, p^{\hat\rho} \left( F_N \right)^i.
\label{eq:Flux_T_g}
\end{equation}
Finally in Eq.~(\ref{eq:Spacetime_T_31_Z}) we have
\begin{equation}
E_{S,Z} = - \frac{\sqrt{\gamma}}{\left( -p_{\hat 0} \right)}  \, p^{\hat\rho} p^{\hat\mu} f \left[ \mathcal{L}_{\hat\mu}   \frac{\partial Z_{\hat\rho}}{\partial t} + \left( \alpha\, {\ell^i}_{\hat\mu} - \beta^i \mathcal{L}_{\hat\mu} \right) \frac{\partial Z_{\hat\rho}}{\partial x^i}\right],
\label{eq:SpacetimeExtra_T_Z}
\end{equation}
the `extra' term that comes from pulling $Z_{\hat\rho} \, p^{\hat\rho}$ inside the time and space derivatives.
For $Z_{\hat\rho}\, p^{\hat\rho} \rightarrow \mathcal{L}_{\hat\rho} \, p^{\hat\rho}$,
\begin{equation}
E_{S,n} = - \frac{\sqrt{\gamma}}{\left( -p_{\hat 0} \right)}  \, p^{\hat\rho} p^{\hat\mu} f \left[ \mathcal{L}_{\hat\mu}   \frac{\partial \mathcal{L}_{\hat\rho}}{\partial t} + \left( \alpha\, {\ell^i}_{\hat\mu} - \beta^i \mathcal{L}_{\hat\mu} \right) \frac{\partial \mathcal{L}_{\hat\rho}}{\partial x^i}\right]
\label{eq:SpacetimeExtra_T_n}
\end{equation}
is the `extra' term in the energy equation,
while for $Z_{\hat\rho}\, p^{\hat\rho} \rightarrow \ell_{j \hat\rho}\, p^{\hat\rho}$
\begin{eqnarray}
\left(E_{S,\gamma}\right)_j &=& - \frac{\sqrt{\gamma}}{\left( -p_{\hat 0} \right)}  \, p^{\hat\rho} p^{\hat\mu} f \nonumber \\
& & \times \left[ \mathcal{L}_{\hat\mu}   \frac{\partial \ell_{j \hat\rho}}{\partial t} + \left( \alpha\, {\ell^i}_{\hat\mu} - \beta^i \mathcal{L}_{\hat\mu} \right) \frac{\partial \ell_{j \hat\rho}}{\partial x^i}\right]
\label{eq:SpacetimeExtra_T_g}
\end{eqnarray}
is the `extra' term in the momentum equation.
(We caution that $E_{S,n} \ne -n_\rho \left(E_S\right)^\rho$ and $\left(E_{S,\gamma}\right)_j \ne -\gamma_{j \rho} \left(E_S\right)^\rho$, where $\left(E_S\right)^\rho$ is given by Eq.~(\ref{eq:SpacetimeExtra_T}), as might plausibly but incorrectly be suggested by the notation.)

Next we turn to the momentum space divergence.
The result of multiplying Eq.~(\ref{eq:Momentum_N_31}) with $Z_{\hat\rho}\, p^{\hat\rho}$ can be cast in the form
\begin{eqnarray}
Z_{\hat\rho} \, p^{\hat\rho} \, M_N &=& \frac{1}{\alpha\sqrt{\gamma}} \frac{\left( -p_{\hat 0} \right)}{\sqrt{\lambda}} \nonumber \\ 
& & \times \frac{\partial}{\partial p^{\tilde\imath}} \left\{
\sqrt{\lambda} \, \frac{Q^{\tilde\imath} \left(-p_{\hat 0}\right)}{p}\! \left[ \left( R_{T,Z} \right)^{\hat 0} + \left( O_{T,Z} \right)^{\hat 0} \right] \right. \nonumber \\
&& \left. + \sqrt{\lambda} \, {U^{\tilde \imath}}_{\hat \imath} \left[ \left( R_{T,Z} \right)^{\hat \imath} + \left( O_{T,Z} \right)^{\hat \imath} \right] \right\} \nonumber \\
& & + \frac{\left(-p_{\hat 0}\right)}{\alpha\sqrt{\gamma}} E_{M,Z}.
\label{eq:Momentum_Z_31}
\end{eqnarray}
Here
\begin{eqnarray}
\left( R_{T,Z} \right)^{\hat\mu} &=& Z_{\hat\rho} \, p^{\hat\rho} \left( R_N \right)^{\hat\mu}, 
\label{eq:Redshift_T_Z} \\
\left( O_{T,Z} \right)^{\hat\mu} &=& Z_{\hat\rho} \, p^{\hat\rho} \left( O_N \right)^{\hat\mu},
\label{eq:Observer_T_Z}
\end{eqnarray} 
or
\begin{eqnarray}
\left( R_{T,n} \right)^{\hat\mu} &=& \mathcal{L}_{\hat\rho} \, p^{\hat\rho} \left( R_N \right)^{\hat\mu}, 
\label{eq:Redshift_T_n} \\
\left( O_{T,n} \right)^{\hat\mu} &=& \mathcal{L}_{\hat\rho} \, p^{\hat\rho} \left( O_N \right)^{\hat\mu},
\label{eq:Observer_T_n}
\end{eqnarray} 
for $Z_{\hat\rho} \, p^{\hat\rho} \rightarrow \mathcal{L}_{\hat\rho} \, p^{\hat\rho} $ (energy equation), and
\begin{eqnarray}
{\left( R_{T,\gamma} \right)^{\hat\mu}}_j &=& \ell_{j \hat\rho} \, p^{\hat\rho} \left( R_N \right)^{\hat\mu}, 
\label{eq:Redshift_T_g}\\
{\left( O_{T,\gamma} \right)^{\hat\mu}}_j &=& \ell_{j \hat\rho} \, p^{\hat\rho} \left( O_N \right)^{\hat\mu},
\label{eq:Observer_T_g}
\end{eqnarray} 
for $Z_{\hat\rho} \, p^{\hat\rho} \rightarrow \ell_{j \hat\rho} \, p^{\hat\rho}$ (momentum equation), 
with $\left( R_N \right)^{\hat\mu}$ and $\left( O_N \right)^{\hat\mu}$ given by Eqs.~(\ref{eq:Redshift_N}) and (\ref{eq:Observer_N}) respectively.
The `extra' term $E_{M,Z}$ in Eq.~(\ref{eq:Momentum_Z_31}) is
\begin{eqnarray}
E_{M,Z} &=& - \left\{
\frac{Q^{\tilde\imath} \left(-p_{\hat 0}\right)}{p}\! \left[ \left( R_{N} \right)^{\hat 0} + \left( O_{N} \right)^{\hat 0} \right] \right. \nonumber \\
&& \left. + {U^{\tilde \imath}}_{\hat \imath} \left[ \left( R_{N} \right)^{\hat \imath} + \left( O_{N} \right)^{\hat \imath} \right] \right\} Z_{\hat\rho} \frac{\partial p^{\hat\rho}}{\partial p^{\tilde\imath}}. 
 \label{eq:ExtraMomentum_T_Z_3} 
\end{eqnarray}
The last two factors can be rewritten as
\begin{eqnarray}
Z_{\hat\rho} \frac{\partial p^{\hat\rho}}{\partial p^{\tilde\imath}} &=& {P^{\hat\jmath}}_{\tilde\imath} Z_{\hat\rho} \frac{\partial p^{\hat\rho}}{\partial p^{\hat\jmath}} \\
&=& \left( \frac{ p^{\hat\jmath} \, Q_{\tilde\imath}}{p} + {U^{\hat \jmath}}_{\tilde \imath}\right) \left[Z_{\hat 0} \frac{p_{\hat\jmath}}{\left(-p_{\hat 0}\right)} + Z_{\hat\jmath} \right],
\end{eqnarray}
in which we have used Eqs.~(\ref{eq:InverseMomentumTransformation}), (\ref{eq:InverseMomentumDecomposition}), and (\ref{eq:MomentumDerivativeComponents}). 
Plugging back into Eq.~(\ref{eq:ExtraMomentum_T_Z_3}) and using Eqs.~(\ref{eq:DeltaDecomposition_pp})-(\ref{eq:DeltaDecomposition_kk}) yields
\begin{eqnarray}
E_{M,Z} &=& - \left\{
\frac{\left(-p_{\hat 0}\right)}{p} \frac{p^{\hat\jmath}}{p} \! \left[ \left( R_{N} \right)^{\hat 0} + \left( O_{N} \right)^{\hat 0} \right] \right. \nonumber \\
&& \left. + {k^{\hat\jmath}}_{\hat \imath} \left[ \left( R_{N} \right)^{\hat \imath} + \left( O_{N} \right)^{\hat \imath} \right] \right\} \left[Z_{\hat 0} \frac{p_{\hat\jmath}}{\left(-p_{\hat 0}\right)} + Z_{\hat\jmath} \right]. \nonumber \\
& &
\end{eqnarray}
This reduces further to 
\begin{eqnarray}
E_{M,Z} &=& 
- \left[ Z_{\hat\rho} - \frac{p^{\hat\jmath} Z_{\hat\jmath}}{p^2}\, p_{\hat\rho} \right] \left[ \left( R_{N} \right)^{\hat\rho} + \left( O_{N} \right)^{\hat\rho} \right]  \\
&=& 
- Z_{\hat\rho} \left[ \left( R_{N} \right)^{\hat\rho} + \left( O_{N} \right)^{\hat\rho} \right]. 
 \label{eq:ExtraMomentum_T_Z_4} 
\end{eqnarray}
The second line follows from the fact that
\begin{equation}
0 = p_{\hat\rho} \, {\Gamma^{\hat\rho}}_{\hat\nu \hat\mu}\, p^{\hat\nu} p^{\hat\mu},
\label{eq:GeodesicIdentityCovariant}
\end{equation}
by virtue of Eqs.~(\ref{eq:MassShellDerivative}) and (\ref{eq:Geodesic_p}), through which
\begin{equation} 
0 = p_{\hat\rho} \left[ \left( R_{N} \right)^{\hat\rho} + \left( O_{N} \right)^{\hat\rho} \right]
\label{eq:GeodesicIdentityReduced}
\end{equation} 
is also implied (see Appendix~\ref{app:ComovingConnection} and Sec.~\ref{sec:Boltzmann}).

Considering the first term in Eq.~(\ref{eq:ExtraMomentum_T_Z_4}), we will see that the contribution of the redshift/aberration terms $\left( R_{N} \right)^{\hat\rho}$ to the `extra' term $E_{M,Z}$ from the momentum space divergence becomes the geometric/gravitational source term arising from the covariant derivative in Eq.~(\ref{eq:SpacetimeDivergence_T}).
For $Z_{\hat\rho}  \rightarrow \mathcal{L}_{\hat\rho}$ (energy equation), we have
\begin{eqnarray}
-G_{T,n} &=& -\mathcal{L}_{\hat\rho} \left( R_{N} \right)^{\hat\rho} \\
&=& \!\frac{\alpha\sqrt{\gamma}}{\left( -p_{\hat 0} \right)} p^{\hat\nu} p^{\hat\mu} f \, {\ell^j}_{\hat\nu} \!\left(\! \frac{ \mathcal{L}_{\hat\mu}  }{\alpha} \frac{\partial \alpha}{\partial x^j} 
\!-\!  {\ell^k}_{\hat\mu} \, K_{jk} \!\right)\!, 
\label{eq:GeometrySource_n}
\end{eqnarray}
in which only the second line of Eq.~(\ref{eq:Redshift_N}) contributes, thanks to Eqs.~(\ref{eq:DeltaDecomposition_nn}) and (\ref{eq:DeltaDecomposition_ng}).
On the other hand, for $Z_{\hat\rho}  \rightarrow \ell_{j \hat\rho} $ (momentum equation) we have
\begin{eqnarray}  
-\left(G_{T,\gamma}\right)_j &=& - \ell_{j \hat\rho}  \left( R_N \right)^{\hat\rho} \\
&=& \frac{\alpha\sqrt{\gamma}}{\left( -p_{\hat 0} \right)}\, p^{\hat\nu} p^{\hat\mu} f 
\left(  \frac{\mathcal{L}_{\hat\nu} \mathcal{L}_{\hat\mu}  }{\alpha} \frac{\partial \alpha}{\partial x^j}  \right. \nonumber  \\
& & \left.
- \ell_{k\hat\nu} \, \mathcal{L}_{\hat\mu} \frac{\partial \beta^k}{\partial x^j} 
-  \frac{{\ell^k}_{\hat\nu} \,{\ell^i}_{\hat\mu}}{2} \frac{\partial \gamma_{ki} }{\partial x^j} \right), 
\label{eq:GeometrySource_g}
\end{eqnarray}
in which only the third line of Eq.~(\ref{eq:Redshift_N}) contributes, 
thanks to Eqs.~(\ref{eq:DeltaDecomposition_gn}) and (\ref{eq:DeltaDecomposition_gg}).

Turning to the second term of Eq.~(\ref{eq:ExtraMomentum_T_Z_4}), we show that
the contribution of the observer corrections $\left( O_{N} \right)^{\hat\rho}$ to the `extra' term $E_{M,Z}$ from the momentum space divergence cancels with the `extra' term $E_{S,Z}$ from the spacetime divergence.
Indeed, for $Z_{\hat\rho}  \rightarrow \mathcal{L}_{\hat\rho} $ (energy equation), only the second line of Eq.~(\ref{eq:Observer_N}) contributes thanks to Eqs.~(\ref{eq:DeltaDecomposition_nn}) and (\ref{eq:DeltaDecomposition_ng}):
\begin{equation}
- \mathcal{L}_{\hat\rho} \left( O_N \right)^{\hat\rho} = - E_{S,n},
\label{eq:ExtraMomentum_T_n_Observer}
\end{equation}
where we have compared the result with Eq.~(\ref{eq:SpacetimeExtra_T_n}).
On the other hand, for $Z_{\hat\rho}  \rightarrow \ell_{j \hat\rho} $ (momentum equation), only the third line of Eq.~(\ref{eq:Observer_N}) contributes thanks to Eqs.~(\ref{eq:DeltaDecomposition_gn}) and (\ref{eq:DeltaDecomposition_gg}):
\begin{equation}
- \ell_{j \hat\rho} \left( O_N \right)^{\hat\rho} = - \left( E_{S,\gamma} \right)_j,
\label{eq:ExtraMomentum_T_g_Observer}
\end{equation}
where we have compared the result with Eq.~(\ref{eq:SpacetimeExtra_T_g}).

We are now in a position to combine the above results into the energy- and momentum-conservative reformulations of the left-hand side of the Boltzmann equation.
In the energy equation,
\begin{eqnarray}
\mathcal{L}_{\hat\rho} \, p^{\hat\rho} \left( S_N + M_N \right) 
&=& -n_\rho\, p^\rho \left( S_N + M_N \right) \\
&=& -n_\rho \left[ \left(S_T\right)^\rho + \left(M_T\right)^\rho \right],
\end{eqnarray} 
where 
\begin{equation}
-n_\rho \left(S_T\right)^\rho = \frac{\left( -p_{\hat 0} \right)}{\alpha\sqrt{\gamma}}\left[ \frac{\partial \left(D_{T,n} \right)}{\partial t} + \frac{\partial \left( F_{T,n} \right)^i }{\partial x^i} -G_{T,n} \right]
\label{eq:SpacetimeDivergence_n}
\end{equation}
and
\begin{eqnarray}
-n_\rho \left(M_T\right)^\rho &=& 
\frac{1}{\alpha\sqrt{\gamma}} \frac{\left( -p_{\hat 0} \right)}{\sqrt{\lambda}} \nonumber \\ 
& & \times \frac{\partial}{\partial p^{\tilde\imath}} \left\{
\sqrt{\lambda} \, \frac{Q^{\tilde\imath} \left(-p_{\hat 0}\right)}{p}\! \left[ \left( R_{T,n} \right)^{\hat 0} + \left( O_{T,n} \right)^{\hat 0} \right] \right. \nonumber \\
&& \left. + \sqrt{\lambda} \, {U^{\tilde \imath}}_{\hat \imath} \left[ \left( R_{T,n} \right)^{\hat \imath} + \left( O_{T,n} \right)^{\hat \imath} \right] \right\}.
\label{eq:MomentumDivergence_n}
\end{eqnarray}
We note the consistency of Eq.~(\ref{eq:SpacetimeDivergence_n}) with Eq.~(A14) of Ref.~\cite{Cardall2012Conservative-31}, with the substitution $T^{\mu\nu} \rightarrow {L^{\mu}}_{\hat\mu} {L^{\nu}}_{\hat\nu}\, p^{\hat\mu} p^{\hat\nu}   f$.   
In the momentum equation, we have
\begin{eqnarray}
\ell_{j \hat\rho} \, p^{\hat\rho} \left( S_N + M_N \right) 
&=& \gamma_{j \rho}\, p^\rho \left( S_N + M_N \right) \\
&=& \gamma_{j \rho} \left[ \left(S_T\right)^\rho + \left(M_T\right)^\rho \right],
\end{eqnarray} 
where 
\begin{equation}
\gamma_{j \rho} \left(S_T\right)^\rho = \frac{\left( -p_{\hat 0} \right)}{\alpha\sqrt{\gamma}}\left[ \frac{\partial \left(D_{T,\gamma} \right)_j}{\partial t} + \frac{\partial {\left( F_{T,\gamma} \right)^i}_j }{\partial x^i} -\left(G_{T,\gamma}\right)_j \right]
\label{eq:SpacetimeDivergence_g}
\end{equation}
and
\begin{eqnarray}
\gamma_{j \rho} \left(M_T\right)^\rho &=& 
\frac{1}{\alpha\sqrt{\gamma}} \frac{\left( -p_{\hat 0} \right)}{\sqrt{\lambda}} \nonumber \\ 
& & \times \frac{\partial}{\partial p^{\tilde\imath}} \left\{
\sqrt{\lambda} \, \frac{Q^{\tilde\imath} \left(-p_{\hat 0}\right)}{p}\! \left[ {\left( R_{T,\gamma} \right)^{\hat 0}}_j + {\left( O_{T,\gamma} \right)^{\hat 0}}_j \right] \right. \nonumber \\
&& \left. + \sqrt{\lambda} \, {U^{\tilde \imath}}_{\hat \imath} \left[ {\left( R_{T,\gamma} \right)^{\hat \imath}}_j + {\left( O_{T,\gamma} \right)^{\hat \imath}}_j \right] \right\}.
\label{eq:MomentumDivergence_g}
\end{eqnarray}
We note the consistency of Eq.~(\ref{eq:SpacetimeDivergence_g}) with Eq.~(A31) of Ref.~\cite{Cardall2012Conservative-31}, with the substitution $T^{\mu\nu} \rightarrow {L^{\mu}}_{\hat\mu} {L^{\nu}}_{\hat\nu}\, p^{\hat\mu} p^{\hat\nu}   f$.

\section{Conclusion \label{sec:Conclusion}}

\begin{table}[b]
\caption{\label{tab:Spacetime}
Some spacetime and fluid variables.}
\begin{ruledtabular}
\begin{tabular}{lll}
$\Sigma_t$ & Spacelike slice labeled by time $t$ & Sec.~\ref{sec:GeometryDescription} \\
$x^\mu$ & Spacetime position & Sec.~\ref{sec:GeometryDescription} \\
$\alpha$ & Lapse function & Eq.~(\ref{eq:LineElement}) \\
$\beta^i$ & Shift vector & Eq.~(\ref{eq:LineElement}) \\
$\gamma_{ij} $ & Three-metric & Eq.~(\ref{eq:LineElement}) \\
$\gamma$ & Determinant of $\gamma_{ij}$ & Eq.~(\ref{eq:MetricDeterminant}) \\
$K_{ij} $ & Extrinsic curvature & Eq.~(\ref{eq:Gradient_n}) \\
$n^\mu$ & Unit normal to $\Sigma_t$; Eulerian four-velocity &   Eq.~(\ref{eq:UnitNormalU}) \\
$u^\mu$ & Lagrangian (i.e. fluid) four-velocity & Eq.~(\ref{eq:FluidVelocityU}) \\
$v^\mu$ & Three-velocity of Lagrangian observers & Eq.~(\ref{eq:ThreeVelocity}) \\
$\Lambda $ & Lorentz factor of Lagrangian observers & Eq.~(\ref{eq:BoostFactor}) \\
$\gamma_{\mu\nu}$ & Projector orthogonal to $n^\mu$ & Eq.~(\ref{eq:OrthogonalProjector}) \\
\end{tabular}
\end{ruledtabular}
\end{table}

\begin{table}[b]
\caption{\label{tab:Tetrad}
Transformation ${L^\mu}_{\hat{\mu}}$ between comoving and lab frames, which also can be regarded as the tetrad of comoving orthonormal basis vectors; and its Eulerian decomposition into  projections normal and tangent to the spacelike slice $\Sigma_t$.}
\begin{ruledtabular}
\begin{tabular}{lll}
${L^\mu}_{\hat{\mu}}$ & Transformation, or comoving tetrad &  Eq.~(\ref{eq:IndexedBasis}) \\
$u^\mu$ & ${L^\mu}_{\hat{0}}$; timelike comoving unit vector & Eq.~(\ref{eq:TimelikeVector}) \\
$w^\mu$ & ${L^\mu}_{\hat{1}}$; spacelike comoving unit vector & Eq.~(\ref{eq:SpacelikeVector_1}) \\
$y^\mu$ & ${L^\mu}_{\hat{2}}$; spacelike comoving unit vector & Eq.~(\ref{eq:SpacelikeVector_2}) \\
$z^\mu$ & ${L^\mu}_{\hat{3}}$; spacelike comoving unit vector & Eq.~(\ref{eq:SpacelikeVector_3}) \\
${L^\mu}_{\hat{\mu}}$ & Eulerian decomposition & Eq.~(\ref{eq:TetradDecomposition}) \\
$u^\mu$ & Eulerian decomposition  & Eq.~(\ref{eq:ThreeVelocity}) \\
$w^\mu$ & Eulerian decomposition  & Eq.~(\ref{eq:wyz_1}) \\
$y^\mu$ & Eulerian decomposition  & Eq.~(\ref{eq:wyz_2}) \\
$z^\mu$ & Eulerian decomposition  & Eq.~(\ref{eq:wyz_3}) \\
$\mathcal{L}_{\hat\mu}$ & Projection of ${L^\mu}_{\hat{\mu}}$ normal to $\Sigma_t$ & Eq.~(\ref{eq:TetradTimelike}) \\
${\ell^\mu}_{\hat\mu} $ & Projection of ${L^\mu}_{\hat{\mu}}$ tangent to $\Sigma_t$ & Eq.~(\ref{eq:TetradSpatial}) \\
$\Lambda$ & Projection of $u^\mu$ normal to $\Sigma_t$ & Eq.~(\ref{eq:BoostFactor}) \\
$A$ & Projection of $w^\mu$ normal to $\Sigma_t$ & Eq.~(\ref{eq:ABC_1}) \\
$B$ & Projection of $y^\mu$ normal to $\Sigma_t$ & Eq.~(\ref{eq:ABC_2}) \\
$C$ & Projection of $z^\mu$ normal to $\Sigma_t$ & Eq.~(\ref{eq:ABC_3}) \\
$\Lambda v^i$ & Projection of $u^\mu$ tangent to $\Sigma_t$ & Eq.~(\ref{eq:v_Spacelike}) \\
$a^i$ & Projection of $w^\mu$ tangent to $\Sigma_t$ & Appendix~\ref{app:TetradSolution} \\
$b^i$ & Projection of $y^\mu$ tangent to $\Sigma_t$ & Appendix~\ref{app:TetradSolution} \\
$c^i$ & Projection of $z^\mu$ tangent to $\Sigma_t$ & Appendix~\ref{app:TetradSolution} \\
\end{tabular}
\end{ruledtabular}
\end{table}

\begin{table}[b]
\caption{\label{tab:Momentum}
Particle momentum; particle distribution function; transformation to comoving curvilinear momentum coordinates, and its decomposition and projections; curvilinear momentum space metric.}
\begin{ruledtabular}
\begin{tabular}{lll}
$p^\mu$ & Coordinate basis &  Sec.~\ref{sec:PhaseSpaceCoordinates} \\
$p^{\hat\mu} $ & Comoving orthonormal basis &  Sec.~\ref{sec:PhaseSpaceCoordinates} \\
$p^{\tilde\imath} $ & Comoving curvilinear basis &  Sec.~\ref{sec:PhaseSpaceCoordinates} \\
$p$ & Three-momentum magnitude & Eq.~(\ref{eq:MomentumMagnitude}) \\
$f(x^\mu, p^{\tilde\imath} )$ & Particle distribution function &  Sec.~\ref{sec:PhaseSpaceCoordinates} \\
${P^{\tilde \imath}}_{\hat \imath}$ & Transformation to curvilinear basis & Eq.~(\ref{eq:MomentumTransformation}) \\
${P^{\tilde \imath}}_{\hat \imath}$ & Decomposition with respect to $p^{\hat\imath}$ & Eq.~(\ref{eq:MomentumDecomposition}) \\
${Q^{\tilde \imath}}$ & Projection of ${P^{\tilde \imath}}_{\hat \imath}$ parallel to $p^{\hat\imath}$ & Eq.~(\ref{eq:Projection_Q}) \\
${U^{\tilde \imath}}_{\hat \imath}$ & Projection of ${P^{\tilde \imath}}_{\hat \imath}$ orthogonal to $p^{\hat\imath}$ & Eq.~(\ref{eq:Projection_U}) \\
${k^{\hat \jmath}}_{\hat \imath}$ & Projector orthogonal to $p^{\hat\imath}$ & Eq.~(\ref{eq:MomentumSpaceProjector}) \\
$\lambda_{\tilde\imath \tilde\jmath}$ & Curvilinear momentum space metric & Eq.~(\ref{eq:LineElementMomentum}) \\
$\lambda$ & Determinant of $\lambda_{\tilde\imath \tilde\jmath}$ & Eq.~(\ref{eq:MomentumTransformationDeterminant}) \\
\end{tabular}
\end{ruledtabular}
\end{table}

\begin{table}[b]
\caption{\label{tab:Equations}
Major entities appearing in Eq.~(\ref{eq:NumberConservationFinal}), the 3+1 general relativistic number-conservative reformulation of the Boltzmann equation; and in Eqs.~(\ref{eq:EnergyConservationFinal}) and (\ref{eq:MomentumConservationFinal}), the 3+1 general relativistic energy- and momentum-conservative reformulations of the Boltzmann equation.}
\begin{ruledtabular}
\begin{tabular}{lll}
$ D_N$ & Conserved number density & Eq.~(\ref{eq:Density_N}) \\
$ D_{T,n}$ & Conserved energy density &  Eq.~(\ref{eq:Density_T_n}) \\
$\left(D_{T,\gamma}\right)_j$ & Conserved momentum density & Eq.~(\ref{eq:Density_T_g}) \\
$\left(F_N\right)^i$ & Number flux & Eq.~(\ref{eq:Flux_N}) \\
$\left(F_{T,n}\right)^i$ & Energy flux & Eq.~(\ref{eq:Flux_T_n}) \\
${\left(F_{T,\gamma}\right)^i}_j$ & Momentum flux & Eq.~(\ref{eq:Flux_T_g}) \\
$ \left(R_N\right)^{\hat\rho} $ & Gravitational shifts, number & Eq.~(\ref{eq:Redshift_N}) \\
$ \left(R_{T,n}\right)^{\hat\rho} $ & Gravitational shifts, energy & Eq.~(\ref{eq:Redshift_T_n}) \\
${\left(R_{T,\gamma}\right)^{\hat\rho}}_j $ & Gravitational shifts, momentum & Eq.~(\ref{eq:Redshift_T_g}) \\
$\left(O_N\right)^{\hat\rho} $ & Observer corrections, number & Eq.~(\ref{eq:Observer_N}) \\
$\left(O_{T,n}\right)^{\hat\rho} $ & Observer corrections, energy & Eq.~(\ref{eq:Observer_T_n}) \\
${\left(O_{T,\gamma} \right)^{\hat\rho}}_j$ & Observer corrections, momentum & Eq.~(\ref{eq:Observer_T_g}) \\
$G_{T,n}$ & Gravitational energy source & Eq.~(\ref{eq:GeometrySource_n}) \\
$\left(G_{T,\gamma}\right)_j$ & Gravitational momentum source & Eq.~(\ref{eq:GeometrySource_g}) \\
$C_N$ & Collision number source & Eq.~(\ref{eq:RHS_N}) \\
$C_{T,n}$ & Collision energy source & Eq.~(\ref{eq:RHS_T_n}) \\
$\left(C_{T,\gamma}\right)_j$ & Collision momentum source & Eq.~(\ref{eq:RHS_T_g}) \\
\end{tabular}
\end{ruledtabular}
\end{table}

We conclude by assembling expressions obtained in Sec.~\ref{sec:31_Specialization} into conservative 3+1 general relativistic reformulations of the Boltzmann equation, and discussing lepton number and four-momentum exchange between neutrinos and the fluid.
Tables~\ref{tab:Spacetime}-\ref{tab:Equations} present overviews of many of the  variables that have been assembled into the major entities appearing in these equations.

The number-conservative 3+1 general relativistic Boltzmann equation is
\begin{eqnarray}
 \frac{\partial \left(D_N\right)}{\partial t} &+& \frac{\partial \left(F_N\right)^i }{\partial x^i} \nonumber \\
&+& \frac{1}{\sqrt{\lambda}} \frac{\partial}{\partial p^{\tilde\imath}} 
\left\{
\sqrt{\lambda} \, \frac{Q^{\tilde\imath} \left(-p_{\hat 0}\right)}{p}\! \left[ \left( R_N \right)^{\hat 0} + \left( O_N \right)^{\hat 0} \right] \right. \nonumber \\
&& \left. + \sqrt{\lambda} \, {U^{\tilde \imath}}_{\hat \imath} \left[ \left( R_N \right)^{\hat \imath} + \left( O_N \right)^{\hat \imath} \right] \right\}
 = C_N.
\label{eq:NumberConservationFinal}
\end{eqnarray}
This is a conservation law (or balance equation) for particle number.
The particle density $D_N$ and spatial flux $\left(F_N\right)^i$ are given by Eqs.~(\ref{eq:Density_N}) and (\ref{eq:Flux_N}) respectively.
The contributions $\left( R_N \right)^{\hat\rho}$ and $\left( O_N \right)^{\hat\rho}$ to the momentum space flux---which represent momentum shifts (redshift and angular aberration) due to geometry/gravitation and acceleration of a Lagrangian observer respectively---are given by Eqs.~(\ref{eq:Redshift_N}) and (\ref{eq:Observer_N}). 
The right-hand side is 
\begin{equation}
C_N = \frac{\alpha \sqrt{\gamma}}{\left( -p_{\hat 0} \right)}\, C[f],
\label{eq:RHS_N}
\end{equation}
where $\alpha$ and $\gamma$ are respectively the lapse function and the determinant of the three-metric characterizing the internal geometry of a spacelike slice (see Sec.~\ref{sec:GeometryDescription}), $\left(-p_{\hat 0} \right)$ is the particle energy in the comoving frame, and $C[f]$ is the invariant collision integral appearing for instance in Eq.~(\ref{eq:BoltzmannComoving}).

The phase space coordinates used in Eq.~(\ref{eq:NumberConservationFinal}) allow particle/fluid interactions to be evaluated in the comoving frame in the context of Eulerian grid-based approaches to multidimensional spatial dependence.
The global `lab frame' spacetime coordinates $t$ and $x^i$ are those associated with the 3+1 formulation of general relativity, in which the line element and metric components $g_{\mu\nu}$ are given by Eq.~(\ref{eq:LineElement}).
In momentum space we apply the transformation ${P^{\tilde\imath}}_{\hat\imath} = \partial p^{\tilde\imath} / \partial p^{\hat\imath}$ from comoving frame orthonormal Cartesian momentum components $p^{\hat\imath}$ to comoving frame curvilinear momentum components $p^{\tilde\imath}$.
This transformation is decomposed into a portion $Q^{\tilde\imath}$ parallel to $p^{\hat\imath}$, and a portion ${U^{\tilde \imath}}_{\hat \imath}$ orthogonal to $p^{\hat\imath}$; see Eq.~(\ref{eq:MomentumDecomposition}).
For the case of massless particles, we note that the particle energy $\left(-p_{\hat 0} \right)$ cancels with the three-momentum magnitude $p = \sqrt{p_{\hat\imath} p^{\hat\imath}}$ in Eq.~(\ref{eq:NumberConservationFinal}). 
In our derivation we introduce a momentum space metric with determinant $\lambda$ (see the `line element' in Eq.~(\ref{eq:LineElementMomentum}), and also Eq.~(\ref{eq:MomentumTransformationDeterminant})).
A particular example of curvilinear momentum space coordinates is the version of momentum space spherical coordinates discussed in the next-to-last paragraph of Sec.~\ref{sec:PhaseSpaceCoordinates}.

Making the Boltzmann equation explicit has required specification of ${L^\mu}_{\hat\mu}$, which can be regarded either as the transformation from a comoving frame orthonormal basis to the lab frame coordinate basis, or as the tetrad of comoving frame basis vectors; see Sec.~\ref{sec:TetradDecomposition}.
The timelike vector ${L^\mu}_{\hat 0} = u^\mu$ is uniquely specified as the four-velocity of a Lagrangian observer riding along with the fluid, but additional conditions must be imposed to fix the orientation of the triad of comoving spatial basis vectors;
we have chosen algebraic conditions that simplify the transformation to an orthonormal frame.
Also, in order to arrive at a useful 3+1 formulation we have found it convenient to perform an `Eulerian decomposition' of ${L^\mu}_{\hat\mu}$ into its `Eulerian projections' $\mathcal{L}_{\hat\mu}$ and ${\ell^i}_{\hat\mu}$ orthogonal and tangent to the spacelike slice respectively.
This geometric approach (in perspective if not notation) enables evaluation of the momentum space divergence while almost completely avoiding explicit encounters with connection coefficients.  

Turning to astrophysical simulations in which neutrinos interact with matter, we note that solution of Eq.~(\ref{eq:NumberConservationFinal})---for $D_N$, and ultimately for the distribution function $f\left( x^\mu, p^{\tilde\imath} \right)$, through Eq.~(\ref{eq:Density_N})---enables calculation of lepton number exchange with the fluid.
Conservation of total electron lepton number is expressed
\begin{equation}
\nabla_\mu \left( N^{\mu}_{e} + N_{\nu_e}^{\mu} - N_{\bar{\nu}_e}^{\mu} \right) = 0,
\label{eq:TotalNumberConservation}
\end{equation}
where $N^\mu_e$ is the net electron number flux vector of the fluid, and $N_{\nu_e}^{\mu}$ and $N_{\nu_e}^{\mu}$ are the number flux vectors of the electron neutrinos and antineutrinos respectively.
From Eq.~(\ref{eq:TotalNumberConservation}) and the discussion in Sec.~\ref{sec:Divergences}, we see that the fluid electron lepton number source term is the momentum space integral of the difference of the $\nu_e$ and $\bar\nu_e$ collision integrals:
\begin{equation}
\nabla_\mu N^{\mu}_{e} = - \int \left( C_{\nu_e}[ f ] - C_{\bar\nu_e}[f] \right) \, d\tilde{\mathbf{p}}. 
\end{equation}
Because the left-hand side takes the conservative form
\begin{equation}
\nabla_\mu N^{\mu}_{e} = \frac{1}{\alpha \sqrt{\gamma}} \frac{\partial}{\partial x^\mu}
\left( \alpha \sqrt{\gamma} \,N^\mu_e \right),
\end{equation}
solving for the neutrino distributions using Eq.~(\ref{eq:NumberConservationFinal})---which is manifestly conservative in both position space and momentum space---renders almost trivial the achievement of machine-precision global lepton number conservation (the position space integral of Eq.~(\ref{eq:TotalNumberConservation})) in numerical simulations.

In addition to the number-conservative formulation in Eq.~(\ref{eq:NumberConservationFinal}), we have also the energy- and momentum-conservative formulations 
\begin{eqnarray}
 \frac{\partial \left(D_{T,n}\right)}{\partial t} &+& \frac{\partial \left(F_{T,n}\right)^i }{\partial x^i} \nonumber \\
&+& \frac{1}{\sqrt{\lambda}} \frac{\partial}{\partial p^{\tilde\imath}} \left\{
\sqrt{\lambda} \, \frac{Q^{\tilde\imath} \left(-p_{\hat 0}\right)}{p}\! \left[ \left( R_{T,n} \right)^{\hat 0} + \left( O_{T,n} \right)^{\hat 0} \right] \right. \nonumber \\
&& \left. + \sqrt{\lambda} \, {U^{\tilde \imath}}_{\hat \imath} \left[ \left( R_{T,n} \right)^{\hat \imath} + \left( O_{T,n} \right)^{\hat \imath} \right] \right\}  \nonumber \\
& & = G_{T,n} + C_{T,n},
\label{eq:EnergyConservationFinal} \\
 \frac{\partial \left(D_{T,\gamma}\right)_j}{\partial t} &+& \frac{\partial {\left(F_{T,\gamma}\right)^i}_j }{\partial x^i} \nonumber \\
&+& \frac{1}{\sqrt{\lambda}} \frac{\partial}{\partial p^{\tilde\imath}} \left\{
\sqrt{\lambda} \, \frac{Q^{\tilde\imath} \left(-p_{\hat 0}\right)}{p}\! \left[ {\left( R_{T,\gamma} \right)^{\hat 0}}_j + {\left( O_{T,\gamma} \right)^{\hat 0}}_j \right] \right. \nonumber \\
&& \left. + \sqrt{\lambda} \, {U^{\tilde \imath}}_{\hat \imath} \left[ {\left( R_{T,\gamma} \right)^{\hat \imath}}_j + {\left( O_{T,\gamma} \right)^{\hat \imath}}_j \right] \right\} \nonumber \\
& & = \left( G_{T,\gamma} \right)_j + \left( C_{T,\gamma} \right)_j.
\label{eq:MomentumConservationFinal}
\end{eqnarray}
The energy and momentum densities $\left(D_{T,n}\right)$ and $\left(D_{T,\gamma}\right)_j$ are given by Eqs.~(\ref{eq:Density_T_n}) and (\ref{eq:Density_T_g}) respectively,
and the corresponding spatial fluxes $\left(F_{T,n}\right)^i$ and ${\left(F_{T,\gamma}\right)^i}_j$ are given by Eqs.~(\ref{eq:Flux_T_n}) and (\ref{eq:Flux_T_g}).
The contributions $\left( R_{T,n} \right)^{\hat\rho}$ and ${\left( R_{T,\gamma} \right)^{\hat\rho}}_j$ to the momentum space flux---which represent momentum shifts (redshift and angular aberration) due to geometry/gravitation---are given by Eqs.~(\ref{eq:Redshift_T_n}) and (\ref{eq:Redshift_T_g}) respectively; 
and the contributions $ \left( O_{T,n} \right)^{\hat\rho}$ and ${\left( O_{T,\gamma} \right)^{\hat\rho}}_j$ due to the acceleration of a Lagrangian observer are given by Eqs.~(\ref{eq:Observer_T_n}) and (\ref{eq:Observer_T_g}).  
The source terms
\begin{eqnarray}
C_{T,n} &=& \frac{\alpha \sqrt{\gamma}}{\left( -p_{\hat 0} \right)}\,\mathcal{L}_{\hat\mu} \, p^{\hat\mu}\, C[f], 
\label{eq:RHS_T_n}\\
\left( C_{T,\gamma} \right)_j &=& \frac{\alpha \sqrt{\gamma}}{\left( -p_{\hat 0} \right)}\, \ell_{j \hat\mu}\, p^{\hat\mu} \, C[f]
\label{eq:RHS_T_g}
\end{eqnarray}
are due to particle interactions.
The energy- and momentum-conservative formulations in Eqs.~(\ref{eq:EnergyConservationFinal}) and (\ref{eq:MomentumConservationFinal}) contain extra source terms $G_{T,n}$ and $\left( G_{T,\gamma} \right)_j$, given by Eqs.~(\ref{eq:GeometrySource_n}) and (\ref{eq:GeometrySource_g}), relative to the number-conservative formulation in Eq.~(\ref{eq:NumberConservationFinal}).
Physically, these source terms represent the exchange of energy and momentum with the gravitational field; 
mathematically, they are a consequence of the fact that Eqs.~(\ref{eq:EnergyConservationFinal}) and (\ref{eq:MomentumConservationFinal}) are projections of the divergence of a rank-two tensor, which produces an additional connection coefficient term relative to the divergence of a vector.
(In our derivation in Sec.~\ref{sec:Reconciliation}, we do not compute these connection coefficients directly; 
instead, $G_{T,n}$ and $\left( G_{T,\gamma} \right)_j$ emerge from the second and third lines of $\left( R_{N} \right)^{\hat\rho}$ in Eq.~(\ref{eq:Redshift_N}) respectively when the momentum projections $\mathcal{L}_{\hat\mu} \, p^{\hat\mu}$ and $\ell_{j \hat\mu}\, p^{\hat\mu}$ orthogonal and tangent to the spacelike slice---which relate Eqs.~(\ref{eq:RHS_T_n}) and (\ref{eq:RHS_T_g}) to Eq.~(\ref{eq:RHS_N})---are pulled into the momentum divergence.)

Consideration of the particle stress-energy tensor shows how the Boltzmann equation is related to the exchange of energy and momentum with the fluid.
While Eq.~(\ref{eq:NumberVector}) gives the particle number flux vector in terms of the distribution function, the particle stress energy tensor is (e.g. Ref.~\cite{Ehlers1971General-Relativ})
\begin{equation}
T^{\nu\mu} = \int  f \,p^\nu p^\mu \, d{\tilde{\mathbf{p}}},
\label{eq:StressEnergy}
\end{equation}
expressed here in terms of the lab frame coordinate basis (note $p^\mu = {L^\mu}_{\hat{\mu}} \,p^{\hat{\mu}}$).
This stress energy obeys
\begin{equation}
\nabla_{\mu} T^{\nu\mu} = \int p^\nu \, C[f] \, d{\tilde{\mathbf{p}}}. 
\label{eq:StressEnergyDivergence}
\end{equation}
In an astrophysical simulation involving neutrino transport, conservation of total energy-momentum is expressed
\begin{equation}
\nabla_\mu \left[ T^{\nu\mu}_{\mathrm{fluid}} + \sum_{a = e, \mu, \tau}  \left( T_{\nu_a}^{\nu\mu} + T_{\bar\nu_a}^{\nu\mu} \right) \right] = 0,
\label{eq:TotalEnergyMomentumConservation}
\end{equation}
where $T^{\nu\mu}_{\mathrm{fluid}}$ is the fluid stress-energy, and the sum is over all flavors of neutrinos.
In light of Eqs.~(\ref{eq:StressEnergyDivergence}) and (\ref{eq:TotalEnergyMomentumConservation}), the fluid energy and momentum source terms again emerge from momentum space integrals of collision integrals:
\begin{eqnarray}
-n_\nu \nabla_\mu T^{\nu\mu}_{\mathrm{fluid}} &=& \mathcal{L}_{\hat\mu} \, q^{\hat\mu}, 
\label{eq:EnergyExchange}\\
\gamma_{j\nu} \nabla_\mu T^{\nu\mu}_{\mathrm{fluid}} &=& \ell_{j \hat\mu} \, q^{\hat\mu},
\label{eq:MomentumExchange}
\end{eqnarray}
where
\begin{equation}
q^{\hat\mu} = - \sum_{a = e, \mu, \tau}  \int  p^{\hat\mu} \left( C_{\nu_a}[ f ] + C_{\bar\nu_a}[f] \right) \, d\tilde{\mathbf{p}}.
\label{eq:FourMomentumSource}
\end{equation} 
(The energy and momentum equations are given by the projections of the stress-energy divergence orthogonal and tangent to the spacelike slice, via contraction with the normal $n_\mu$ to the spacelike slice and the orthogonal projector $\gamma_{\mu\nu} = g_{\mu\nu} + n_\mu n_\nu$.)
Therefore knowledge of the distribution functions obtained by solution of the number-conservative Eq.~(\ref{eq:NumberConservationFinal}) also yields the fluid energy and momentum sources, via Eqs.~(\ref{eq:EnergyExchange})-(\ref{eq:FourMomentumSource}). 

Finally, we point out that while Eqs.~(\ref{eq:EnergyConservationFinal}) and (\ref{eq:MomentumConservationFinal}) are redundant with Eq.~(\ref{eq:NumberConservationFinal}) and therefore do not need to independently be solved, they illuminate two challenges to maintenance of global four-momentum conservation in numerical simulations.

First, while the left-hand sides of Eqs.~(\ref{eq:EnergyConservationFinal}) and (\ref{eq:MomentumConservationFinal}) are transparently related to surface integrals in phase space, the presence of extra connection coefficient terms in the divergences appearing in Eq.~(\ref{eq:TotalEnergyMomentumConservation}) prevent trivial machine-precision four-momentum conservation in discretized models. 
These extra source terms---$G_{T,n}$ and $\left( G_{T,\gamma} \right)_j$ in Eqs.~(\ref{eq:EnergyConservationFinal}) and (\ref{eq:MomentumConservationFinal}), and analogous terms on the left-hand sides of Eqs.~(\ref{eq:EnergyExchange}) and (\ref{eq:MomentumExchange})---represent energy and momentum exchange between the forms of stress energy present (fluid and neutrinos) and the gravitational field as embodied in the spacetime geometry.
Even where {\em global} energy and momentum can be defined in general relativity---for instance in asymptotically flat spacetimes---conversion to a {\em local} conservation law, via definitions of gravitational energy and momentum densities and fluxes, is complicated at best (one possibility being use of a gravitational stress-energy pseudotensor \cite{Landau1975The-classical-t}).

Second, even if the gravitational source terms were not present in Eqs.~(\ref{eq:EnergyConservationFinal}) and (\ref{eq:MomentumConservationFinal}), a straightforward discretization of Eq.~(\ref{eq:NumberConservationFinal}) would not be consistent with straightforward discretizations of Eqs.~(\ref{eq:EnergyConservationFinal}) and (\ref{eq:MomentumConservationFinal}). 
Presentation and testing of a full discretization is beyond the scope of this paper.
But as was done by Liebend\"{o}rfer et al. \cite{Liebendorfer2004A-Finite-Differ} in the spherically symmetric case, analysis of the relationship between the number- and four-momentum conservative formulations examined in Sec.~\ref{sec:Reconciliation} can illuminate the search for discretizations that provide good energy conservation at modest resolution in addition to the precise lepton number conservation that comes naturally via Eq.~(\ref{eq:NumberConservationFinal}).
A few initial comments on this sort of consistent discretization of the spatially multidimensional angular moments formalism were also made in Appendix~D of Ref.~\cite{Cardall2012Conservative-31}.
Consistent discretization of the spatially multidimensional Boltzmann equation will be algebraically more complicated than either of these other two cases, as can be seen for instance by considering the manipulations between Eqs.~(\ref{eq:ExtraMomentum_T_Z_3}) and (\ref{eq:ExtraMomentum_T_Z_4}).
Our decomposition in Eq.~(\ref{eq:Momentum_N_31}) of the momentum space transformation ${P^{\tilde\imath}}_{\hat\imath}$ into a portion $Q^{\tilde\imath}$ parallel to $p^{\hat\imath}$, and a portion ${U^{\tilde \imath}}_{\hat \imath}$ orthogonal to $p^{\hat\imath}$, is a helpful step.
This allows us to introduce, via Eq.~(\ref{eq:GeodesicIdentity}), the comoving-frame timelike components $\left( R_N \right)^{\hat 0}$ and $\left( O_N \right)^{\hat 0}$ into Eqs.~(\ref{eq:Redshift_N}) and (\ref{eq:Observer_N}).
The presence of all four components facilitates more transparent analysis of the relationship between the number- and energy-momentum-conservative formulations of the Boltzmann equation, by allowing use of Eqs.~(\ref{eq:DeltaDecomposition_nn})-(\ref{eq:DeltaDecomposition_gg}) to project out the different pieces of Eqs.~(\ref{eq:Redshift_N}) and (\ref{eq:Observer_N}) relevant to the energy- and momentum-conservative equations.
However, it is not clear to us whether it will be possible in the discretized case to impose satisfaction of Eq.~(\ref{eq:GeodesicIdentityReduced}), which follows from Eq.~(\ref{eq:GeodesicIdentityCovariant}). 
Use of this nontrivial identity is the final step we take in arriving at Eq.~(\ref{eq:ExtraMomentum_T_Z_4}).
The extent to which a discrete analogue of Eq.~(\ref{eq:GeodesicIdentityReduced}) deviates from its vanishing analytic value---and what, if anything, might be done about it---remain to be seen.

In addition to the challenges faced when attempting to construct
energy-momentum conservative discretizations of Eq.~(\ref{eq:NumberConservationFinal}), a
numerical method should also preserve positivity of the distribution
function, which is often ensured with the use of limiters and a suitable
condition on the time step \cite{Zhang2010On-maximum-prin}.  (For the case of
fermions, the distribution function should also remain bounded by 1.)
We will also consider these difficulties in the future when we develop
specific discretizations.

In spite of these remaining challenges to energy-momentum conservation, the number-conservative 3+1 general relativistic Boltzmann equation we present in Eq.~(\ref{eq:NumberConservationFinal}) is a significant practical step forward towards numerical implementation of multidimensional relativistic Boltzmann transport with coordinate basis spacetime position components and comoving frame curvilinear momentum space coordinates.

\appendix

\section{The `extra' momentum derivative term \label{app:ExtraMomentumTerm}}

In this appendix we compute the `extra' momentum derivative term $ f \,  \mathcal{D}_{\tilde\imath} \left(
{P^{\tilde\imath}}_{\hat\imath} {\Gamma^{\hat\imath}}_{\hat\nu \hat\mu}\, p^{\hat\nu} p^{\hat\mu} \right)$, i.e. the fourth term on the left-hand side of Eq.~(\ref{eq:ConservativeExtra}).
Three factors in this expression are affected by the momentum derivative: ${P^{\tilde\imath}}_{\hat\imath}$, which is the only one in which momentum space connection coefficients come into play; and the factors $p^{\hat\nu}$ and $p^{\hat\mu}$.
In the momentum derivatives, $p^{\hat 0}$ must be considered a function of $p^{\hat\imath}$ through Eq.~(\ref{eq:MassShell}), so that
\begin{equation}
\frac{\partial p^{\hat 0}}{\partial p^{\hat\imath}} = \frac{p_{\hat\imath}}{\left(-p_{\hat 0}\right)}, \ \ \  \frac{\partial p^{\hat \jmath}}{\partial p^{\hat\imath}} = {\delta^{\hat\jmath}}_{\hat\imath}
\label{eq:MomentumDerivativeComponents}
\end{equation}
are the derivatives of the time and space momentum components respectively.

First we show that
\begin{equation}
\mathcal{D}_{\tilde\imath} {P^{\tilde\imath}}_{\hat\imath} = 0. 
\end{equation}
The result follows quickly with the help of Eq.~(\ref{eq:ConnectionMomentum}):
\begin{eqnarray}
\mathcal{D}_{\tilde\imath} {P^{\tilde\imath}}_{\hat\imath} &=& \frac{\partial {P^{\tilde\imath}}_{\hat\imath} }{\partial p^{\tilde \imath}}  + {\Pi^{\tilde\imath}}_{\tilde\jmath \tilde \imath} {{P^{\tilde\jmath}}_{\hat\imath}} \\
&=& 
\frac{\partial {P^{\tilde\imath}}_{\hat\imath} }{\partial p^{\tilde \imath}}  
+\left({P^{\tilde \imath}}_{\hat \ell} {P^{\hat k}}_{\tilde \imath}
\frac{\partial {P^{\hat\ell}}_{\tilde{\jmath}}}{\partial p^{\hat k}} \right)
{{P^{\tilde\jmath}}_{\hat\imath}} \\
&=& \frac{\partial {P^{\tilde\imath}}_{\hat\imath} }{\partial p^{\tilde \imath}} 
- {P^{\tilde \imath}}_{\hat \ell} {P^{\hat k}}_{\tilde \imath} {P^{\hat\ell}}_{\tilde{\jmath}}
\frac{\partial {{P^{\tilde\jmath}}_{\hat\imath}}}{\partial p^{\hat k}} \\
&=& \frac{\partial {P^{\tilde\imath}}_{\hat\imath} }{\partial p^{\tilde \imath}} 
- {\delta^{\tilde \imath}}_{\tilde\jmath}
\frac{\partial {{P^{\tilde\jmath}}_{\hat\imath}}}{\partial p^{\tilde\imath}} = 0,
\end{eqnarray}
where we used the identity
\begin{equation}
0 = \frac{\partial {P^{\hat\ell}}_{\tilde{\jmath}}}{\partial p^{\hat k}}
{{P^{\tilde\jmath}}_{\hat\imath}} + {P^{\hat\ell}}_{\tilde{\jmath}} \frac{\partial {P^{\tilde\jmath}}_{\hat\imath}}{\partial p^{\hat k}} 
\label{eq:PDerivativeReversal}
\end{equation}
as well as Eq.~(\ref{eq:InverseMomentumTransformation}).

Next we turn to the first term in Eq.~(\ref{eq:LastTwoMomentumDerivatives}).
Using Eq.~(\ref{eq:MomentumDerivativeComponents}), 
\begin{eqnarray}
 f \, {\Gamma^{\hat\imath}}_{\hat\nu \hat\mu} \frac{\partial p^{\hat\nu}}{\partial p^{\hat\imath}} p^{\hat\mu} 
 &=& f \left[ \frac{1}{\left(-p_{\hat 0}\right)}{\Gamma^{\hat\imath}}_{\hat 0 \hat\mu}\, p_{\hat\imath} p^{\hat\mu} + {\Gamma^{\hat\imath}}_{\hat\imath \hat\mu}\, p^{\hat\mu} \right]  \\
&=& f \left[ \frac{1}{\left(-p_{\hat 0}\right)}{\Gamma^{\hat\nu}}_{\hat 0  \hat\mu}\, p_{\hat\nu} p^{\hat\mu} + {\Gamma^{\hat\nu}}_{\hat\nu \hat\mu}\, p^{\hat\mu} \right], \nonumber \\
& & \label{eq:FirstMomentumDerivativeTerm}
\end{eqnarray}
where we have added and subtracted ${\Gamma^{\hat 0}}_{\hat 0 \hat\mu} p^{\hat\mu}$ to obtain the second line.

Focus on the first term of Eq.~(\ref{eq:FirstMomentumDerivativeTerm}).
Using the lowered-index version
\begin{equation}
\frac{dp_{\hat\mu}}{d\lambda} =  {\Gamma^{\hat\nu}}_{\hat\mu \hat\rho} \, p_{\hat\nu} p^{\hat\rho}
  \label{eq:Geodesic_p_D}
\end{equation}
of the second geodesic equation, the first term of Eq.~(\ref{eq:FirstMomentumDerivativeTerm}) is
\begin{eqnarray}
 \frac{f}{\left(-p_{\hat 0}\right)}{\Gamma^{\hat\nu}}_{\hat 0  \hat\mu}\, p_{\hat\nu} p^{\hat\mu} &=&  \frac{f}{\left(-p_{\hat 0}\right)} \frac{dp_{\hat 0}}{d\lambda} \\
&=& - \frac{f}{\left(-p_{\hat 0}\right)} \frac{dp^{\hat 0}}{d\lambda}.
\end{eqnarray}
On the right-hand side we use Eq.~(\ref{eq:Geodesic_p}) and the fact that Eq.~(\ref{eq:MassShell}) implies 
\begin{equation}
0 = \mathrm{d} \left( p_{\hat\mu} p^{\hat\mu} \right) = 2\, p_{\hat\mu}\, \mathrm{d}p^{\hat\mu},
\label{eq:MassShellDerivative}
\end{equation}
where here ``$\mathrm{d}$'' represents any derivative operator,
and find
\begin{eqnarray}
\frac{f}{\left(-p_{\hat 0}\right)}{\Gamma^{\hat\nu}}_{\hat 0  \hat\mu}\, p_{\hat\nu} p^{\hat\mu} 
&=& \frac{f}{\left(-p_{\hat 0}\right)^2} \,p_{\hat 0} \frac{dp^{\hat 0}}{d\lambda} \\
&=& - \frac{f}{\left(-p_{\hat 0}\right)^2} \,p_{\hat \imath} \frac{dp^{\hat \imath}}{d\lambda} \\
&=&  \frac{f}{\left(-p_{\hat 0}\right)^2} \,p_{\hat \imath}\, {\Gamma^{\hat\imath}}_{\hat\nu \hat\mu} \,p^{\hat\nu} p^{\hat\mu}.
\end{eqnarray}
We rewrite the right-hand side using Eq.~(\ref{eq:MomentumDerivativeComponents}) to obtain
\begin{eqnarray}
\frac{f}{\left(-p_{\hat 0}\right)}{\Gamma^{\hat\nu}}_{\hat 0  \hat\mu} \,p_{\hat\nu} p^{\hat\mu} 
&\!=\!&\!
f \,{\Gamma^{\hat\imath}}_{\hat\nu \hat\mu} \,p^{\hat\nu} p^{\hat\mu} \frac{1}{\left(-p_{\hat 0}\right)} \frac{\partial \left(-p_{\hat 0}\right)}{\partial p^{\hat\imath}} \\
&\!=\!&\! f \,{\Gamma^{\hat\imath}}_{\hat\nu \hat\mu} \,p^{\hat\nu} p^{\hat\mu} \frac{{P^{\tilde\imath}}_{\hat\imath}}{\left(-p_{\hat 0}\right)} \frac{\partial \left(-p_{\hat 0}\right)}{\partial p^{\tilde\imath}}\!.
\end{eqnarray}
We massage the right-hand side further using the fact that $p^{\hat 0}$ is a momentum (three-)space scalar, and find
\begin{eqnarray} 
\frac{f}{\left(-p_{\hat 0}\right)}&{\Gamma^{\hat\nu}}&_{\hat 0  \hat\mu} \,p_{\hat\nu} p^{\hat\mu} \nonumber\\
&=& f \,{\Gamma^{\hat\imath}}_{\hat\nu \hat\mu} \,p^{\hat\nu} p^{\hat\mu} \frac{{P^{\tilde\imath}}_{\hat\imath}}{\left(-p_{\hat 0}\right)} \mathcal{D}_{\tilde\imath}\left(-p_{\hat 0}\right) \\
&=& -f \,{\Gamma^{\hat\imath}}_{\hat\nu \hat\mu} \,p^{\hat\nu} p^{\hat\mu} {P^{\tilde\imath}}_{\hat\imath} \left(-p_{\hat 0}\right) \mathcal{D}_{\tilde\imath}\left[\frac{1}{\left(-p_{\hat 0}\right)}\right]\!.
\label{eq:FirstMomentumDerivativeTermFirst}
\end{eqnarray} 
The sign reversal obtained in the last line will prove important.

Turning next to the second term of Eq.~(\ref{eq:FirstMomentumDerivativeTerm}), we show that
\begin{equation}
{\Gamma^{\hat\nu}}_{\hat\nu \hat\mu} = 0.
\label{eq:ConnectionComovingTrace_1}
\end{equation}
Invoking Eq.~(\ref{eq:ConnectionComoving}), we have
\begin{eqnarray}
{\Gamma^{\hat{\nu}}}_{\hat{\nu}\hat{\mu}}
&=&  {L^{\hat{\nu}}}_{{\nu}} {L^\rho}_{\hat{\nu}} {L^\mu}_{\hat{\mu}} \,{\Gamma^\nu}_{\rho\mu} 
+ {L^{\hat{\nu}}}_{{\nu}} {L^\mu}_{\hat{\mu}} 
\frac{\partial {L^\nu}_{\hat{\nu}}}{\partial x^\mu} \\
&=& {L^\mu}_{\hat{\mu}} \left( {\Gamma^\nu}_{\nu\mu} 
+ {L^{\hat{\nu}}}_{{\nu}} \frac{\partial {L^\nu}_{\hat{\nu}}}{\partial x^\mu} \right). 
\end{eqnarray}
Because the coordinate basis connection coefficients are torsion-free (i.e. symmetric in the lower indices),
\begin{equation}
 {\Gamma^\nu}_{\nu\mu}  =  {\Gamma^\nu}_{\mu\nu} =  \frac{1}{\sqrt{-g}} \frac{\partial \sqrt{-g}}{\partial x^\mu} = \frac{\partial \ln \sqrt{-g}}{\partial x^\mu}
\end{equation}
(cf. Eq.~(\ref{eq:GammaContraction})).
But using the matrix identities (e.g. Ref.~\cite{Misner1973Gravitation})
\begin{equation}
\mathrm{Tr}\left(A^{-1} \frac{\partial A}{\partial z} \right) = \frac{\partial}{\partial z}\mathrm{Tr}\left(\ln A\right)
\end{equation}
and
\begin{equation}
\mathrm{Tr}\left(\ln A \right) = \ln \left( \det A \right)
\end{equation}
together with Eq.~(\ref{eq:TransformationDeterminant}), we also have
\begin{equation}
{L^{\hat{\nu}}}_{{\nu}} \frac{\partial {L^\nu}_{\hat{\nu}}}{\partial x^\mu} = \frac{\partial \ln (-g)^{-1/2}}{\partial x^\mu} = -\frac{\partial \ln \sqrt{-g}}{\partial x^\mu},
\end{equation}
and Eq.~(\ref{eq:ConnectionComovingTrace_1}) is confirmed.

Combining Eqs.~(\ref{eq:FirstMomentumDerivativeTermFirst}) and (\ref{eq:ConnectionComovingTrace_1}) in Eq.~(\ref{eq:FirstMomentumDerivativeTerm}), we have
\begin{eqnarray}
f  &{\Gamma^{\hat\imath}}&_{\hat\nu \hat\mu} \frac{\partial p^{\hat\nu}}{\partial p^{\hat\imath}} p^{\hat\mu} \nonumber \\
 &=& -f \,{\Gamma^{\hat\imath}}_{\hat\nu \hat\mu} \,p^{\hat\nu} p^{\hat\mu} {P^{\tilde\imath}}_{\hat\imath} \left(-p_{\hat 0}\right) \mathcal{D}_{\tilde\imath}\left[\frac{1}{\left(-p_{\hat 0}\right)}\right]
\label{eq:FirstMomentumDerivativeTermResultAppendix} 
\end{eqnarray}
for the first term in Eq.~(\ref{eq:LastTwoMomentumDerivatives}).

Next we turn to the second term in Eq.~(\ref{eq:LastTwoMomentumDerivatives}), which is more straightforward.
Using Eq.~(\ref{eq:MomentumDerivativeComponents}), 
\begin{eqnarray}
 f \, {\Gamma^{\hat\imath}}_{\hat\nu \hat\mu} p^{\hat\nu}  \frac{\partial p^{\hat\mu}}{\partial p^{\hat\imath}} 
 &=& f \left[ \frac{1}{\left(-p_{\hat 0}\right)}{\Gamma^{\hat\imath}}_{\hat\nu \hat 0}\, p_{\hat\imath} p^{\hat\nu} + {\Gamma^{\hat\imath}}_{\hat\nu \hat\imath}\, p^{\hat\nu} \right]  \\
&=& f \left[ \frac{1}{\left(-p_{\hat 0}\right)}{\Gamma^{\hat\mu}}_{\hat\nu  \hat 0}\, p_{\hat\mu} p^{\hat\nu} + {\Gamma^{\hat\mu}}_{\hat\nu \hat\mu}\, p^{\hat\nu} \right], \nonumber \\
& & \label{eq:SecondMomentumDerivativeTerm}
\end{eqnarray}
where we have added and subtracted ${\Gamma^{\hat 0}}_{\hat\nu \hat 0} p^{\hat\nu}$ to obtain the second line.
The first term in Eq.~(\ref{eq:SecondMomentumDerivativeTerm}) vanishes, because
\begin{equation}
{\Gamma^{\hat\mu}}_{\hat\nu  \hat 0}\, p_{\hat\mu} p^{\hat\nu} = p_{\hat\mu} \nabla_{\hat 0}\, p^{\hat\mu} - p_{\hat\mu}\, \partial_{\hat 0}\, p^{\hat\mu} = 0,
\end{equation}
with both terms in the middle independently vanishing thanks to Eq.~(\ref{eq:MassShellDerivative}).
As for the second term in Eq.~(\ref{eq:SecondMomentumDerivativeTerm}), we have
\begin{eqnarray}
{\Gamma^{\hat\mu}}_{\hat\nu \hat\mu}
&=&  {L^{\hat{\mu}}}_{{\mu}} {L^\nu}_{\hat{\nu}} {L^\rho}_{\hat{\mu}} \,{\Gamma^\mu}_{\nu\rho} 
+ {L^{\hat{\mu}}}_{{\mu}} {L^\rho}_{\hat{\mu}} 
\frac{\partial {L^\mu}_{\hat{\nu}}}{\partial x^\rho} \\
&=& {L^\nu}_{\hat{\nu}} \,{\Gamma^\mu}_{\nu\mu} +  \frac{\partial {L^\mu}_{\hat{\nu}}}{\partial x^\mu}.
\end{eqnarray}
Combining these results, we have 
\begin{equation}
f \, {\Gamma^{\hat\imath}}_{\hat\nu \hat\mu} p^{\hat\nu}  \frac{\partial p^{\hat\mu}}{\partial p^{\hat\imath}} 
 = f p^{\hat\nu} \left( {L^\nu}_{\hat{\nu}} \,{\Gamma^\mu}_{\nu\mu} +  \frac{\partial {L^\mu}_{\hat{\nu}}}{\partial x^\mu} \right)
\label{eq:SecondMomentumDerivativeTermResultAppendix} 
\end{equation}
for the second term in Eq.~(\ref{eq:LastTwoMomentumDerivatives}).

\section{Tetrad solution \label{app:TetradSolution}}

In this appendix we give the equations that complete the specification of the comoving frame tetrad $u^\mu, w^\mu, y^\mu, z^\mu$ of orthonormal basis vectors, according to the algebraic conditions outlined in Sec.~\ref{sec:TetradDecomposition}.
The timelike basis vector $u^\mu$ is the Lagrangian observer four-velocity, which is also the fluid four-velocity given by the evolution of the fluid and geometry.
In Eqs.~(\ref{eq:wyz_1})-(\ref{eq:wyz_3}), the spatial basis vectors $w^\mu, y^\mu, z^\mu$ are decomposed into portions orthogonal---$A n^\mu, B n^\mu, C n^\mu$---and tangent---$a^\mu, b^\mu, c^\mu$---to the spacelike slice.
The orthogonal components $A, B, C$ are given by Eqs.~(\ref{eq:ABC_1})-(\ref{eq:ABC_3}); and in the lab frame coordinate basis, $a^\mu, b^\mu, c^\mu$ are spacelike (Eq.~(\ref{eq:abc_Spacelike})), so that the nine components $a^i, b^i, c^i$ remain to be determined. 
For starters, we specify
\begin{eqnarray} 
b^1 &=& 0, \label{eq:b_Vanish} \\
c^1 &=& c^2 = 0. \label{eq:c_Vanish}
\end{eqnarray}
The single component $c^3$ is given by
\begin{equation}
\gamma_{33} \left( c^3 \right)^2 = 1 + \left( v_3 c^3 \right)^2,
\end{equation}
which follows from Eqs.~(\ref{eq:ABC_3}), (\ref{eq:abc_Normalization_3}), and (\ref{eq:c_Vanish}).
Once $c^3$ is known, the non-vanishing components $b^2$ and $b^3$ are determined by Eqs.~(\ref{eq:abc_Normalization_2}) and (\ref{eq:abc_NonOrthogonality_3}), which become
\begin{equation}  
\gamma_{22}\left(b^2\right)^2 + 2 \gamma_{23}\left(b^2 b^3\right) + \gamma_{33}\left(b^3\right)^2 
=  1 +  \left( v_2 b^2 + v_3 b^3 \right)^2
\end{equation}
and
\begin{equation}
\gamma_{23} b^2 c^3 + \gamma_{33} b^3 c^3 = \left( v_2 b^2 + v_3 b^3 \right)  \left( v_3 c^3 \right)
\end{equation}
when Eqs.~(\ref{eq:ABC_2})-(\ref{eq:ABC_3}), (\ref{eq:b_Vanish})-(\ref{eq:c_Vanish}) are taken into account.
Now that $b^i$ and $c^i$ are specified, the three components of $a^i$ are given by Eq.~(\ref{eq:abc_Normalization_1}),
\begin{eqnarray}
\gamma_{11}\left(a^1\right)^2 &+& 2 \gamma_{12}\left(a^1 a^2\right) + \gamma_{22}\left(a^2\right)^2 \nonumber \\
&+& 2 \gamma_{23}\left(a^2 a^3\right) + \gamma_{33}\left(a^3\right)^2 \nonumber \\
& & =  1 +  \left( v_1 a^1 + v_2 a^2 + v_3 a^3 \right)^2,
\label{eq:a1}
\end{eqnarray} 
by Eq.~(\ref{eq:abc_NonOrthogonality_1}),
\begin{eqnarray}
\gamma_{12} a^1 b^2 &+& \gamma_{22} a^2 b^2 +  \gamma_{32} a^3 b^2  \nonumber \\
+ \gamma_{13} a^1 b^3 &+& \gamma_{23} a^2 b^3 + \gamma_{33} a^3 b^3 \nonumber \\
&=&\! \left( v_1 a^1 + v_2 a^2 + v_3 a^3 \right) \! \left( v_2 b^2 + v_3 b^3 \right)\!,
\label{eq:a2}
\end{eqnarray}
and by Eq.~(\ref{eq:abc_NonOrthogonality_2}),
\begin{eqnarray}
\gamma_{13} a^1 c^3 &+& \gamma_{23} a^2 c^3 + \gamma_{33} a^3 c^3 \nonumber \\
&=& \left( v_1 a^1 + v_2 a^2 + v_3 a^3 \right)  \left( v_3 c^3 \right),
\label{eq:a3}
\end{eqnarray}
where Eqs.~(\ref{eq:ABC_1})-(\ref{eq:ABC_3}), (\ref{eq:b_Vanish})-(\ref{eq:c_Vanish}) are taken into account.

\section{Comoving frame connection coefficients \label{app:ComovingConnection}}

In this appendix we derive, in the context of the 3+1 approach, expressions for the comoving frame connection coefficients ${\Gamma^{\hat{\imath}}}_{\hat{\nu}\hat{\mu}}$---and their contraction with the momentum variable transformation ${P^{\tilde \imath}}_{\hat \imath}$---appearing in the momentum space divergence in Eq.~(\ref{eq:MomentumDivergencePartial}).
We begin with the expression
\begin{equation}
{\Gamma^{\hat{\rho}}}_{\hat{\nu}\hat{\mu}} = {L^{\hat\rho}}_\nu {L^\mu}_{\hat\mu} \left(\nabla_\mu {L^\nu}_{\hat\nu} \right)
\label{eq:RicciRotationSpatial}
\end{equation}
from Eq.~(\ref{eq:RicciRotation}).
Key to our derivation is use of the Eulerian decompositions of the tetrad ${L^\mu}_{\hat\mu}$ and its inverse ${L^{\hat\mu}}_\mu$ given in Eqs.~(\ref{eq:TetradDecomposition}) and (\ref{eq:TetradDecompositionInverse}),
keeping in mind that their projections ${\ell^{\mu}}_{\hat\mu}$ and ${\ell^{\hat\mu}}_ \mu$ are tangent to the spacelike slice (see Eqs.~(\ref{eq:TetradSpatial}), (\ref{eq:TetradSpatialInverse}), (\ref{eq:v_Spacelike}), and (\ref{eq:abc_Spacelike})).

Some relations involving derivatives of the unit normal $n^\mu$ will prove useful.
The gradient of the unit normal is related to the extrinsic curvature and lapse function by \cite{Gourgoulhon200731-Formalism-an}
\begin{equation}
\nabla_\mu n_\nu = -K_{\nu\mu} - \frac{n_\mu}{\alpha} \frac{\partial \alpha}{\partial x^\nu}.
\label{eq:Gradient_n}
\end{equation}
Because 
\begin{equation}
n^\nu \nabla_\mu n_\nu = \nabla_\mu ( n^\nu n_\nu ) / 2 = 0,
\label{eq:n_Gradient_n}
\end{equation}
and because $K_{\mu\nu}$ is tangent to the spacelike slice, i.e. spacelike in the lab frame coordinate basis, 
\begin{equation}
n^\mu K_{\mu\nu} = n^\nu K_{\mu\nu} = 0, 
\end{equation}
the nonvanishing projections of Eq.~(\ref{eq:Gradient_n}) are
\begin{eqnarray}
n^\mu \nabla_\mu n_\nu &=& \frac{1}{\alpha} \frac{\partial \alpha}{\partial x^\nu},
\label{eq:nGradientTime} \\
{\gamma^\mu}_i {\gamma^\nu}_j \nabla_\mu n_\nu &=& -K_{ij}
\label{eq:nGradientSpace}
\end{eqnarray}
in the lab frame coordinate basis.
Equation~(\ref{eq:nGradientTime}) relates the four-acceleration of an Eulerian observer to the gradient of the lapse function.
Equation~(\ref{eq:nGradientSpace}) relates the spatial part of the gradient of the unit normal to the extrinsic curvature, expressing the fact that the direction of the normal varies with the warp of the slice as embedded in spacetime.
Another relation valid in the lab frame coordinate basis for vectors $z^\mu$ tangent to the spacelike slice ($z^0 = 0$) is
\begin{equation}
z_\mu \frac{\partial n^\mu}{\partial x^\nu} = -\frac{z_i}{\alpha} \frac{\partial \beta^i}{\partial x^\nu}. \ \ \  (z^\mu \ \mathrm{spacelike})
\label{eq:nGradientU}
\end{equation}
This follows from writing 
\begin{eqnarray}
z_\mu \frac{\partial n^\mu }{ \partial x^\nu } &=& z_0 \frac{ \partial n^0 }{ \partial x^\nu } + z_i \frac{ \partial n^i }{ \partial x^\nu } \\
&=&  g_{0i} z^i \frac{\partial n^0 }{ \partial x^\nu } + z_i \frac{ \partial n^i }{ \partial x^\nu }
\end{eqnarray}
and using $g_{0i} = \beta_i$ and Eq.~(\ref{eq:UnitNormalU}).

We proceed with the calculation in three parts, based on the terms
\begin{equation}
\nabla_\mu {L^\nu}_{\hat\nu} = \mathcal{L}_{\hat\nu} \left( \nabla_\mu n^\nu \right) + n^\nu \left( \partial_\mu \mathcal{L}_{\hat\nu} \right) + \left( \nabla_\mu {\ell^\nu}_{\hat\nu} \right)
\label{eq:TetradGradient}
\end{equation}
obtained by using the Eulerian decomposition in Eq.~(\ref{eq:TetradDecomposition}) in the rightmost factor of Eq.~(\ref{eq:RicciRotationSpatial}).

We consider first the contribution
\begin{equation}
{L^{\hat\rho}}_\nu {L^\mu}_{\hat\mu}\, \mathcal{L}_{\hat\nu} \left( \nabla_\mu n^\nu \right) = {\ell^{\hat\rho}}_\nu \left( n^\mu \mathcal{L}_{\hat\mu} + {\ell^\mu}_{\hat\mu} \right)  \mathcal{L}_{\hat\nu} \left( \nabla_\mu n^\nu \right)
\end{equation}
to Eq.~(\ref{eq:RicciRotationSpatial}) from the first term in Eq.~(\ref{eq:TetradGradient}), where we have used Eq.~(\ref{eq:n_Gradient_n}).
Using Eq.~(\ref{eq:Gradient_n}) we have
\begin{equation}
{L^{\hat\rho}}_\nu {L^\mu}_{\hat\mu}\, \mathcal{L}_{\hat\nu} \left( \nabla_\mu n^\nu \right) = \ell^{\hat\rho j} \mathcal{L}_{\hat\nu} \left( \frac{\mathcal{L}_{\hat\mu}}{\alpha} \frac{\partial\alpha}{\partial x^j} - {\ell^k}_{\hat\mu} K_{jk} \right).
\label{eq:ConnectionComoving_1} 
\end{equation}
for this contribution.

Next is the contribution
\begin{equation}
{L^{\hat\rho}}_\nu {L^\mu}_{\hat\mu}\, n^\nu \left( \partial_\mu \mathcal{L}_{\hat\nu} \right) = -\mathcal{L}^{\hat\rho} \left( n^\mu \mathcal{L}_{\hat\mu} + {\ell^\mu}_{\hat\mu} \right) \left( \partial_\mu \mathcal{L}_{\hat\nu} \right)
\end{equation}
to Eq.~(\ref{eq:RicciRotationSpatial}) from the second term in Eq.~(\ref{eq:TetradGradient}), where we have used the fact that ${\ell^{\hat\rho}}_\nu$ is spacelike.
Using Eq.~(\ref{eq:UnitNormalU}), we have
\begin{eqnarray}
{L^{\hat\rho}}_\nu &{L^\mu}_{\hat\mu}& n^\nu \left( \partial_\mu \mathcal{L}_{\hat\nu} \right) \nonumber \\
&=& - \frac{\mathcal{L}^{\hat\rho}}{\alpha} \left[ \mathcal{L}_{\hat\mu} \frac{\partial \mathcal{L}_{\hat\nu}}{\partial t} + \left(\alpha {\ell^i}_{\hat\mu} - \beta^i \mathcal{L}_{\hat\mu} \right) \frac{\partial \mathcal{L}_{\hat\nu}}{\partial x^i}\right]
\label{eq:ConnectionComoving_2} 
\end{eqnarray}
for this contribution. 

Finally, the contribution
\begin{equation}
{L^{\hat\rho}}_\nu {L^\mu}_{\hat\mu} \left( \nabla_\mu {\ell^\nu}_{\hat\nu} \right) =  
\left( \mathcal{L}^{\hat\rho} n_\nu + {\ell^{\hat\rho}}_\nu \right)
\left( n^\mu \mathcal{L}_{\hat\mu} + {\ell^\mu}_{\hat\mu} \right) \left( \nabla_\mu {\ell^\nu}_{\hat\nu} \right)
\end{equation}
to Eq.~(\ref{eq:RicciRotationSpatial}) from the second term in Eq.~(\ref{eq:TetradGradient}) is more complicated.
Using the identity $0 = \nabla_\mu \left(n_\nu {\ell^\nu}_{\hat\nu} \right) = \left( \nabla_\mu n_\nu\right) {\ell^\nu}_{\hat\nu} + n_\nu \left( \nabla_\mu {\ell^\nu}_{\hat\nu} \right)$ along with Eq.~(\ref{eq:Gradient_n}), two of the terms are
\begin{eqnarray}
\left( \mathcal{L}^{\hat\rho} n_\nu \right)  \left( \right.&n^\mu&\left. \mathcal{L}_{\hat\mu} + {\ell^\mu}_{\hat\mu} \right) \left( \nabla_\mu {\ell^\nu}_{\hat\nu} \right) \nonumber \\
&=& - \mathcal{L}^{\hat\rho}\, {\ell^\nu}_{\hat\nu}\left( n^\mu \mathcal{L}_{\hat\mu} + {\ell^\mu}_{\hat\mu} \right) \left( \nabla_\mu n_\nu \right) \\
&=& -\mathcal{L}^{\hat\rho} \,{\ell^j}_{\hat\nu} \left( \frac{\mathcal{L}_{\hat\mu}}{\alpha} \frac{\partial\alpha}{\partial x^j} - {\ell^k}_{\hat\mu} K_{jk} \right).
\label{eq:ConnectionComoving_3a} 
\end{eqnarray}
Another term gives
\begin{eqnarray}
{\ell^{\hat\rho}}_\nu \left( \right. &n^\mu& \left. \mathcal{L}_{\hat\mu} \right) \left( \nabla_\mu {\ell^\nu}_{\hat\nu} \right) \nonumber \\
&=& {\ell^{\hat\rho \nu}}\, n^\mu \, \mathcal{L}_{\hat\mu} \left(\partial_\mu \ell_{\nu\hat\nu} - {\Gamma^\rho}_{\nu\mu} \ell_{\rho\hat\nu}\right)  \\
&=& {\ell^{\hat\rho \nu}}\, n^\mu \, \mathcal{L}_{\hat\mu} \,\partial_\mu \ell_{\nu\hat\nu} \nonumber \\
& & - {\ell^{\hat\rho \nu}} \, \mathcal{L}_{\hat\mu}\,\ell_{\rho\hat\nu} \left( \nabla_\nu n^\rho - \partial_\nu n^\rho \right) \\
&=& \frac{\ell^{\hat\rho k} \mathcal{L}_{\hat\mu}}{\alpha}\left( \frac{\partial \ell_{k \hat\nu}}{\partial t}   -   \beta^j  \frac{\partial \ell_{k \hat\nu}}{\partial x^j} \right) \nonumber \\
& & + \ell^{\hat\rho j} \mathcal{L}_{\hat\mu} \left({\ell^k}_{\hat\nu} K_{jk} - \frac{\ell_{k\hat\nu}}{\alpha} \frac{\partial \beta^k}{\partial x^j} \right),
\label{eq:ConnectionComoving_3b} 
\end{eqnarray}
where now Eq.~(\ref{eq:nGradientU}) has been used in addition to Eq.~(\ref{eq:Gradient_n}).
The final piece is
\begin{eqnarray}
{\ell^{\hat\rho}}_\nu &{\ell^\mu}_{\hat\mu} &  \left( \nabla_\mu {\ell^\nu}_{\hat\nu} \right) \nonumber \\
&=& {\ell^{\hat\rho\nu}} {\ell^\mu}_{\hat\mu} \left(\partial_\mu \ell_{\nu\hat\nu} -  {\Gamma^\rho}_{\nu\mu} {\ell_{\rho\hat\nu}} \right) \\
&\rightarrow& {\ell^{\hat\rho k}} {\ell^j}_{\hat\mu} \left(\frac{\partial \ell_{k\hat\nu}}{\partial x^j} - \frac{{\ell^i}_{\hat\nu}}{2} \frac{\partial \gamma_{ij}}{\partial x^k} \right),
\label{eq:ConnectionComoving_3c} 
\end{eqnarray}
in which we have anticipated that the two terms from the connection coefficients antisymmetric in $i$ and $j$ will cancel upon contraction with the product ${\ell^j}_{\hat\mu} {\ell^i}_{\hat\nu}$ (which becomes symmetric in $i$ and $j$ upon contraction with $p^{\hat\mu} p^{\hat\nu}$). 

In summary, the comoving frame connection coefficients ${\Gamma^{\hat{\rho}}}_{\hat{\nu}\hat{\mu}}$ are given by the sum of Eqs.~(\ref{eq:ConnectionComoving_1}), (\ref{eq:ConnectionComoving_2}), (\ref{eq:ConnectionComoving_3a}), (\ref{eq:ConnectionComoving_3b}), and (\ref{eq:ConnectionComoving_3c}).  
We note that two of the terms involving $K_{jk}$ cancel upon contraction with $p^{\hat\mu} p^{\hat\nu}$.

In the momentum space divergence in Eq.~(\ref{eq:MomentumDivergencePartial}), the comoving frame connection coefficients ${\Gamma^{\hat{\imath}}}_{\hat{\nu}\hat{\mu}} $ appear contracted with the momentum space coordinate transformation ${P^{\tilde \imath}}_{\hat \imath}$ and two factors of momentum.
It will prove convenient to use the decomposition of ${P^{\tilde \imath}}_{\hat \imath}$ in Eq.~(\ref{eq:MomentumDecomposition}):
\begin{equation}
{P^{\tilde \imath}}_{\hat \imath} {\Gamma^{\hat{\imath}}}_{\hat{\nu}\hat{\mu}}  \, p^{\hat\nu} p^{\hat\mu} 
= \left( \frac{Q^{\tilde\imath} \, p_{\hat\imath}}{p} + {U^{\tilde \imath}}_{\hat \imath}\right) {\Gamma^{\hat{\imath}}}_{\hat{\nu}\hat{\mu}}  \, p^{\hat\nu} p^{\hat\mu}. 
\label{eq:PGamma}
\end{equation}
The reason this is useful is that it allows us to use the identity
\begin{equation}
p_{\hat\imath}\, {\Gamma^{\hat{\imath}}}_{\hat{\nu}\hat{\mu}}  \, p^{\hat\nu} p^{\hat\mu}
= \left(-p_{\hat 0}\right) {\Gamma^{\hat{0}}}_{\hat{\nu}\hat{\mu}}  \, p^{\hat\nu} p^{\hat\mu},
\label{eq:GeodesicIdentity}
\end{equation}
which follows from Eqs.~(\ref{eq:Geodesic_p}) and (\ref{eq:MassShellDerivative}).
Thus Eq.~(\ref{eq:PGamma}) becomes
\begin{equation}
{P^{\tilde \imath}}_{\hat \imath} {\Gamma^{\hat{\imath}}}_{\hat{\nu}\hat{\mu}}  \, p^{\hat\nu} p^{\hat\mu} 
= \left[ \frac{Q^{\tilde\imath} \, \left(-p_{\hat 0}\right)}{p}\, {\Gamma^{\hat{0}}}_{\hat{\nu}\hat{\mu}} + {U^{\tilde \imath}}_{\hat \imath} \,{\Gamma^{\hat{\imath}}}_{\hat{\nu}\hat{\mu}} \right]  \, p^{\hat\nu} p^{\hat\mu}. 
\label{eq:PGamma2}
\end{equation}
It is in this combination that we use the comoving frame connection coefficients---the sum of Eqs.~(\ref{eq:ConnectionComoving_1}), (\ref{eq:ConnectionComoving_2}), (\ref{eq:ConnectionComoving_3a}), (\ref{eq:ConnectionComoving_3b}), and (\ref{eq:ConnectionComoving_3c})---in the momentum space divergence in Sec.~\ref{sec:Boltzmann}.

\begin{acknowledgments}
This research was supported by the Office of Advanced Scientific Computing Research and the Office of Nuclear Physics, U.S. Department of Energy. 
\end{acknowledgments}


\def\apjs{Astrophys. J. Suppl. Ser. }
\def\mnras{Mon. Not. Roy. Ast. Soc. }
\def\aap{Astron. Astrophys. }
\def\jqsrt{J. Quant. Spectrosc. Radiat. Transfer }
\def\apss{Astrophys. Sp. Sci. }


\begin{thebibliography}{54}%
\makeatletter
\providecommand \@ifxundefined [1]{%
 \@ifx{#1\undefined}
}%
\providecommand \@ifnum [1]{%
 \ifnum #1\expandafter \@firstoftwo
 \else \expandafter \@secondoftwo
 \fi
}%
\providecommand \@ifx [1]{%
 \ifx #1\expandafter \@firstoftwo
 \else \expandafter \@secondoftwo
 \fi
}%
\providecommand \natexlab [1]{#1}%
\providecommand \enquote  [1]{``#1''}%
\providecommand \bibnamefont  [1]{#1}%
\providecommand \bibfnamefont [1]{#1}%
\providecommand \citenamefont [1]{#1}%
\providecommand \href@noop [0]{\@secondoftwo}%
\providecommand \href [0]{\begingroup \@sanitize@url \@href}%
\providecommand \@href[1]{\@@startlink{#1}\@@href}%
\providecommand \@@href[1]{\endgroup#1\@@endlink}%
\providecommand \@sanitize@url [0]{\catcode `\\12\catcode `\$12\catcode
  `\&12\catcode `\#12\catcode `\^12\catcode `\_12\catcode `\%12\relax}%
\providecommand \@@startlink[1]{}%
\providecommand \@@endlink[0]{}%
\providecommand \url  [0]{\begingroup\@sanitize@url \@url }%
\providecommand \@url [1]{\endgroup\@href {#1}{\urlprefix }}%
\providecommand \urlprefix  [0]{URL }%
\providecommand \Eprint [0]{\href }%
\providecommand \doibase [0]{http://dx.doi.org/}%
\providecommand \selectlanguage [0]{\@gobble}%
\providecommand \bibinfo  [0]{\@secondoftwo}%
\providecommand \bibfield  [0]{\@secondoftwo}%
\providecommand \translation [1]{[#1]}%
\providecommand \BibitemOpen [0]{}%
\providecommand \bibitemStop [0]{}%
\providecommand \bibitemNoStop [0]{.\EOS\space}%
\providecommand \EOS [0]{\spacefactor3000\relax}%
\providecommand \BibitemShut  [1]{\csname bibitem#1\endcsname}%
\let\auto@bib@innerbib\@empty
\bibitem [{\citenamefont {Mezzacappa}(2005)}]{Mezzacappa2005ASCERTAINING-TH}%
  \BibitemOpen
  \bibfield  {author} {\bibinfo {author} {\bibfnamefont {A.}~\bibnamefont
  {Mezzacappa}},\ }\href@noop {} {\bibfield  {journal} {\bibinfo  {journal}
  {Annu. Rev. Nucl. Part. Sci.}\ }\textbf {\bibinfo {volume} {55}},\ \bibinfo
  {pages} {467} (\bibinfo {year} {2005})}\BibitemShut {NoStop}%
\bibitem [{\citenamefont {{Kotake}}\ \emph {et~al.}(2006)\citenamefont
  {{Kotake}}, \citenamefont {{Sato}},\ and\ \citenamefont
  {{Takahashi}}}]{Kotake2006Explosion-mecha}%
  \BibitemOpen
  \bibfield  {author} {\bibinfo {author} {\bibfnamefont {K.}~\bibnamefont
  {{Kotake}}}, \bibinfo {author} {\bibfnamefont {K.}~\bibnamefont {{Sato}}}, \
  and\ \bibinfo {author} {\bibfnamefont {K.}~\bibnamefont {{Takahashi}}},\
  }\href@noop {} {\bibfield  {journal} {\bibinfo  {journal} {Rep. Prog. Phys.}\
  }\textbf {\bibinfo {volume} {69}},\ \bibinfo {pages} {971} (\bibinfo {year}
  {2006})}\BibitemShut {NoStop}%
\bibitem [{\citenamefont {{Kotake}}\ \emph
  {et~al.}(2012{\natexlab{a}})\citenamefont {{Kotake}}, \citenamefont
  {{Takiwaki}}, \citenamefont {{Suwa}}, \citenamefont {{Iwakami Nakano}},
  \citenamefont {{Kawagoe}}, \citenamefont {{Masada}},\ and\ \citenamefont
  {{Fujimoto}}}]{Kotake2012Multimessengers}%
  \BibitemOpen
  \bibfield  {author} {\bibinfo {author} {\bibfnamefont {K.}~\bibnamefont
  {{Kotake}}}, \bibinfo {author} {\bibfnamefont {T.}~\bibnamefont
  {{Takiwaki}}}, \bibinfo {author} {\bibfnamefont {Y.}~\bibnamefont {{Suwa}}},
  \bibinfo {author} {\bibfnamefont {W.}~\bibnamefont {{Iwakami Nakano}}},
  \bibinfo {author} {\bibfnamefont {S.}~\bibnamefont {{Kawagoe}}}, \bibinfo
  {author} {\bibfnamefont {Y.}~\bibnamefont {{Masada}}}, \ and\ \bibinfo
  {author} {\bibfnamefont {S.-i.}\ \bibnamefont {{Fujimoto}}},\ }\href@noop {}
  {\bibfield  {journal} {\bibinfo  {journal} {Advances in Astronomy}\ }\textbf
  {\bibinfo {volume} {2012}},\ \bibinfo {pages} {428757} (\bibinfo {year}
  {2012}{\natexlab{a}})}\BibitemShut {NoStop}%
\bibitem [{\citenamefont {{Kotake}}\ \emph
  {et~al.}(2012{\natexlab{b}})\citenamefont {{Kotake}}, \citenamefont
  {{Sumiyoshi}}, \citenamefont {{Yamada}}, \citenamefont {{Takiwaki}},
  \citenamefont {{Kuroda}}, \citenamefont {{Suwa}},\ and\ \citenamefont
  {{Nagakura}}}]{Kotake2012Core-Collapse-S}%
  \BibitemOpen
  \bibfield  {author} {\bibinfo {author} {\bibfnamefont {K.}~\bibnamefont
  {{Kotake}}}, \bibinfo {author} {\bibfnamefont {K.}~\bibnamefont
  {{Sumiyoshi}}}, \bibinfo {author} {\bibfnamefont {S.}~\bibnamefont
  {{Yamada}}}, \bibinfo {author} {\bibfnamefont {T.}~\bibnamefont
  {{Takiwaki}}}, \bibinfo {author} {\bibfnamefont {T.}~\bibnamefont
  {{Kuroda}}}, \bibinfo {author} {\bibfnamefont {Y.}~\bibnamefont {{Suwa}}}, \
  and\ \bibinfo {author} {\bibfnamefont {H.}~\bibnamefont {{Nagakura}}},\
  }\href@noop {} {\bibfield  {journal} {\bibinfo  {journal} {Prog. Theor. Exp.
  Phys.}\ }\textbf {\bibinfo {volume} {2012}},\ \bibinfo {pages} {01A301}
  (\bibinfo {year} {2012}{\natexlab{b}})}\BibitemShut {NoStop}%
\bibitem [{\citenamefont {{Janka}}(2012)}]{Janka2012Explosion-Mecha}%
  \BibitemOpen
  \bibfield  {author} {\bibinfo {author} {\bibfnamefont {H.-T.}\ \bibnamefont
  {{Janka}}},\ }\href@noop {} {\bibfield  {journal} {\bibinfo  {journal} {Annu.
  Rev. Nucl. Part. Sci.}\ }\textbf {\bibinfo {volume} {62}},\ \bibinfo {pages}
  {407} (\bibinfo {year} {2012})}\BibitemShut {NoStop}%
\bibitem [{\citenamefont {{Burrows}}(2013)}]{Burrows2012Perspectives-on}%
  \BibitemOpen
  \bibfield  {author} {\bibinfo {author} {\bibfnamefont {A.}~\bibnamefont
  {{Burrows}}},\ }\href@noop {} {\bibfield  {journal} {\bibinfo  {journal}
  {Rev. Mod. Phys.}\ }\textbf {\bibinfo {volume} {85}},\ \bibinfo {pages} {245}
  (\bibinfo {year} {2013})}\BibitemShut {NoStop}%
\bibitem [{\citenamefont {{Janka}}\ \emph {et~al.}(2012)\citenamefont
  {{Janka}}, \citenamefont {{Hanke}}, \citenamefont {{H{\"u}depohl}},
  \citenamefont {{Marek}}, \citenamefont {{M{\"u}ller}},\ and\ \citenamefont
  {{Obergaulinger}}}]{Janka2012Core-collapse-s}%
  \BibitemOpen
  \bibfield  {author} {\bibinfo {author} {\bibfnamefont {H.-T.}\ \bibnamefont
  {{Janka}}}, \bibinfo {author} {\bibfnamefont {F.}~\bibnamefont {{Hanke}}},
  \bibinfo {author} {\bibfnamefont {L.}~\bibnamefont {{H{\"u}depohl}}},
  \bibinfo {author} {\bibfnamefont {A.}~\bibnamefont {{Marek}}}, \bibinfo
  {author} {\bibfnamefont {B.}~\bibnamefont {{M{\"u}ller}}}, \ and\ \bibinfo
  {author} {\bibfnamefont {M.}~\bibnamefont {{Obergaulinger}}},\ }\href@noop {}
  {\bibfield  {journal} {\bibinfo  {journal} {Prog. Theor. Exp. Phys.}\
  }\textbf {\bibinfo {volume} {2012}},\ \bibinfo {pages} {01A309} (\bibinfo
  {year} {2012})}\BibitemShut {NoStop}%
\bibitem [{Note1()}]{Note1}%
  \BibitemOpen
  \bibinfo {note} {Calculation of the emerging neutrino signals---of intrinsic
  interest as an observational probe of the core-collapse supernova
  environment, and of the properties of the neutrinos themselves---definitely
  requires treatment of the quantum effects induced by neutrino mass and flavor
  mixing \cite
  {Dasgupta2010Physics-and-Ast,Raffelt2011New-opportuniti,Dighe2011Signatures-of-s}.
  Recent explorations suggest that flavor mixing does not impact the explosion
  mechansim \cite
  {Chakraborty2011No-Collective-N,Suwa2011Impacts-of-Coll,Dasgupta2012Role-of-collect,Sarikas2012Suppression-of-,Saviano2012Stability-analy,Sarikas2012Supernova-neutr,Pejcha2012Effect-of-colle},
  but consensus on the impacts of flavor mixing in supernovae has a fickle
  history, and future more definitive simulations that include neutrino
  transport with quantum kinetics could surprise us with flavor mixing effects
  on the explosion mechanism as well.}\BibitemShut {Stop}%
\bibitem [{\citenamefont {{Lentz}}\ \emph {et~al.}(2012)\citenamefont
  {{Lentz}}, \citenamefont {{Mezzacappa}}, \citenamefont {{Bronson Messer}},
  \citenamefont {{Liebend{\"o}rfer}}, \citenamefont {{Hix}},\ and\
  \citenamefont {{Bruenn}}}]{Lentz2012On-the-Requirem}%
  \BibitemOpen
  \bibfield  {author} {\bibinfo {author} {\bibfnamefont {E.~J.}\ \bibnamefont
  {{Lentz}}}, \bibinfo {author} {\bibfnamefont {A.}~\bibnamefont
  {{Mezzacappa}}}, \bibinfo {author} {\bibfnamefont {O.~E.}\ \bibnamefont
  {{Bronson Messer}}}, \bibinfo {author} {\bibfnamefont {M.}~\bibnamefont
  {{Liebend{\"o}rfer}}}, \bibinfo {author} {\bibfnamefont {W.~R.}\ \bibnamefont
  {{Hix}}}, \ and\ \bibinfo {author} {\bibfnamefont {S.~W.}\ \bibnamefont
  {{Bruenn}}},\ }\href@noop {} {\bibfield  {journal} {\bibinfo  {journal}
  {\apj}\ }\textbf {\bibinfo {volume} {747}} (\bibinfo {year}
  {2012})}\BibitemShut {NoStop}%
\bibitem [{\citenamefont {Lindquist}(1966)}]{Lindquist1966Relativistic-Tr}%
  \BibitemOpen
  \bibfield  {author} {\bibinfo {author} {\bibfnamefont {R.~W.}\ \bibnamefont
  {Lindquist}},\ }\href@noop {} {\bibfield  {journal} {\bibinfo  {journal}
  {Ann. Phys. (NY)}\ }\textbf {\bibinfo {volume} {37}},\ \bibinfo {pages} {487}
  (\bibinfo {year} {1966})}\BibitemShut {NoStop}%
\bibitem [{\citenamefont {{Castor}}(1972)}]{Castor1972Radiative-Trans}%
  \BibitemOpen
  \bibfield  {author} {\bibinfo {author} {\bibfnamefont {J.~I.}\ \bibnamefont
  {{Castor}}},\ }\href@noop {} {\bibfield  {journal} {\bibinfo  {journal}
  {\apj}\ }\textbf {\bibinfo {volume} {178}},\ \bibinfo {pages} {779} (\bibinfo
  {year} {1972})}\BibitemShut {NoStop}%
\bibitem [{\citenamefont {{Mihalas}}(1980)}]{Mihalas1980Solution-of-the}%
  \BibitemOpen
  \bibfield  {author} {\bibinfo {author} {\bibfnamefont {D.}~\bibnamefont
  {{Mihalas}}},\ }\href@noop {} {\bibfield  {journal} {\bibinfo  {journal}
  {\apj}\ }\textbf {\bibinfo {volume} {237}},\ \bibinfo {pages} {574} (\bibinfo
  {year} {1980})}\BibitemShut {NoStop}%
\bibitem [{\citenamefont {{Mihalas}}(1981)}]{Mihalas1981A-Comment-on-Ra}%
  \BibitemOpen
  \bibfield  {author} {\bibinfo {author} {\bibfnamefont {D.}~\bibnamefont
  {{Mihalas}}},\ }\href@noop {} {\bibfield  {journal} {\bibinfo  {journal}
  {\apj}\ }\textbf {\bibinfo {volume} {250}},\ \bibinfo {pages} {373} (\bibinfo
  {year} {1981})}\BibitemShut {NoStop}%
\bibitem [{\citenamefont {{Buchler}}(1983)}]{Buchler1983Radiation-trans}%
  \BibitemOpen
  \bibfield  {author} {\bibinfo {author} {\bibfnamefont {J.~R.}\ \bibnamefont
  {{Buchler}}},\ }\href@noop {} {\bibfield  {journal} {\bibinfo  {journal}
  {\jqsrt}\ }\textbf {\bibinfo {volume} {30}},\ \bibinfo {pages} {395}
  (\bibinfo {year} {1983})}\BibitemShut {NoStop}%
\bibitem [{\citenamefont {{Buchler}}(1986)}]{Buchler1986Radiation-trans}%
  \BibitemOpen
  \bibfield  {author} {\bibinfo {author} {\bibfnamefont {J.~R.}\ \bibnamefont
  {{Buchler}}},\ }\href@noop {} {\bibfield  {journal} {\bibinfo  {journal}
  {\jqsrt}\ }\textbf {\bibinfo {volume} {36}},\ \bibinfo {pages} {441}
  (\bibinfo {year} {1986})}\BibitemShut {NoStop}%
\bibitem [{\citenamefont {{Kaneko}}\ \emph {et~al.}(1984)\citenamefont
  {{Kaneko}}, \citenamefont {{Morita}},\ and\ \citenamefont
  {{Maekawa}}}]{Kaneko1984The-comoving-fr}%
  \BibitemOpen
  \bibfield  {author} {\bibinfo {author} {\bibfnamefont {N.}~\bibnamefont
  {{Kaneko}}}, \bibinfo {author} {\bibfnamefont {K.}~\bibnamefont {{Morita}}},
  \ and\ \bibinfo {author} {\bibfnamefont {M.}~\bibnamefont {{Maekawa}}},\
  }\href@noop {} {\bibfield  {journal} {\bibinfo  {journal} {\apss}\ }\textbf
  {\bibinfo {volume} {107}},\ \bibinfo {pages} {333} (\bibinfo {year}
  {1984})}\BibitemShut {NoStop}%
\bibitem [{\citenamefont {{Munier}}\ and\ \citenamefont
  {{Weaver}}(1986{\natexlab{a}})}]{Munier1986Radiation-trans}%
  \BibitemOpen
  \bibfield  {author} {\bibinfo {author} {\bibfnamefont {A.}~\bibnamefont
  {{Munier}}}\ and\ \bibinfo {author} {\bibfnamefont {R.}~\bibnamefont
  {{Weaver}}},\ }\href@noop {} {\bibfield  {journal} {\bibinfo  {journal}
  {Comp. Phys. Rep.}\ }\textbf {\bibinfo {volume} {3}},\ \bibinfo {pages} {127}
  (\bibinfo {year} {1986}{\natexlab{a}})}\BibitemShut {NoStop}%
\bibitem [{\citenamefont {{Munier}}\ and\ \citenamefont
  {{Weaver}}(1986{\natexlab{b}})}]{Munier1986Radiation-trans2}%
  \BibitemOpen
  \bibfield  {author} {\bibinfo {author} {\bibfnamefont {A.}~\bibnamefont
  {{Munier}}}\ and\ \bibinfo {author} {\bibfnamefont {R.}~\bibnamefont
  {{Weaver}}},\ }\href@noop {} {\bibfield  {journal} {\bibinfo  {journal}
  {Comp. Phys. Rep.}\ }\textbf {\bibinfo {volume} {3}},\ \bibinfo {pages} {165}
  (\bibinfo {year} {1986}{\natexlab{b}})}\BibitemShut {NoStop}%
\bibitem [{\citenamefont {{Riffert}}(1986)}]{Riffert1986A-general-Euler}%
  \BibitemOpen
  \bibfield  {author} {\bibinfo {author} {\bibfnamefont {H.}~\bibnamefont
  {{Riffert}}},\ }\href@noop {} {\bibfield  {journal} {\bibinfo  {journal}
  {\apj}\ }\textbf {\bibinfo {volume} {310}},\ \bibinfo {pages} {729} (\bibinfo
  {year} {1986})}\BibitemShut {NoStop}%
\bibitem [{\citenamefont {{Mezzacappa}}\ and\ \citenamefont
  {{Matzner}}(1989)}]{Mezzacappa1989Computer-simula}%
  \BibitemOpen
  \bibfield  {author} {\bibinfo {author} {\bibfnamefont {A.}~\bibnamefont
  {{Mezzacappa}}}\ and\ \bibinfo {author} {\bibfnamefont {R.~A.}\ \bibnamefont
  {{Matzner}}},\ }\href@noop {} {\bibfield  {journal} {\bibinfo  {journal}
  {\apj}\ }\textbf {\bibinfo {volume} {343}},\ \bibinfo {pages} {853} (\bibinfo
  {year} {1989})}\BibitemShut {NoStop}%
\bibitem [{\citenamefont {{Misner}}\ \emph {et~al.}(1973)\citenamefont
  {{Misner}}, \citenamefont {{Thorne}},\ and\ \citenamefont
  {{Wheeler}}}]{Misner1973Gravitation}%
  \BibitemOpen
  \bibfield  {author} {\bibinfo {author} {\bibfnamefont {C.~W.}\ \bibnamefont
  {{Misner}}}, \bibinfo {author} {\bibfnamefont {K.~S.}\ \bibnamefont
  {{Thorne}}}, \ and\ \bibinfo {author} {\bibfnamefont {J.~A.}\ \bibnamefont
  {{Wheeler}}},\ }\href@noop {} {\emph {\bibinfo {title} {{Gravitation}}}}\
  (\bibinfo  {publisher} {W.H.~Freeman and Co.},\ \bibinfo {address} {San
  Francisco},\ \bibinfo {year} {1973})\BibitemShut {NoStop}%
\bibitem [{\citenamefont {{Schinder}}(1988)}]{Schinder1988General-relativ}%
  \BibitemOpen
  \bibfield  {author} {\bibinfo {author} {\bibfnamefont {P.~J.}\ \bibnamefont
  {{Schinder}}},\ }\href@noop {} {\bibfield  {journal} {\bibinfo  {journal}
  {\prd}\ }\textbf {\bibinfo {volume} {38}},\ \bibinfo {pages} {1673} (\bibinfo
  {year} {1988})}\BibitemShut {NoStop}%
\bibitem [{\citenamefont {{Mezzacappa}}\ and\ \citenamefont
  {{Bruenn}}(1993{\natexlab{a}})}]{Mezzacappa1993Type-II-superno}%
  \BibitemOpen
  \bibfield  {author} {\bibinfo {author} {\bibfnamefont {A.}~\bibnamefont
  {{Mezzacappa}}}\ and\ \bibinfo {author} {\bibfnamefont {S.~W.}\ \bibnamefont
  {{Bruenn}}},\ }\href@noop {} {\bibfield  {journal} {\bibinfo  {journal}
  {\apj}\ }\textbf {\bibinfo {volume} {405}},\ \bibinfo {pages} {637} (\bibinfo
  {year} {1993}{\natexlab{a}})}\BibitemShut {NoStop}%
\bibitem [{\citenamefont {{Mezzacappa}}\ and\ \citenamefont
  {{Bruenn}}(1993{\natexlab{b}})}]{Mezzacappa1993A-numerical-met}%
  \BibitemOpen
  \bibfield  {author} {\bibinfo {author} {\bibfnamefont {A.}~\bibnamefont
  {{Mezzacappa}}}\ and\ \bibinfo {author} {\bibfnamefont {S.~W.}\ \bibnamefont
  {{Bruenn}}},\ }\href@noop {} {\bibfield  {journal} {\bibinfo  {journal}
  {\apj}\ }\textbf {\bibinfo {volume} {405}},\ \bibinfo {pages} {669} (\bibinfo
  {year} {1993}{\natexlab{b}})}\BibitemShut {NoStop}%
\bibitem [{\citenamefont {{Bruenn}}(1985)}]{Bruenn1985Stellar-core-co}%
  \BibitemOpen
  \bibfield  {author} {\bibinfo {author} {\bibfnamefont {S.~W.}\ \bibnamefont
  {{Bruenn}}},\ }\href@noop {} {\bibfield  {journal} {\bibinfo  {journal}
  {\apjs}\ }\textbf {\bibinfo {volume} {58}},\ \bibinfo {pages} {771} (\bibinfo
  {year} {1985})}\BibitemShut {NoStop}%
\bibitem [{\citenamefont {{Liebend{\"o}rfer}}\ \emph
  {et~al.}(2004)\citenamefont {{Liebend{\"o}rfer}}, \citenamefont {{Messer}},
  \citenamefont {{Mezzacappa}}, \citenamefont {{Bruenn}}, \citenamefont
  {{Cardall}},\ and\ \citenamefont
  {{Thielemann}}}]{Liebendorfer2004A-Finite-Differ}%
  \BibitemOpen
  \bibfield  {author} {\bibinfo {author} {\bibfnamefont {M.}~\bibnamefont
  {{Liebend{\"o}rfer}}}, \bibinfo {author} {\bibfnamefont {O.~E.~B.}\
  \bibnamefont {{Messer}}}, \bibinfo {author} {\bibfnamefont {A.}~\bibnamefont
  {{Mezzacappa}}}, \bibinfo {author} {\bibfnamefont {S.~W.}\ \bibnamefont
  {{Bruenn}}}, \bibinfo {author} {\bibfnamefont {C.~Y.}\ \bibnamefont
  {{Cardall}}}, \ and\ \bibinfo {author} {\bibfnamefont {F.}~\bibnamefont
  {{Thielemann}}},\ }\href@noop {} {\bibfield  {journal} {\bibinfo  {journal}
  {\apjs}\ }\textbf {\bibinfo {volume} {150}},\ \bibinfo {pages} {263}
  (\bibinfo {year} {2004})}\BibitemShut {NoStop}%
\bibitem [{\citenamefont {Cardall}\ and\ \citenamefont
  {Mezzacappa}(2003)}]{Cardall2003Conservative-fo}%
  \BibitemOpen
  \bibfield  {author} {\bibinfo {author} {\bibfnamefont {C.~Y.}\ \bibnamefont
  {Cardall}}\ and\ \bibinfo {author} {\bibfnamefont {A.}~\bibnamefont
  {Mezzacappa}},\ }\href@noop {} {\bibfield  {journal} {\bibinfo  {journal}
  {Phys. Rev. D}\ }\textbf {\bibinfo {volume} {68}},\ \bibinfo {pages} {023006}
  (\bibinfo {year} {2003})}\BibitemShut {NoStop}%
\bibitem [{\citenamefont {{Sumiyoshi}}\ and\ \citenamefont
  {{Yamada}}(2012)}]{Sumiyoshi2012Neutrino-Transf}%
  \BibitemOpen
  \bibfield  {author} {\bibinfo {author} {\bibfnamefont {K.}~\bibnamefont
  {{Sumiyoshi}}}\ and\ \bibinfo {author} {\bibfnamefont {S.}~\bibnamefont
  {{Yamada}}},\ }\href@noop {} {\bibfield  {journal} {\bibinfo  {journal}
  {\apjs}\ }\textbf {\bibinfo {volume} {199}},\ \bibinfo {pages} {17} (\bibinfo
  {year} {2012})}\BibitemShut {NoStop}%
\bibitem [{\citenamefont {M{\"u}ller}\ \emph {et~al.}(2010)\citenamefont
  {M{\"u}ller}, \citenamefont {Janka},\ and\ \citenamefont
  {Dimmelmeier}}]{Muller2010A-New-Multi-dim}%
  \BibitemOpen
  \bibfield  {author} {\bibinfo {author} {\bibfnamefont {B.}~\bibnamefont
  {M{\"u}ller}}, \bibinfo {author} {\bibfnamefont {H.-T.}\ \bibnamefont
  {Janka}}, \ and\ \bibinfo {author} {\bibfnamefont {H.}~\bibnamefont
  {Dimmelmeier}},\ }\href@noop {} {\bibfield  {journal} {\bibinfo  {journal}
  {\apjs}\ }\textbf {\bibinfo {volume} {189}},\ \bibinfo {pages} {104}
  (\bibinfo {year} {2010})}\BibitemShut {NoStop}%
\bibitem [{\citenamefont {Endeve}\ \emph {et~al.}(2012)\citenamefont {Endeve},
  \citenamefont {{Cardall}},\ and\ \citenamefont
  {{Mezzacappa}}}]{Endeve2012Conservative-Mu}%
  \BibitemOpen
  \bibfield  {author} {\bibinfo {author} {\bibfnamefont {E.}~\bibnamefont
  {Endeve}}, \bibinfo {author} {\bibfnamefont {C.~Y.}\ \bibnamefont
  {{Cardall}}}, \ and\ \bibinfo {author} {\bibfnamefont {A.}~\bibnamefont
  {{Mezzacappa}}},\ }\href@noop {} {\bibfield  {journal} {\bibinfo  {journal}
  {{ArXiv e-prints}}\ ,\ \bibinfo {pages} {1212.4064}} (\bibinfo {year}
  {2012})}\BibitemShut {NoStop}%
\bibitem [{\citenamefont {{Hanke}}\ \emph {et~al.}(2013)\citenamefont
  {{Hanke}}, \citenamefont {{Mueller}}, \citenamefont {{Wongwathanarat}},
  \citenamefont {{Marek}},\ and\ \citenamefont
  {{Janka}}}]{Hanke2013SASI-Activity-i}%
  \BibitemOpen
  \bibfield  {author} {\bibinfo {author} {\bibfnamefont {F.}~\bibnamefont
  {{Hanke}}}, \bibinfo {author} {\bibfnamefont {B.}~\bibnamefont {{Mueller}}},
  \bibinfo {author} {\bibfnamefont {A.}~\bibnamefont {{Wongwathanarat}}},
  \bibinfo {author} {\bibfnamefont {A.}~\bibnamefont {{Marek}}}, \ and\
  \bibinfo {author} {\bibfnamefont {H.-T.}\ \bibnamefont {{Janka}}},\
  }\href@noop {} {\bibfield  {journal} {\bibinfo  {journal} {{ArXiv e-prints}}\
  ,\ \bibinfo {pages} {1303.6269}} (\bibinfo {year} {2013})}\BibitemShut
  {NoStop}%
\bibitem [{\citenamefont {{Takiwaki}}\ \emph {et~al.}(2012)\citenamefont
  {{Takiwaki}}, \citenamefont {{Kotake}},\ and\ \citenamefont
  {{Suwa}}}]{Takiwaki2012Three-dimension}%
  \BibitemOpen
  \bibfield  {author} {\bibinfo {author} {\bibfnamefont {T.}~\bibnamefont
  {{Takiwaki}}}, \bibinfo {author} {\bibfnamefont {K.}~\bibnamefont
  {{Kotake}}}, \ and\ \bibinfo {author} {\bibfnamefont {Y.}~\bibnamefont
  {{Suwa}}},\ }\href@noop {} {\bibfield  {journal} {\bibinfo  {journal} {\apj}\
  }\textbf {\bibinfo {volume} {749}},\ \bibinfo {pages} {98} (\bibinfo {year}
  {2012})}\BibitemShut {NoStop}%
\bibitem [{\citenamefont {{Shibata}}\ \emph {et~al.}(2011)\citenamefont
  {{Shibata}}, \citenamefont {{Kiuchi}}, \citenamefont {{Sekiguchi}},\ and\
  \citenamefont {{Suwa}}}]{Shibata2011Truncated-Momen}%
  \BibitemOpen
  \bibfield  {author} {\bibinfo {author} {\bibfnamefont {M.}~\bibnamefont
  {{Shibata}}}, \bibinfo {author} {\bibfnamefont {K.}~\bibnamefont {{Kiuchi}}},
  \bibinfo {author} {\bibfnamefont {Y.}~\bibnamefont {{Sekiguchi}}}, \ and\
  \bibinfo {author} {\bibfnamefont {Y.}~\bibnamefont {{Suwa}}},\ }\href@noop {}
  {\bibfield  {journal} {\bibinfo  {journal} {Prog. Theor. Phys.}\ }\textbf
  {\bibinfo {volume} {125}},\ \bibinfo {pages} {1255} (\bibinfo {year}
  {2011})}\BibitemShut {NoStop}%
\bibitem [{\citenamefont {{Kuroda}}\ \emph {et~al.}(2012)\citenamefont
  {{Kuroda}}, \citenamefont {{Kotake}},\ and\ \citenamefont
  {{Takiwaki}}}]{Kuroda2012Fully-General-R}%
  \BibitemOpen
  \bibfield  {author} {\bibinfo {author} {\bibfnamefont {T.}~\bibnamefont
  {{Kuroda}}}, \bibinfo {author} {\bibfnamefont {K.}~\bibnamefont {{Kotake}}},
  \ and\ \bibinfo {author} {\bibfnamefont {T.}~\bibnamefont {{Takiwaki}}},\
  }\href@noop {} {\bibfield  {journal} {\bibinfo  {journal} {\apj}\ }\textbf
  {\bibinfo {volume} {755}},\ \bibinfo {pages} {11} (\bibinfo {year}
  {2012})}\BibitemShut {NoStop}%
\bibitem [{\citenamefont {{Cardall}}\ \emph {et~al.}(2013)\citenamefont
  {{Cardall}}, \citenamefont {{Endeve}},\ and\ \citenamefont
  {{Mezzacappa}}}]{Cardall2012Conservative-31}%
  \BibitemOpen
  \bibfield  {author} {\bibinfo {author} {\bibfnamefont {C.~Y.}\ \bibnamefont
  {{Cardall}}}, \bibinfo {author} {\bibfnamefont {E.}~\bibnamefont {{Endeve}}},
  \ and\ \bibinfo {author} {\bibfnamefont {A.}~\bibnamefont {{Mezzacappa}}},\
  }\href@noop {} {\bibfield  {journal} {\bibinfo  {journal} {\prd}\ }\textbf
  {\bibinfo {volume} {87}},\ \bibinfo {pages} {103004} (\bibinfo {year}
  {2013})}\BibitemShut {NoStop}%
\bibitem [{\citenamefont {Ehlers}(1971)}]{Ehlers1971General-Relativ}%
  \BibitemOpen
  \bibfield  {author} {\bibinfo {author} {\bibfnamefont {J.}~\bibnamefont
  {Ehlers}},\ }in\ \href@noop {} {\emph {\bibinfo {booktitle} {Proceedings of
  the International School of Physics ``Enrico Fermi'' Course XLVII: General
  Relativity and Cosmology}}},\ \bibinfo {editor} {edited by\ \bibinfo {editor}
  {\bibfnamefont {R.~K.}\ \bibnamefont {Sachs}}}\ (\bibinfo  {publisher}
  {Academic Press},\ \bibinfo {address} {New York},\ \bibinfo {year} {1971})\
  pp.\ \bibinfo {pages} {1--70}\BibitemShut {NoStop}%
\bibitem [{\citenamefont {Israel}(1972)}]{Israel1972The-Relativisti}%
  \BibitemOpen
  \bibfield  {author} {\bibinfo {author} {\bibfnamefont {W.}~\bibnamefont
  {Israel}},\ }in\ \href@noop {} {\emph {\bibinfo {booktitle} {General
  Relativity: Papers in Honour of J. L. Synge}}},\ \bibinfo {editor} {edited
  by\ \bibinfo {editor} {\bibfnamefont {L.}~\bibnamefont {O'Raifeartaigh}}}\
  (\bibinfo  {publisher} {Clarendon},\ \bibinfo {address} {Oxford},\ \bibinfo
  {year} {1972})\ pp.\ \bibinfo {pages} {201--241}\BibitemShut {NoStop}%
\bibitem [{Note2()}]{Note2}%
  \BibitemOpen
  \bibinfo {note} {This expression does not at first appear symmetric in
  $\protect \mathaccentV {tilde}07E\jmath $ and $\protect \mathaccentV
  {tilde}07Ek$ as expected of ${\Pi ^{\protect \mathaccentV {tilde}07E\imath
  }}_{\protect \mathaccentV {tilde}07E\jmath \protect \mathaccentV
  {tilde}07Ek}$; but note that when the curvilinear $p^{\protect \mathaccentV
  {tilde}07E\imath }$ are regarded as the independent variables, ${P^{\protect
  \mathaccentV {hat}05Ek}}_{\protect \mathaccentV {tilde}07Ek} \left ( \partial
  {P^{\protect \mathaccentV {hat}05E\imath }}_{\protect \mathaccentV
  {tilde}07E{\jmath }} / \partial p^{\protect \mathaccentV {hat}05Ek}\right ) =
  \partial {P^{\protect \mathaccentV {hat}05E\imath }}_{\protect \mathaccentV
  {tilde}07E{\jmath }} / \partial p^{\protect \mathaccentV {tilde}07Ek} =
  \partial ^2 p^{\protect \mathaccentV {hat}05E\imath } / \partial p^{\protect
  \mathaccentV {tilde}07Ek}\partial p^{\protect \mathaccentV {tilde}07E\jmath
  }$}\BibitemShut {NoStop}%
\bibitem [{\citenamefont {{York}}(1983)}]{York1983The-initial-val}%
  \BibitemOpen
  \bibfield  {author} {\bibinfo {author} {\bibfnamefont {J.~W.}\ \bibnamefont
  {{York}}, \bibfnamefont {Jr.}},\ }in\ \href@noop {} {\emph {\bibinfo
  {booktitle} {Gravitational Radiation}}},\ \bibinfo {editor} {edited by\
  \bibinfo {editor} {\bibfnamefont {N.}~\bibnamefont {{Deruelle}}}\ and\
  \bibinfo {editor} {\bibfnamefont {T.}~\bibnamefont {{Piran}}}}\ (\bibinfo
  {publisher} {North-Holland},\ \bibinfo {address} {Amsterdam},\ \bibinfo
  {year} {1983})\ pp.\ \bibinfo {pages} {175--201}\BibitemShut {NoStop}%
\bibitem [{\citenamefont
  {{Gourgoulhon}}(2007)}]{Gourgoulhon200731-Formalism-an}%
  \BibitemOpen
  \bibfield  {author} {\bibinfo {author} {\bibfnamefont {E.}~\bibnamefont
  {{Gourgoulhon}}},\ }\href@noop {} {\bibfield  {journal} {\bibinfo  {journal}
  {arXiv:gr-qc/0703035}\ } (\bibinfo {year} {2007})}\BibitemShut {NoStop}%
\bibitem [{\citenamefont {{Alcubierre}}(2008)}]{Alcubierre2008Introduction-to}%
  \BibitemOpen
  \bibfield  {author} {\bibinfo {author} {\bibfnamefont {M.}~\bibnamefont
  {{Alcubierre}}},\ }\href@noop {} {\emph {\bibinfo {title} {{Introduction to
  3+1 Numerical Relativity}}}}\ (\bibinfo  {publisher} {Oxford University
  Press},\ \bibinfo {year} {2008})\BibitemShut {NoStop}%
\bibitem [{\citenamefont {{Baumgarte}}\ and\ \citenamefont
  {{Shapiro}}(2010)}]{Baumgarte2010Numerical-Relat}%
  \BibitemOpen
  \bibfield  {author} {\bibinfo {author} {\bibfnamefont {T.~W.}\ \bibnamefont
  {{Baumgarte}}}\ and\ \bibinfo {author} {\bibfnamefont {S.~L.}\ \bibnamefont
  {{Shapiro}}},\ }\href@noop {} {\emph {\bibinfo {title} {{Numerical
  Relativity: Solving Einstein's Equations on the Computer}}}}\ (\bibinfo
  {publisher} {Cambridge University Press},\ \bibinfo {year}
  {2010})\BibitemShut {NoStop}%
\bibitem [{\citenamefont {{Landau}}\ and\ \citenamefont
  {{Lifshitz}}(1975)}]{Landau1975The-classical-t}%
  \BibitemOpen
  \bibfield  {author} {\bibinfo {author} {\bibfnamefont {L.~D.}\ \bibnamefont
  {{Landau}}}\ and\ \bibinfo {author} {\bibfnamefont {E.~M.}\ \bibnamefont
  {{Lifshitz}}},\ }\href@noop {} {\emph {\bibinfo {title} {{The classical
  theory of fields}}}}\ (\bibinfo  {publisher} {Oxford: Pergamon Press},\
  \bibinfo {year} {1975})\BibitemShut {NoStop}%
\bibitem [{\citenamefont {Zhang}\ and\ \citenamefont
  {Shu}(2010)}]{Zhang2010On-maximum-prin}%
  \BibitemOpen
  \bibfield  {author} {\bibinfo {author} {\bibfnamefont {X.}~\bibnamefont
  {Zhang}}\ and\ \bibinfo {author} {\bibfnamefont {C.-W.}\ \bibnamefont
  {Shu}},\ }\href@noop {} {\bibfield  {journal} {\bibinfo  {journal} {J. Comp.
  Phys.}\ }\textbf {\bibinfo {volume} {229}},\ \bibinfo {pages} {3091}
  (\bibinfo {year} {2010})}\BibitemShut {NoStop}%
\bibitem [{\citenamefont {Dasgupta}(2010)}]{Dasgupta2010Physics-and-Ast}%
  \BibitemOpen
  \bibfield  {author} {\bibinfo {author} {\bibfnamefont {B.}~\bibnamefont
  {Dasgupta}},\ }\href@noop {} {\bibfield  {journal} {\bibinfo  {journal}
  {Proc. of Sci.}\ }\textbf {\bibinfo {volume} {ICHEP 2010}},\ \bibinfo {pages}
  {294} (\bibinfo {year} {2010})}\BibitemShut {NoStop}%
\bibitem [{\citenamefont {Raffelt}(2011)}]{Raffelt2011New-opportuniti}%
  \BibitemOpen
  \bibfield  {author} {\bibinfo {author} {\bibfnamefont {G.~G.}\ \bibnamefont
  {Raffelt}},\ }\href@noop {} {\bibfield  {journal} {\bibinfo  {journal} {Nucl.
  Phys. B (Proc. Suppl.)}\ }\textbf {\bibinfo {volume} {217}},\ \bibinfo
  {pages} {95} (\bibinfo {year} {2011})}\BibitemShut {NoStop}%
\bibitem [{\citenamefont {Dighe}(2011)}]{Dighe2011Signatures-of-s}%
  \BibitemOpen
  \bibfield  {author} {\bibinfo {author} {\bibfnamefont {A.}~\bibnamefont
  {Dighe}},\ }in\ \href@noop {} {\emph {\bibinfo {booktitle} {Proceedings of
  the Hamburg Neutrinos from Supernova Explosions, HA$\nu$SE 2011, July 19-23,
  2011, Hamburg, Germany}}},\ \bibinfo {series and number} {DESY Proceedings
  Series},\ \bibinfo {editor} {edited by\ \bibinfo {editor} {\bibfnamefont
  {A.}~\bibnamefont {Mirizzi}}, \bibinfo {editor} {\bibfnamefont {P.~D.}\
  \bibnamefont {Serpico}}, \ and\ \bibinfo {editor} {\bibfnamefont
  {G.}~\bibnamefont {Sigl}}}\ (\bibinfo  {publisher} {Verlag Deutsches
  Elektronen-Synchrotron},\ \bibinfo {address} {Hamburg},\ \bibinfo {year}
  {2011})\ pp.\ \bibinfo {pages} {101--112}\BibitemShut {NoStop}%
\bibitem [{\citenamefont {{Chakraborty}}\ \emph {et~al.}(2011)\citenamefont
  {{Chakraborty}}, \citenamefont {{Fischer}}, \citenamefont {{Mirizzi}},
  \citenamefont {{Saviano}},\ and\ \citenamefont
  {{Tom{\`a}s}}}]{Chakraborty2011No-Collective-N}%
  \BibitemOpen
  \bibfield  {author} {\bibinfo {author} {\bibfnamefont {S.}~\bibnamefont
  {{Chakraborty}}}, \bibinfo {author} {\bibfnamefont {T.}~\bibnamefont
  {{Fischer}}}, \bibinfo {author} {\bibfnamefont {A.}~\bibnamefont
  {{Mirizzi}}}, \bibinfo {author} {\bibfnamefont {N.}~\bibnamefont
  {{Saviano}}}, \ and\ \bibinfo {author} {\bibfnamefont {R.}~\bibnamefont
  {{Tom{\`a}s}}},\ }\href@noop {} {\bibfield  {journal} {\bibinfo  {journal}
  {\prl}\ }\textbf {\bibinfo {volume} {107}},\ \bibinfo {pages} {151101}
  (\bibinfo {year} {2011})}\BibitemShut {NoStop}%
\bibitem [{\citenamefont {{Suwa}}\ \emph {et~al.}(2011)\citenamefont {{Suwa}},
  \citenamefont {{Kotake}}, \citenamefont {{Takiwaki}}, \citenamefont
  {{Liebend{\"o}rfer}},\ and\ \citenamefont
  {{Sato}}}]{Suwa2011Impacts-of-Coll}%
  \BibitemOpen
  \bibfield  {author} {\bibinfo {author} {\bibfnamefont {Y.}~\bibnamefont
  {{Suwa}}}, \bibinfo {author} {\bibfnamefont {K.}~\bibnamefont {{Kotake}}},
  \bibinfo {author} {\bibfnamefont {T.}~\bibnamefont {{Takiwaki}}}, \bibinfo
  {author} {\bibfnamefont {M.}~\bibnamefont {{Liebend{\"o}rfer}}}, \ and\
  \bibinfo {author} {\bibfnamefont {K.}~\bibnamefont {{Sato}}},\ }\href@noop {}
  {\bibfield  {journal} {\bibinfo  {journal} {\apj}\ }\textbf {\bibinfo
  {volume} {738}},\ \bibinfo {pages} {165} (\bibinfo {year}
  {2011})}\BibitemShut {NoStop}%
\bibitem [{\citenamefont {{Dasgupta}}\ \emph {et~al.}(2012)\citenamefont
  {{Dasgupta}}, \citenamefont {{O'Connor}},\ and\ \citenamefont
  {{Ott}}}]{Dasgupta2012Role-of-collect}%
  \BibitemOpen
  \bibfield  {author} {\bibinfo {author} {\bibfnamefont {B.}~\bibnamefont
  {{Dasgupta}}}, \bibinfo {author} {\bibfnamefont {E.~P.}\ \bibnamefont
  {{O'Connor}}}, \ and\ \bibinfo {author} {\bibfnamefont {C.~D.}\ \bibnamefont
  {{Ott}}},\ }\href@noop {} {\bibfield  {journal} {\bibinfo  {journal} {\prd}\
  }\textbf {\bibinfo {volume} {85}},\ \bibinfo {pages} {065008} (\bibinfo
  {year} {2012})}\BibitemShut {NoStop}%
\bibitem [{\citenamefont {Sarikas}\ \emph
  {et~al.}(2012{\natexlab{a}})\citenamefont {Sarikas}, \citenamefont {Raffelt},
  \citenamefont {H\"udepohl},\ and\ \citenamefont
  {Janka}}]{Sarikas2012Suppression-of-}%
  \BibitemOpen
  \bibfield  {author} {\bibinfo {author} {\bibfnamefont {S.}~\bibnamefont
  {Sarikas}}, \bibinfo {author} {\bibfnamefont {G.~G.}\ \bibnamefont
  {Raffelt}}, \bibinfo {author} {\bibfnamefont {L.}~\bibnamefont {H\"udepohl}},
  \ and\ \bibinfo {author} {\bibfnamefont {H.-T.}\ \bibnamefont {Janka}},\
  }\href@noop {} {\bibfield  {journal} {\bibinfo  {journal} {\prl}\ }\textbf
  {\bibinfo {volume} {108}},\ \bibinfo {pages} {061101} (\bibinfo {year}
  {2012}{\natexlab{a}})}\BibitemShut {NoStop}%
\bibitem [{\citenamefont {{Saviano}}\ \emph {et~al.}(2012)\citenamefont
  {{Saviano}}, \citenamefont {{Chakraborty}}, \citenamefont {{Fischer}},\ and\
  \citenamefont {{Mirizzi}}}]{Saviano2012Stability-analy}%
  \BibitemOpen
  \bibfield  {author} {\bibinfo {author} {\bibfnamefont {N.}~\bibnamefont
  {{Saviano}}}, \bibinfo {author} {\bibfnamefont {S.}~\bibnamefont
  {{Chakraborty}}}, \bibinfo {author} {\bibfnamefont {T.}~\bibnamefont
  {{Fischer}}}, \ and\ \bibinfo {author} {\bibfnamefont {A.}~\bibnamefont
  {{Mirizzi}}},\ }\href@noop {} {\bibfield  {journal} {\bibinfo  {journal}
  {\prd}\ }\textbf {\bibinfo {volume} {85}},\ \bibinfo {pages} {113002}
  (\bibinfo {year} {2012})}\BibitemShut {NoStop}%
\bibitem [{\citenamefont {Sarikas}\ \emph
  {et~al.}(2012{\natexlab{b}})\citenamefont {Sarikas}, \citenamefont
  {Tamborra}, \citenamefont {Raffelt}, \citenamefont {H\"udepohl},\ and\
  \citenamefont {Janka}}]{Sarikas2012Supernova-neutr}%
  \BibitemOpen
  \bibfield  {author} {\bibinfo {author} {\bibfnamefont {S.}~\bibnamefont
  {Sarikas}}, \bibinfo {author} {\bibfnamefont {I.}~\bibnamefont {Tamborra}},
  \bibinfo {author} {\bibfnamefont {G.}~\bibnamefont {Raffelt}}, \bibinfo
  {author} {\bibfnamefont {L.}~\bibnamefont {H\"udepohl}}, \ and\ \bibinfo
  {author} {\bibfnamefont {H.-T.}\ \bibnamefont {Janka}},\ }\href@noop {}
  {\bibfield  {journal} {\bibinfo  {journal} {\prd}\ }\textbf {\bibinfo
  {volume} {85}},\ \bibinfo {pages} {113007} (\bibinfo {year}
  {2012}{\natexlab{b}})}\BibitemShut {NoStop}%
\bibitem [{\citenamefont {{Pejcha}}\ \emph {et~al.}(2012)\citenamefont
  {{Pejcha}}, \citenamefont {{Dasgupta}},\ and\ \citenamefont
  {{Thompson}}}]{Pejcha2012Effect-of-colle}%
  \BibitemOpen
  \bibfield  {author} {\bibinfo {author} {\bibfnamefont {O.}~\bibnamefont
  {{Pejcha}}}, \bibinfo {author} {\bibfnamefont {B.}~\bibnamefont
  {{Dasgupta}}}, \ and\ \bibinfo {author} {\bibfnamefont {T.~A.}\ \bibnamefont
  {{Thompson}}},\ }\href@noop {} {\bibfield  {journal} {\bibinfo  {journal}
  {\mnras}\ }\textbf {\bibinfo {volume} {425}},\ \bibinfo {pages} {1083}
  (\bibinfo {year} {2012})}\BibitemShut {NoStop}%
\end{thebibliography}

%

\end{document}